\documentclass[iop,twocolappendix]{emulateapj}

\usepackage{natbib}
\usepackage{graphicx}
\usepackage{multirow}
\usepackage{mathtools}
\usepackage{makecell}
\usepackage{rotating}

\usepackage{hyperref}
\hypersetup{
    colorlinks=true,
    citecolor=blue,
    linkcolor=blue,
    filecolor=magenta,      
    urlcolor=cyan,
}

\newcommand{\msun}{\mbox{M}_{\odot}}
\newcommand{\mstar}{$\mbox{M}_*$}

\newcommand{\hii}{H~\textsc{ii}}
\newcommand{\one}{~\textsc{i}}
\newcommand{\ii}{~\textsc{ii}}
\newcommand{\iii}{~\textsc{iii}}

\newcommand{\ebvgas}{E(B-V)$_{\text{gas}}$}
\newcommand{\ebvstars}{E(B-V)$_{\text{stars}}$}
\newcommand{\betauv}{$\beta_{\text{UV}}$}
\newcommand{\mto}{$M_{\text{TO}}$}

\newcommand{\mugas}{$\mu_{\text{gas}}$}
\newcommand{\zin}{$Z_{\text{in}}$}
\newcommand{\zout}{$Z_{\text{out}}$}
\newcommand{\zism}{$Z_{\text{ISM}}$}
\newcommand{\zetain}{$\zeta_{\text{in}}$}
\newcommand{\zetaout}{$\zeta_{\text{out}}$}
\newcommand{\etain}{$\eta_{\text{in}}$}
\newcommand{\etaout}{$\eta_{\text{out}}$}
\newcommand{\mdotin}{$\dot{M}_{\text{in}}$}
\newcommand{\mdotout}{$\dot{M}_{\text{out}}$}

\newcommand{\vcirc}{$v_{\text{circ}}$}
\newcommand{\etaoutten}{$\eta_{\text{out,10}}$}

\shorttitle{The Mass-Metallicity Relation at $z=0-3.3$}
\shortauthors{Sanders et al.}

\begin{document}

\title{The MOSDEF Survey: The Evolution of the Mass-Metallicity Relation from $\lowercase{z}=0$ to $\lowercase{z}\sim3.3$\altaffilmark{*}}
\altaffiltext{*}{Based on data obtained at the
W.M. Keck Observatory, which is operated as a scientific partnership among the California Institute of Technology, the
University of California, and NASA, and was made possible by the generous financial support of the W.M. Keck Foundation.
}

\author{Ryan L. Sanders\altaffilmark{1,2}} \altaffiltext{1}{Department of Physics and Astronomy, University of California, Davis, One Shields Ave, Davis, CA 95616, USA} \altaffiltext{2}{Hubble Fellow}

\author{Alice E. Shapley\altaffilmark{3}} \altaffiltext{3}{Department of Physics \& Astronomy, University of California, Los Angeles, 430 Portola Plaza, Los Angeles, CA 90095, USA}

\author{Tucker Jones\altaffilmark{1}}

\author{Naveen A. Reddy\altaffilmark{4,5}} \altaffiltext{4}{Department of Physics \& Astronomy, University of California, Riverside, 900 University Avenue, Riverside, CA 92521, USA} \altaffiltext{5}{Alfred P. Sloan Fellow}

\author{Mariska Kriek\altaffilmark{6}} \altaffiltext{6}{Astronomy Department, University of California, Berkeley, CA 94720, USA}

\author{Brian Siana\altaffilmark{4}}

\author{Alison L. Coil\altaffilmark{7}} \altaffiltext{7}{Center for Astrophysics and Space Sciences, University of California, San Diego, 9500 Gilman Dr., La Jolla, CA 92093-0424, USA}

\author{Bahram Mobasher\altaffilmark{4}}

\author{Irene Shivaei\altaffilmark{8,2}} \altaffiltext{8}{Department of Astronomy/Steward Observatory, 933 North Cherry Ave, Rm N204, Tucson, AZ, 85721-0065, USA}

\author{Romeel Dav\'{e}\altaffilmark{9}} \altaffiltext{9}{Institute for Astronomy, Unversity of Edinburgh, James Clerk Maxwell Building, Peter Guthrie Tait Road, Edinburgh, EH9 3FD, United Kingdom}

\author{Mojegan Azadi\altaffilmark{10}} \altaffiltext{10}{Harvard-Smithsonian Center for Astrophysics, 60 Garden Street, Cambridge, MA 02138, USA}

\author{Sedona H. Price\altaffilmark{11}} \altaffiltext{11}{Max-Planck-Institut f{\"u}r extraterrestrische Physik, Postfach 1312, Garching, 85741, Germany}

\author{Gene Leung\altaffilmark{7}}

\author{William R. Freeman\altaffilmark{4}}

\author{Tara Fetherolf\altaffilmark{4}}

\author{Laura de Groot\altaffilmark{12}} \altaffiltext{12}{Department of Physics, The College of Wooster, 1189 Beall Avenue, Wooster, OH 44691, USA}

\author{Tom Zick\altaffilmark{6}}

\author{Guillermo Barro\altaffilmark{13}} \altaffiltext{13}{Department of Physics, University of the Pacific, 3601 Pacific Ave, Stockton, CA 95211, USA}

\email{email: rlsand@ucdavis.edu}

\begin{abstract}
We investigate the evolution of galaxy gas-phase metallicity (O/H) over the range $z=0-3.3$ using samples of $\sim300$ galaxies at $z\sim2.3$
 and $\sim150$ galaxies at $z\sim3.3$ from the MOSDEF survey.
This analysis crucially utilizes different metallicity calibrations at $z\sim0$ and $z>1$ to account for evolving ISM conditions.
We find significant correlations between O/H and stellar mass (\mstar) at $z\sim2.3$ and $z\sim3.3$.
The low-mass power law slope of the mass-metallicity relation
 is remarkably invariant over $z=0-3.3$, such that O/H$\propto$$M_*^{0.30}$ at all redshifts in this range.
At fixed \mstar, O/H decreases with increasing redshift as dlog(O/H)/d$z=-0.11\pm0.02$.
We find no evidence that the fundamental metallicity relation between \mstar, O/H, and star-formation rate (SFR)
 evolves out to $z\sim3.3$.
We employ analytic chemical evolution models to place constraints on the mass and metal loading factors of galactic outflows.
The efficiency of metal removal increases toward lower \mstar\ at fixed redshift, and toward higher redshift at fixed \mstar.
These models suggest that the slope of the mass-metallicity relation is primarily set by the scaling of the outflow metal loading factor with \mstar,
 not by the change in gas fraction as a function of \mstar.
The evolution toward lower O/H at fixed \mstar\ with increasing redshift is driven by both higher gas fraction
 (leading to stronger dilution of ISM metals) and higher metal removal efficiency.
These results suggest that the processes governing the smooth baryonic growth of galaxies via gas flows and star formation
 hold in the same form over at least the past 12~Gyr.
\end{abstract}




\section{Introduction}\label{sec:intro}

The metallicity of the interstellar medium (ISM) of galaxies is a powerful tool with which to understand the
 baryonic processes that govern the secular growth of galaxies.
Gas-phase metallicity is closely related to past and current star formation (the nucleosynthetic origin of metals),
 the gas reservoir, and gas flows including accretion from the intergalactic medium (IGM) and circumgalactic medium (CGM),
 outflows driven by feedback from supernovae (SNe) and accreting black holes, and recycling of material
 from past outflows.
Characterizing how metallicity scales with global galaxy properties including stellar mass (\mstar)
 and star-formation rate (SFR) over a range of redshifts can constrain the scaling of gas accretion
 and outflow rates with these properties, providing insight into galaxy growth throughout cosmic history.

The relation between the gas-phase oxygen abundance (O/H) and \mstar, referred to as the mass-metallicity relation (MZR),
 has been extensively studied in the
 local universe, where O/H and \mstar\ are found to be positively correlated over five decades in \mstar\
 \citep[e.g.,][]{leq79,tre04,lee06,kew08,man10,ber12,and13,bla19,cur20b}.
At $z=0$, the MZR is generally described by a power law at low masses ($\lesssim10^{10}~\msun$) that begins
 to flatten toward an aysmptotic value in metallicity at high masses.

The MZR has been observed out to $z\sim3.5$ and evolves such that O/H decreases with increasing redshift
 at fixed \mstar\ \citep{sav05,erb06,mai08,man09,zah11,zah14a,zah14b,wuy12,wuy16,bel13,kul13,hen13,cul14,mai14,ste14,tro14,kap15,kap16,ly15,ly16,san15,san18,san20,hun16a,ono16,suz17}.
At $z>4$, the commonly used rest-optical metallicity indicators redshift out of atmospheric transmission windows
 that are accessible from the ground (i.e., beyond 2.5~$\mu$m).
The metallicity of only a single galaxy at $z=4.4$ has been measured from a rest-optical line ratio \citep{sha17}.
Attempts have been made to constrain the MZR at redshifts above $z=4$ through rest-UV metal absorption lines at $z=4-5$ \citep{fai16}
 and far-IR [O\iii]~88~$\mu$m emission detected with ALMA at $z\sim8$ \citep{jon20},
 but these methods currently suffer from large systematic uncertainties in metallicity, precluding useful comparisons
 with results from rest-optical line ratios at lower redshift.
A robust picture of MZR evolution at $z>4$ will require spectra from the James Webb Space Telescope (\textit{JWST})
 that can access wavelengths beyond 2.5~$\mu$m.

In past studies, the evolution of the MZR has been found to be slow out to $z\sim2.5$ where O/H is $\sim0.3$~dex lower than
 at $z\sim0$ at fixed \mstar\ \citep[e.g.,][]{erb06,ste14,san15}.
Rapid metallicity evolution has been inferred above $z\sim3$, with metallicity dropping $0.3-0.4$~dex between $z\sim2.5$
 and $z\sim3.5$ despite only 1~Gyr of cosmic time passing between these redshifts \citep{mai08,man09,tro14,ono16}.
Such fast evolution between $z\sim2.5$ and $z\sim3.5$ is not observed in numerical simulations of galaxy formation and evolution,
 which instead find a smooth decline in metallicity at fixed \mstar\ out to $z\sim6$ \citep[e.g.,][]{ma16,dav17,der17,tor19}.
There is thus tension between previous constraints on the MZR at $z>3$ and models of hierarchical galaxy formation.
\citet{suz17} found very little MZR evolution between $z\sim2$ and $z\sim3.2$, but their comparison relies on metallicities derived using
 different indicators and calibrations at each redshift and the accompanying systematic effects are unclear.
 
The $z=0$ MZR has been found to have a secondary dependence on SFR such that there is a three parameter relation
 among \mstar, SFR, and O/H, known as the fundamental metallicity relation \citep[FMR; e.g.,][]{ell08,man10,lar10,yat12,cre19,cur20b}.
In the FMR, O/H decreases with increasing SFR at fixed \mstar.
The FMR is closely connected to a relation among \mstar, O/H, and gas fraction in which O/H and gas fraction are
 anti-correlated at fixed \mstar\ \citep{bot13,bot16a,bot16b,bro18}.
The FMR was proposed to be independent of redshift out to $z\sim2.5$ \citep{man10}.
Due to small samples, low-S/N measurements, and biases in metallicity estimates, early work yielded inconclusive
 results regarding the redshift invariance of the FMR and whether the high-redshift MZR displayed any secondary dependence
 on SFR \citep[e.g.,][]{wuy12,wuy14,bel13,sto13,ste14,cul14,zah14b,sal15,san15,yab15,gra16,kas17}.
With improved data sets, recent work has found that the MZR does depend secondarily on SFR at $z\sim2.3$ \citep{san18}
 and the FMR holds out to $z\sim2.5$, though a small offset of $\sim0.1$ dex from the local FMR is seen in some studies
 \citep{san18,cre19,cur20b}. 
Galaxies at $z>3$ do not appear to follow the FMR, with metallicities $\sim0.3-0.6$~dex below the metallicity predicted
 by the local relation \citep{tro14,ono16}.

A class of chemical evolution models known as ``bathtub'' or ``equilibrium'' models has shown success in reproducing
 the observed MZR and its evolution, as well as the FMR \citep[e.g.,][]{fin08,pee11,dav12,lil13}.
These models operate on the principle of conservation
 of baryonic mass in galaxies, establishing a balance between the mass inflow and outflow rates, SFR, rate of returning stellar
 material back into the ISM, and rate of change of the total gas mass (some models assume the latter quantity is negligible;
 \citealt{dav12}).
Galaxies satisfying this balance between gas flows and internal gas processing are said to be in equilibrium.
In this theoretical framework, the MZR arises because gas fractions are higher and/or material is more efficiently removed
 by outflows at lower \mstar\ \citep{tre04,dav12,lil13}.
Other secondary effects may come into play as well, including variations with \mstar\ of the stellar initial mass function (IMF)
 that affect metal yields \citep{kop07} and the metallicity of accreted gas through galactic fountains and outflow recycling
 \citep{dav11,ang17}.
These equilibrium models provide a way to utilize MZR and FMR observations to constrain gas accretion and outflow rates.

The shape and normalization of the MZR and FMR are sensitive to the method used to derive metallicities.
Given the difficulty of measuring faint O recombination lines or auroral emission lines (e.g., [O\iii]$\lambda$4363)
 that are required to employ the most robust metallicity derivation techniques,
 the use of calibrations between ratios of strong emission lines and metallicity is the most practical approach to
 measure metallicity scaling relations for large and representative samples spanning wide ranges in \mstar\ and SFR
 \citep[e.g.,][]{kew02,pet04,mai08,cur17}.
\citet{kew08} showed that the form of the $z\sim0$ mass-metallicity relation varies widely in both high-mass asymptotic O/H
 and low-mass slope based on the choice of strong-line metallicity calibration.
A robust translation between strong-line ratio and O/H is therefore critical to any analysis of metallicity scaling relations.

The problem of calibration choice is further complicated when investigating the evolution of the MZR and FMR over a wide range
 of redshifts.
Star-forming galaxies at $z\sim2$ have been shown to follow different excitation sequences from those of their $z\sim0$ counterparts
 and local \hii\ regions, most notably in the [N\ii] BPT diagram \citep[e.g.,][]{ste14,ste16,sha15,san16a,kas17,str17,str18,top20a,top20b,run20}.
There is a consensus that the excitation properties of $z>1$ galaxies signify that high-redshift \hii\ regions
 have a set of ionized gas physical properties that is distinct from those of $z=0$ \hii\ regions.
The relation between strong-line ratios and metallicity depends sensitively on these same physical properties \citep[e.g.,][]{kew13},
 thus it is probable that metallicity calibrations evolve with redshift.
Nevertheless, it remains the overwhelmingly common practice to apply $z=0$ metallicity calibrations to $z>1$ galaxies.
A robust analysis of the evolution of the MZR and FMR must take into account the evolution of metallicity calibrations
 accordingly by applying appropriate calibrations at each redshift.

In this work, we investigate the evolution of the MZR and FMR over $z=0-3.3$ using large samples of
 representative star-forming galaxies at $z\sim2.3$ and $z\sim3.3$ from the MOSDEF survey.
In addition to a significantly larger sample size at $z>3$, our analysis includes several key improvements over
 past studies, including more robust dust corrections for $z>3$ galaxies calibrated to Balmer decrement measurements at $z\sim2.3$,
 metallicities derived from a uniform set of emission lines that is the same for samples at all redshifts,
 and, for the first time, the application of different metallicity calibrations to samples in the local and high-redshift universe
 to reflect evolving ionized gas conditions in star-forming regions.
We combine our improved constraints on MZR evolution with analytic chemical evolution models to infer the roles of
 metal-enriched outflows and gas fractions in controlling the slope and evolution of the MZR.

This paper is organized as follows.
In Section~\ref{sec:data}, we describe the measurements, samples, and derived quantities.
We report the methods for deriving metallicities in Section~\ref{sec:metallicity}.
We characterize the MZR at $z\sim2.3$ and $z\sim3.3$ and investigate the evolution
 of the FMR in Section~\ref{sec:results}.
We interpret our results using analytic chemical evolution models in Section~\ref{sec:models}, placing constraints
 on the metal loading factor of outflows and investigating which physical mechanisms govern the slope and evolution
 of the MZR.
We discuss our results in Section~\ref{sec:discussion}, comparing to past high-redshift MZR and FMR studies
 and considering the implications of our models for the evolution of the outflow mass loading factor and its
 scaling with stellar mass.
Finally, in Section~\ref{sec:summary}, we summarize our conclusions.
Throughout, we assume a standard $\Lambda$CDM cosmology with H$_0$=70~km~s$^{-1}$~Mpc$^{-1}$, $\Omega_{\text{m}}$=0.3, and
 $\Omega_{\Lambda}$=0.7.
Magnitudes are in the AB system \citep{oke83} and wavelengths are given in air.
The term metallicity refers to the gas-phase oxygen abundance unless otherwise stated.

\section{Data, Measurements, \& Derived Quantities}\label{sec:data}

\subsection{The MOSDEF survey}

Our high-redshift galaxy samples are drawn from the MOSDEF survey, a 4-year program that used
 the Multi-Object Spectrometer For Infrared Exploration \citep[MOSFIRE;][]{mcl12} on the 10~m Keck~I
 telescope to obtain rest-frame optical spectra of galaxies at $z=1.4-3.8$ \citep{kri15}.
Galaxies were targeted in three redshift ranges: $1.37\leq z\leq1.70$, $2.09\leq z\leq2.61$, and $2.95\leq z\leq3.80$.
In these redshift intervals, strong rest-optical emission lines fall within windows of near-infrared atmospheric transmission.
Here, we focus on the higher two redshift bins.
At $z\sim2.3$ (3.3), [O\ii]$\lambda\lambda$3726,3729 and [Ne\iii]$\lambda$3869 fall in the $J$ ($H$) band;
 H$\beta$ and [O\iii]$\lambda\lambda$4959,5007 fall in the $H$ ($K$) band; and
 H$\alpha$, [N\ii]$\lambda\lambda$6548,6584, and [S\ii]$\lambda\lambda$6716,6731 fall in the $K$ band
 (these lines are not covered at $z\sim3.3$).
Targets were drawn from the 3D-HST survey photometric catalogs \citep{bra12a,ske14,mom16},
selected based on $H$-band (rest-frame optical) magnitude as measured from $HST$/WFC3 F160W imaging
 ($H_{\text{AB}}<24.5$ (25.0) at $z\sim2.3$ (3.3)) and redshift (spectroscopic or $HST$ grism when available,
 otherwise photometric).
The $H$-band magnitude limit corresponds to an approximate stellar mass limit of log($M_*/\msun)\sim9.0$
 that is constant across the three redshift bins.
The completed survey targeted $\sim1,500$ galaxies and measured $\sim1,300$ redshifts, with approximately
 half of the sample at $z\sim2.3$ and one quarter at $z\sim3.3$.
For a detailed description of the MOSDEF survey design and data reduction, see \citet{kri15}.

\subsection{Measurements and derived quantities}

\subsubsection{Emission lines fluxes and redshifts}

We utilize measurements of redshifts and emission line fluxes from extracted 1D science spectra that
 have been corrected for slit losses, as described in \citet{kri15}.
The absolute flux calibration of slit-loss corrected science spectra is accurate to better than
 18\% on average with a 16\% uncertainty, and the relative calibration between filters is
 biased less than 13\% with an uncertainty of 18\%.
The MOSDEF line measurements thus provide robust line ratios even when the lines fall in different filters
 (e.g., [O\iii]/[O\ii], H$\alpha$/H$\beta$) and total line fluxes for calculating SFRs.

\subsubsection{Stellar masses and emission-line corrected photometry}\label{sec:sedfitting}

Stellar masses were determined using the extensive broadband photometry in the CANDELS fields \citep{gro11,koe11}
 spanning observed-frame optical to mid-infrared (rest-frame UV to near-IR),
 as cataloged by the 3D-HST survey team \citep{ske14,mom16}.
Because galaxies at $z>2$ commonly have large emission-line equivalent widths \citep[EW$_{\text{obs}}\gtrsim300$~\AA;][]{red18},
 it is important to correct photometric measurements for the contribution from emission lines before fitting with stellar-continuum-only models.

Photometry in rest-optical filters was corrected using the following method.
For each MOSDEF target with a secure spectroscopic redshift and at least one emission line detected at S/N$\ge$3,
 a model emission-line-only spectrum was created by summing the best-fit Gaussian profiles of all emission lines
 with S/N$\ge$3.
This model spectrum was passed through the transmission curves of all filters covering the rest-frame optical to determine
 the flux contributed by emission lines in each filter, and this flux was subtracted from the original photometric measurements.
For each filter, if the difference between the original and corrected photometry was $>1\sigma$ based on the original photometric
 uncertainty, then the corrected photometry is used.  No correction is made if the difference is $<=1\sigma$.
Uncertainties on MOSDEF emission-line fluxes are propagated into uncertainties on corrected photometry.

Emission-line corrected photometry for each target was fit using flexible stellar population synthesis models
 \citep{con09} and the spectral energy distribution (SED) fitting code FAST \citep{kri09}.
Constant star-formation histories, solar stellar metallicities, the \citet{cal00} attenuation curve,
 and a \citet{cha03} initial mass function (IMF) are assumed for all galaxies in the sample.
We investigate the effects of varying the SED fitting assumptions on our results in Sec.~\ref{sec:mzrsys}.
This SED fitting procedure yields stellar masses, \ebvstars, SFR(SED), and a best-fit model of the stellar continuum.
Hydrogen Balmer recombination line fluxes are corrected for the effects of stellar Balmer absorption by
 measuring the absorption line flux from the best-fit SED model and applying a correction
 equal to the total absorption flux multiplied by an emission filling fraction of 0.36 (0.23) for
 H$\alpha$ (H$\beta$) \citep{red18}.
Typical Balmer absorption corrections are $\lesssim1\%$ ($\lesssim3\%$) for H$\alpha$ (H$\beta$).

\subsubsection{Reddening correction}\label{sec:reddening}

Dust-corrected line fluxes are required for both SFR and metallicity calculations.
When both H$\alpha$ and H$\beta$ are detected with S/N$\ge$3, \ebvgas\ is calculated using the Balmer decrement
 assuming an intrinsic ratio of H$\alpha$/H$\beta$=2.86 \citep{ost06} and the Milky Way extinction curve
 \citep{car89}.
A nebular attenuation curve derived directly from $z\sim2$ MOSDEF data is consistent with the Milky Way curve, suggesting
 this curve is an appropriate assumption (Reddy et al., submitted).
However, H$\alpha$ is not covered for galaxies at $z>2.65$ and H$\beta$ is not always detected for galaxies in the $z\sim2.3$ bin.
An alternative dust correction method that does not require detections of multiple Balmer lines is needed for these targets.

It is common practice to estimate \ebvgas\ from the stellar continuum reddening derived from SED fitting, either assuming nebular
 reddening is larger than stellar reddening as found in local starbursts (\ebvgas=\ebvstars/0.44; \citealt{cal00})
 and low-metallicity high-redshift galaxies \citep{shi20}
 or that the two are equal as found by several studies at $z>1$ \citep{erb06b,kas13,pan15,red15,pug16}.
Nebular reddening of high-redshift galaxies has also been estimated from the rest-UV slope at 1600~\AA, \betauv, by assuming a relation
 between \betauv\ and A$_{\text{UV,stars}}$ \citep[e.g.,][]{meu99,cal00,red15,red18b,shi20}, converting to \ebvstars\ using a reddening law, and again assuming
 a relation between \ebvgas\ and \ebvstars\ \citep[as in][]{ono16}.

Here, we instead use a sample of $\sim300$ MOSDEF star-forming galaxies at $z\sim2.3$ with H$\alpha$ and H$\beta$ detections to calibrate
 a relation between SFR and continuum reddening inferred using SED fitting and \ebvgas\ based on the Balmer decrement,
 leveraging correlations among these properties \citep{red15,shi20}.
The derivation of this calibration can be found in Appendix~\ref{app:dust}, and the resulting relation is
\begin{equation}\label{eq:newebvgas}
\begin{multlined}
\text{E(B-V)}_{\text{gas}} = \text{E(B-V)}_{\text{stars}} - 0.604 \\ + 0.538\times[\log(\text{SFR(SED)}) - 0.20\times(z - 2.3)] .
\end{multlined}
\end{equation}
This method reliably recovers the Balmer decrement \ebvgas\ with a mean offset of 0.02 magnitudes and an intrinsic scatter of 0.23
 magnitudes that shows no bias as a function of \mstar\ or SFR, and outperforms the other methods discussed above
 (see Appendix~\ref{app:dust}).\footnote{Note that the calibration in equation~\ref{eq:newebvgas} is only applicable when \ebvstars\ and
 SFR(SED) have been derived under the same set of assumptions for SED fitting as described in Sec.~\ref{sec:sedfitting}, in particular
 assuming a \citet{cal00} attenuation law.  In Appendix~\ref{app:dust}, we provide an alternate form applicable when an SMC extinction law
 \citep{gor03} is instead assumed.}

For targets without Balmer decrement measurements (i.e., $z\sim2.3$ galaxies with undetected H$\beta$ and all $z\sim3.3$ galaxies),
 we estimate \ebvgas\ using equation~\ref{eq:newebvgas}.
When estimated in this way, the uncertainty on \ebvgas\ includes the intrinsic calibration scatter.
Results at $z\sim2.3$ are indistinguishable within the uncertainties if we limit the sample to only galaxies with H$\alpha$ and H$\beta$
 detections or use the new SED-based \ebvgas\ method for all galaxies (including those with measured Balmer decrements).
Emission-line ratios are corrected for reddening using \ebvgas\ and assuming a \citet{car89} Milky Way extinction law.

\subsubsection{Star-formation rates}

Star-formation rates are derived from dust-corrected Balmer emission-line luminosities (H$\alpha$ if available, otherwise H$\beta$)
 using the H$\alpha$ conversion of \citet{hao11}, renormalized to a \citet{cha03} IMF \citep{shi15}.
When H$\beta$ is the only Balmer line detected, an intrinsic ratio of H$\alpha$/H$\beta$=2.86 is assumed.
SFRs at $z\sim3.3$ are derived from H$\beta$, while H$\alpha$ is used at $z\sim2.3$.
Throughout this paper, star-formation rates are those derived from Balmer emission lines unless specifically noted otherwise.

\subsection{Galaxy samples}

\subsubsection{MOSDEF samples at $z\sim2.3$ and $z\sim3.3$}

We selected samples of star-forming galaxies (SFGs) at $z\sim2.3$ and $z\sim3.3$ from the MOSDEF survey.
We required a robust spectroscopic redshift as measured from the MOSFIRE spectrum.
AGN were identified by their X-ray and infrared properties \citep{coi15,aza17,aza18,leu19} and rejected,
 and we further removed galaxies with log([N\ii]/H$\alpha$)$>$$-0.3$ that have a high probability of
 being dominated by AGN emission.
We did not make any additional cuts based on position in the [N\ii] BPT diagram because of the evolution
 of the star-forming sequence at $z\sim2$ towards the \citet{kau03} delineation between $z\sim0$ SFGs and AGN \citep[e.g.,][]{ste14,sha15,san16a,str17}.
The total samples of MOSDEF star-forming galaxies number 523 at $2.09\le z\le2.61$ with a median stellar mass
 of log($M_*/\msun)=9.97$ and $z_{\text{med}}=2.29$; and 245 at $2.95\le z\le3.80$ with a median stellar mass
 of log($M_*/\msun)=9.89$ and $z_{\text{med}}=3.23$.

From these parent samples of MOSDEF SFGs, we selected a sample of individual galaxies with metallicity measurements and
 a sample from which we will produce composite spectra.
The minimum requirement to obtain a metallicity estimate using our methodology is a detection of both [O\ii] and [O\iii]$\lambda$5007
 (see Section~\ref{sec:metallicity} for details on metallicity calculations).
We thus selected individual galaxies from the SFG parent samples by requiring that both [O\ii] and [O\iii]$\lambda$5007 are detected with
 S/N$\ge$3, yielding individual metallicity samples of 265 and 130 galaxies at $z\sim2.3$ and 3.3, respectively.
Galaxies that additionally have detections of H$\beta$ (68\% (66\%) at $z\sim2.3$ (3.3)) and [Ne\iii] (20\% (35\%) at $z\sim2.3$ (3.3))
 will have more robust metallicity determinations.
The individual metallicity sample at $z\sim2.3$ (3.3) has a median stellar mass of log($M_*/\msun)=9.85$ (9.66)
 and $z_{\text{med}}=2.27$ (3.23), where the typical stellar masses are slightly lower than in the parent SFG samples because
 [O\iii] is intrinsically weak at high \mstar\ and [O\ii] is increasingly affected by dust as \mstar\ increases.

\begin{figure*}
 \includegraphics[width=\textwidth]{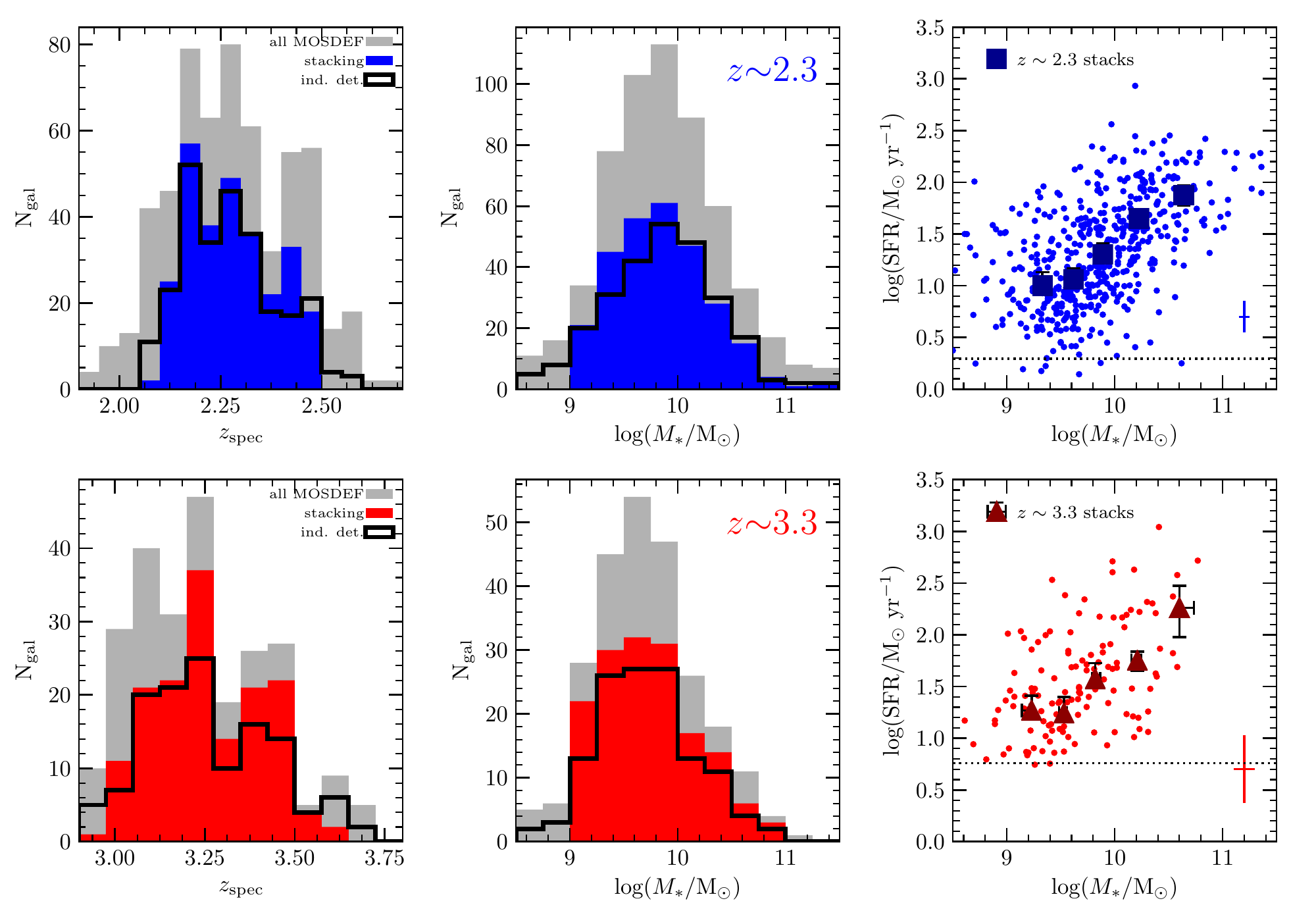}
 \centering
 \caption{
Redshift histogram (left), stellar mass distribution (middle), and SFR vs.\ \mstar\ for MOSDEF star-forming galaxies at $z\sim2.3$ (top) and $z\sim3.3$ (bottom).
In the left and middle panels, the gray histogram represents all MOSDEF star-forming galaxies in each redshift bin, the black outline shows the subset of individual galaxies with metallicity measurements, and the filled color histogram (blue for $z\sim2.3$, red for $z\sim3.3$) denotes the stacking sample.
In the right panel, individual galaxies are shown as colored dots, while values inferred from stacked spectra in stellar mass bins are presented as colored squares/triangles with error bars.
The median uncertainty on \mstar\ and SFR of the individual galaxies is displayed in the lower right corner.
The SFR corresponding to the MOSDEF 3$\sigma$ detection limit of H$\alpha$ at $z\sim2.3$ and H$\beta$ at $z\sim3.3$ is shown by the dotted line in the top and bottom panels, respectively.
}\label{fig:mosdefsamples}
\end{figure*}

The redshift and stellar mass distributions are shown in the left and middle panels of Figure~\ref{fig:mosdefsamples}
 for the MOSDEF SFG parent sample (gray) and the individual metallicity sample (black).
The sample with sufficient emission-line detections for metallicity estimates displays a similar redshift and mass
 distribution to that of the full sample of MOSDEF SFGs at both $z\sim2.3$ and $z\sim3.3$.
The selection of the stacking samples is described in Section~\ref{sec:stacking} below.

\subsubsection{Sample at $z\sim0$}\label{sec:z0sample}

For a local comparison sample, we employ measurements from the composite spectra of $\sim200,000$
 Sloan Digital Sky Survey \citep[SDSS;][]{yor00}
 galaxies at $z\sim0.08$ from \citet[][hereafter AM13]{and13},
 binned both in stellar mass alone and in \mstar\ and SFR.
These stacked spectra have direct-method metallicity measurements from [O\iii]$\lambda$4363 and [O\ii]$\lambda\lambda$7320,7330
 at log($M_*/\msun)<10.5$, and strong-line measurements over $7.5<\log(M_*/\msun)<11.5$.
The stellar masses and SFRs of both the individual SDSS galaxies and \citet{and13} stacks have been shifted to a \citet{cha03}
 IMF, and the SFRs have also been renormalized to the \citet{hao11} H$\alpha$ calibration.
SFR and \mstar\ for stacked spectra are taken to be the median SFR and \mstar\ of the individual galaxies in each bin.

Recent work has demonstrated the importance of accounting for contributions from diffuse ionized gas (DIG)
 to the total emission-line fluxes in integrated galaxy spectra at $z\sim0$ \citep[e.g.,][]{zha17,san17,val19}.
Correcting for DIG contamination is particularly important for gas-phase metallicity studies because DIG emission
 enhances low-ionization lines in galaxy spectra, biasing metallicity estimates high.
The \citet{and13} stacks were corrected for DIG contamination following \citet{san17}.
Note that we have not corrected the H$\alpha$-based SFRs for DIG, though this correction can significantly lower
 the SFR of low-sSFR ($M_*/\mbox{SFR}\lesssim0.01$~Gyr$^{-1}$) galaxies by a factor of $\sim2-3$ \citep{val19}.
High-redshift samples were not corrected for DIG because it is not expected to contribute significantly to their total line
 emission due to the high star-formation rate surface densities \citep[$\Sigma_{\text{SFR}}$;][]{san17,sha19}.

\subsection{Composite spectra}\label{sec:stacking}

We created composite spectra in bins of stellar mass to measure sample averages in a way that includes
 galaxies for which not all lines of interest were detected.
Stacking samples were selected from the MOSDEF SFG parent samples at $z\sim2.3$ and $z\sim3.3$ by further requiring
 detection of [O\iii]$\lambda$5007 at S/N$\ge$3 and spectral coverage of [O\ii], [Ne\iii], H$\beta$, and [O\iii]$\lambda$5007
 (the four strong lines with common coverage between $z\sim2.3$ and $z\sim3.3$).
While not required for selection, the $z\sim2.3$ stacks additionally have coverage of H$\alpha$, [N\ii], and [S\ii].
We note that [Ne\iii] is not required for a metallicity determination, but provides an additional independent line ratio to improve
 abundance constraints.
The stacking sample size increases by only 4\% if [Ne\iii] coverage is not required.
A detection of [O\iii]$\lambda$5007 is required in order to normalize the spectra prior to stacking to ensure that galaxies
 with the brightest lines (i.e., highest SFRs) do not dominate the stacks.\footnote{Ideally,
 spectra would be normalized by a Balmer line instead since [O\iii] flux is sensitive to both SFR and metallicity.
This is not feasible since the strongest Balmer line accessible at $z\sim3.3$ is H$\beta$, and requiring S/N$\ge$3
 for H$\beta$ reduces the $z\sim3.3$ stacking sample size by 40\%. However, we have checked that line ratios in the
 $z\sim2.3$ stacks do not change significantly within the uncertainties when normalizing by H$\alpha$ instead of [O\iii].}
This requirement does not significantly bias the stacking sample since [O\iii]$\lambda$5007 is one of the brightest lines in high-redshift
 galaxy spectra (i.e., almost always detected if a MOSDEF redshift was measured).
We additionally removed objects for which one of the lines of interest is close enough to the edge of the bandpass that the continuum
 is not sufficiently sampled on both sides of the line centroid and targets with double-peaked or otherwise significantly non-Gaussian
 line profiles.
This selection results in a $z\sim2.3$ (3.3) stacking sample of 280 (155) star-forming galaxies with $z_{\text{med}}=2.28$ (3.24)
 and median stellar mass of log($M_*/\msun)=9.96$ (9.89).

We divided the stacking samples into 4 bins of stellar mass.
The MOSDEF survey has a high spectroscopic success rate ($\sim$85\%) at $9.0\le\log(M_*/\msun)\le10.5$ \citep{kri15},
signifying that the MOSDEF sample is highly complete and representative of the typical galaxy population over this mass range
 given the rest-optical magnitude-limited nature of the parent sample.
Below log($M_*/\msun)=9.0$, both the number of targets and spectroscopic success rate drops off sharply as these low-mass
 galaxies are fainter than the $H$-band magnitude cut.
At log($M_*/\msun)>10.5$, targeted galaxies are fewer because of the rarity of such massive systems in the volume probed,
 but the spectroscopic success rate also drops to $\sim$60\%.
As discussed in \citet{kri15}, this lower success rate is at least partially caused by a significantly lower success rate
 for red star-forming galaxies, potentially leading to a bias against metal-rich systems at high masses.

For these reasons, we divided galaxies into 4 bins in \mstar\ over the range $9.0\le\log(M_*/\msun)\le10.5$, separated such that
 an approximately equal number of galaxies falls in each bin.
We created a fifth high-mass bin that contains all galaxies at log($M_*/\msun)>10.5$.
The number of galaxies in each bin are given in Table~\ref{tab:stacks}.
We consider the 4 bins at 10$^{9.0-10.5} \msun$ to be the ``core'' stacking sample where MOSDEF is highly complete and representative,
 and focus our analysis on stacks in this mass range.
We also show and discuss the high-mass ($M_*>10^{10.5} \msun$) stacks, but acknowledge the potential bias against high-metallicity
 systems in this regime.

\begin{table*}
 \centering
 \caption{Properties of stacked spectra in bins of \mstar\ for the $z\sim2.3$ and $z\sim3.3$ samples.
 }\label{tab:stacks}
 \setlength{\tabcolsep}{4.5pt}
 \renewcommand{\arraystretch}{1.5}
 \begin{tabular}{ l l l l l l l l l l l }
   \hline\hline
   $\log{\left(\frac{M_*}{\mbox{M}_\odot}\right)}$$^a$ &
 $N$$^b$ &
 log$\left(\frac{\text{SFR}}{\text{M}_\odot/\text{yr}}\right)$ &
 log$\left(\frac{[\text{O}\ \textsc{iii}]}{\text{H}\beta}\right)$ &
 log$\left(\frac{[\text{O}\ \textsc{ii}]}{\text{H}\beta}\right)$ &
 log$\left(\text{O}_{32}\right)$ &
 log$\left(\text{R}_{32}\right)$ &
 log$\left(\frac{[\text{Ne}\ \textsc{iii}]}{[\text{O}\ \textsc{ii}]}\right)$ &
 log$\left(\frac{[\text{N}\ \textsc{ii}]}{\text{H}\alpha}\right)$ &
 log(O3N2) & 12+log$\left(\frac{\text{O}}{\text{H}}\right)$  \\
   \hline\hline
   \multicolumn{11}{c}{$z\sim2.3$ stacks in bins of \mstar} \\
   \hline
   $9.33^{+0.01}_{-0.05}$ & 65 & $1.00^{+0.13}_{-0.01}$ & $0.68^{+0.01}_{-0.04}$ & $0.47^{+0.05}_{-0.03}$ & $0.21^{+0.01}_{-0.08}$ & $0.97^{+0.01}_{-0.03}$ & $-0.79^{+0.05}_{-0.06}$ & $-1.22^{+0.08}_{-0.05}$ & $1.90^{+0.05}_{-0.12}$ & $8.30^{+0.02}_{-0.02}$ \\
   $9.62^{+0.03}_{-0.01}$ & 65 & $1.06^{+0.11}_{-0.05}$ & $0.56^{+0.04}_{-0.01}$ & $0.51^{+0.05}_{-0.03}$ & $0.05^{+0.04}_{-0.04}$ & $0.91^{+0.04}_{-0.02}$ & $-0.95^{+0.09}_{-0.07}$ & $-1.11^{+0.07}_{-0.06}$ & $1.67^{+0.08}_{-0.08}$ & $8.41^{+0.02}_{-0.02}$ \\
   $9.89^{+0.02}_{-0.02}$ & 65 & $1.30^{+0.11}_{-0.03}$ & $0.53^{+0.03}_{-0.04}$ & $0.63^{+0.03}_{-0.04}$ & $-0.11^{+0.04}_{-0.04}$ & $0.95^{+0.02}_{-0.03}$ & $-0.98^{+0.01}_{-0.16}$ & $-0.93^{+0.04}_{-0.05}$ & $1.46^{+0.06}_{-0.07}$ & $8.48^{+0.02}_{-0.02}$ \\
   $10.23^{+0.02}_{-0.02}$ & 64 & $1.65^{+0.05}_{-0.07}$ & $0.39^{+0.03}_{-0.02}$ & $0.64^{+0.01}_{-0.05}$ & $-0.25^{+0.05}_{-0.02}$ & $0.88^{+0.02}_{-0.04}$ & $-1.19^{+0.13}_{-0.03}$ & $-0.74^{+0.04}_{-0.03}$ & $1.13^{+0.04}_{-0.05}$ & $8.57^{+0.01}_{-0.01}$ \\
   $10.64^{+0.03}_{-0.06}$ & 21 & $1.87^{+0.09}_{-0.10}$ & $0.26^{+0.07}_{-0.06}$ & $0.62^{+0.10}_{-0.06}$ & $-0.36^{+0.10}_{-0.11}$ & $0.82^{+0.07}_{-0.04}$ & $-1.19^{+0.25}_{-0.23}$ & $-0.63^{+0.08}_{-0.01}$ & $0.89^{+0.06}_{-0.13}$ & $8.64^{+0.03}_{-0.04}$ \\
   \hline\hline
   \multicolumn{11}{c}{$z\sim3.3$ stacks in bins of \mstar} \\
   \hline
   $9.23^{+-0.02}_{-0.09}$ & 37 & $1.27^{+0.14}_{-0.06}$ & $0.74^{+0.05}_{-0.04}$ & $0.29^{+0.11}_{-0.04}$ & $0.45^{+0.06}_{-0.10}$ & $0.97^{+0.04}_{-0.02}$ & $-0.66^{+0.05}_{-0.17}$ & --- & --- & $8.19^{+0.03}_{-0.04}$ \\
   $9.53^{+0.02}_{-0.05}$ & 37 & $1.24^{+0.16}_{-0.02}$ & $0.69^{+0.06}_{-0.02}$ & $0.44^{+0.08}_{-0.05}$ & $0.25^{+0.06}_{-0.05}$ & $0.97^{+0.06}_{-0.03}$ & $-0.82^{+0.05}_{-0.15}$ & --- & --- & $8.29^{+0.03}_{-0.02}$ \\
   $9.82^{+0.05}_{-0.03}$ & 36 & $1.57^{+0.15}_{-0.08}$ & $0.65^{+0.03}_{-0.07}$ & $0.50^{+0.07}_{-0.05}$ & $0.15^{+0.03}_{-0.09}$ & $0.96^{+0.03}_{-0.05}$ & $-0.92^{+0.09}_{-0.11}$ & --- & --- & $8.35^{+0.03}_{-0.02}$ \\
   $10.21^{+0.04}_{-0.05}$ & 36 & $1.75^{+0.09}_{-0.10}$ & $0.53^{+0.07}_{-0.03}$ & $0.65^{+0.05}_{-0.07}$ & $-0.13^{+0.09}_{-0.03}$ & $0.95^{+0.05}_{-0.05}$ & $-1.01^{+0.09}_{-0.10}$ & --- & --- & $8.48^{+0.02}_{-0.02}$ \\
   $10.60^{+0.13}_{-0.01}$ & 9 & $2.26^{+0.21}_{-0.28}$ & $0.35^{+0.20}_{-0.14}$ & $0.62^{+0.30}_{-0.16}$ & $-0.27^{+0.13}_{-0.17}$ & $0.87^{+0.25}_{-0.11}$ & $-0.87^{+0.21}_{-0.20}$ & --- & --- & $8.54^{+0.07}_{-0.06}$ \\
   \hline\hline
 \end{tabular}
 \begin{flushleft}
 $^{a}$ {Median stellar mass of galaxies in each bin.}
 $^{b}$ {Number of galaxies in each bin.}
 \end{flushleft}
\end{table*}

The composite spectra were created following the methods outlined in \citet{san18}.
Briefly, the individual spectra were shifted into the rest frame and luminosity density units using the spectroscopic redshift,
 dust-corrected
 according to their \ebvgas\ assuming a \citet{car89} extinction curve (where we use \ebvgas\ from the Balmer decrement when available,
 otherwise \ebvgas\ from equation~\ref{eq:newebvgas}), normalized by the dust-corrected [O\iii]$\lambda$5007 luminosity,
 and resampled onto a uniform wavelength grid.
Individual spectra were then combined by taking the median at each wavelength element\footnote{We do not apply any weighting
 when combining the spectra. Note that inverse-variance weighting, while maximizing S/N, gives higher weight in the emission lines
 to high-SFR objects and thus potentially biases results from stacks.}
 and multiplied by the median [O\iii]$\lambda$5007 luminosity.
Emission-line luminosities were measured from the stacked spectra using the same method as for the individual galaxies.
Balmer absorption corrections were estimated from the median correction applied to the galaxies in each bin.

Uncertainties on the line luminosities and line ratios were estimated using a Monte Carlo method in which we bootstrap resampled
 the galaxies in the stacking sample, perturbed the masses, \ebvgas, and science spectra according to their uncertainties, repopulated
 the mass bins using the original boundaries in \mstar, stacked
 the perturbed spectra according to the method described above, and remeasured the line luminosities and line ratios.
The uncertainties on each property measured from the stacks are inferred from the 68th-percentile width of the distribution
 resulting from 100 realizations.
In this way, errors on properties measured from the stacks include measurement errors and sample variance.
Using random subsets of a sample of galaxies with detections of [O\ii], [Ne\iii], H$\beta$, and [O\iii], we have verified that
 this stacking method reproduces the median line ratios of the input samples to better than 0.05~dex. 
Table~\ref{tab:stacks} presents the \mstar, SFR, and line ratios of the $z\sim2.3$ and $z\sim3.3$ stacked spectra.

\subsection{Representativeness of samples}

A sample that is biased in SFR relative to the mean SFR-\mstar\ relation
 will yield a biased MZR because of the existence of the FMR at both
 $z\sim0$ and $z>1$ \citep{man10,san18}.
At fixed \mstar, if a sample has a higher than average SFR, then O/H will be lower than average, and vice versa.
It is therefore imperative that samples have a SFR distribution that is representative at each stellar mass, i.e., that
 they lie on the ``star-forming main sequence'' at each redshift.

The right panels of Figure~\ref{fig:mosdefsamples} display SFR vs.\ \mstar\ for stacked spectra and individual galaxies with S/N$\ge$3
 in at least one Balmer line at $z\sim2.3$ (top) and $z\sim3.3$ (bottom).
A clear correlation is present among individual galaxies and stacks.
The dotted lines display the SFR corresponding to the H$\alpha$ and H$\beta$ 3$\sigma$ detection threshold of
 MOSDEF at $z\sim2.3$ and $z\sim3.3$, respectively.
At log($M_*/\msun)>9.5$, the distribution of individual galaxies lies above this threshold at both redshifts.
Below log($M_*/\msun)=9.5$, individual galaxies begin to fall below the Balmer line detection limit.
Stacking is meant to include galaxies falling below the detection threshold.
The lowest-mass bin at $z\sim3.3$ has a slightly higher SFR than the next bin higher in mass, and at $z\sim2.3$
 the relation displays  a slight flattening at the low-mass end (though not highly significant relative to the uncertainties).
Assuming that the star-forming main sequence is a monotonically increasing power law, these data indicate that the
 $z\sim2.3$ and $z\sim3.3$ stacking samples are missing the lowest-SFR galaxies at log($M_*/\msun)<9.5$ despite
 not requiring H$\beta$ detections, biasing the SFR high in this lowest-mass bin \citep[see also][]{shi15}.

\begin{figure}
 \includegraphics[width=\columnwidth]{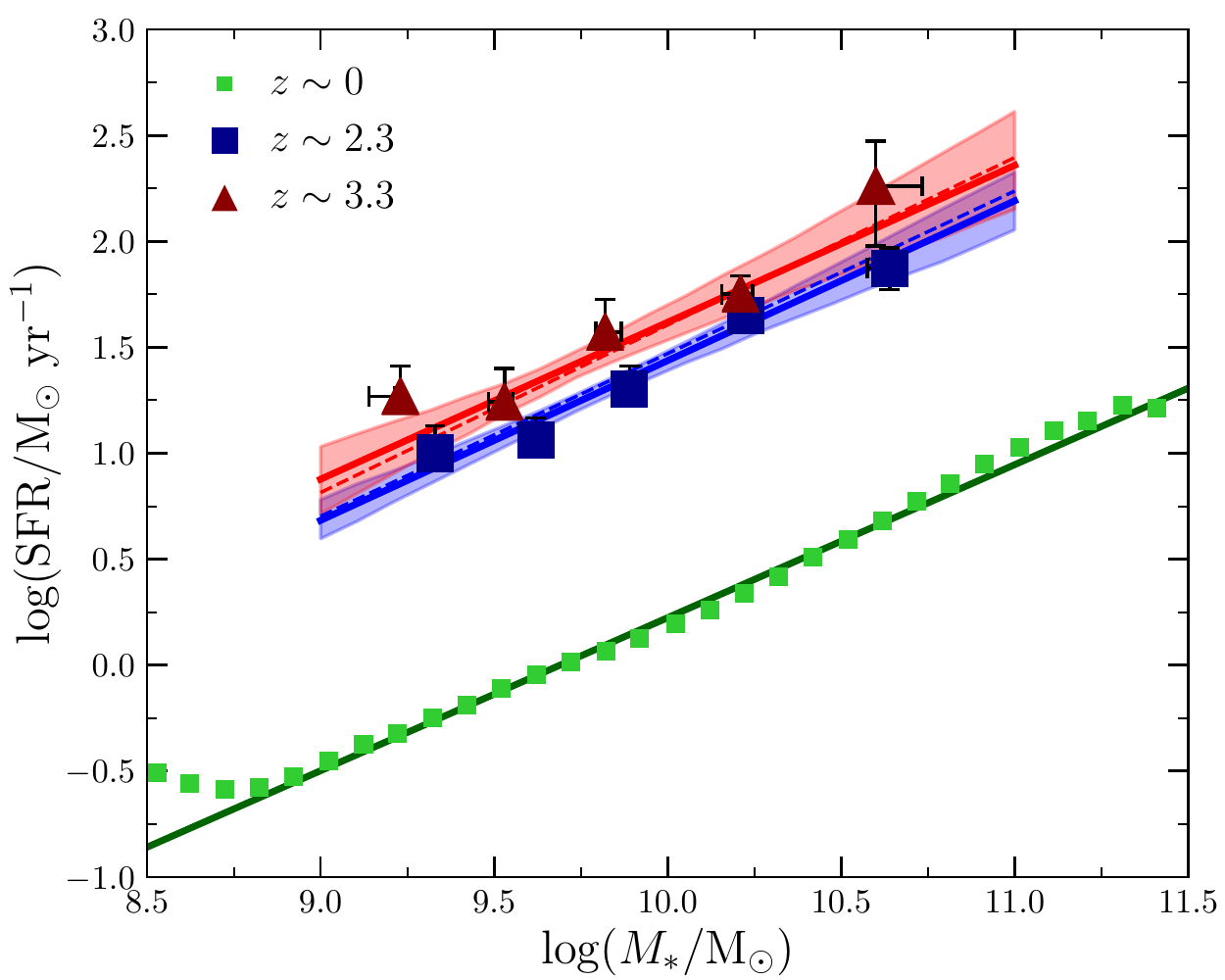}
 \centering
 \caption{
SFR-\mstar\ relation at $z\sim0$ (green), $z\sim2.3$ (blue), and $z\sim3.3$ (red).
Squares denote values for stacked spectra in bins of stellar mass.
Solid lines show the best-fit power-law relations to the stacks, color-coded by redshift.
The shaded regions display the 1$\sigma$ uncertainty intervals around
 the best-fit lines for $z\sim2.3$ and $z\sim3.3$.
SFR-\mstar\ relations from the parameterization of Speagle et al.\ (2014) at $z=2.3$ and $z=3.3$ are given by
 the dashed lines, displaying excellent agreement with our best-fit relations.
}\label{fig:sfrmstar}
\end{figure}

Figure~\ref{fig:sfrmstar} shows SFR vs.\ \mstar\ for the $z\sim0$, $z\sim2.3$, and $z\sim3.3$ samples, with only stacks
 displayed at high redshifts.
Visually, the relation between SFR and \mstar\ displays a similar slope at all redshifts, with the SFR at fixed \mstar\ increasing
 as a function of redshift.
The $z\sim0$ sample is biased high in SFR below log($M_*/\msun)=8.7$, where SFR begins increasing with decreasing \mstar.
This bias is likely due to the emission-line selection and the depth of SDSS spectroscopy.
According to the FMR, if a sample is increasingly biased in SFR with decreasing \mstar, the measured slope of the mass-metallicity
 relation will be artifically steepened.
This SFR bias likely explains the very steep low-mass slope of $\text{O/H}\propto M_*^{0.64}$ obtained by \citet{and13}.
We only use \citet{and13} stacks with log($M_*/\msun)>8.7$ in our analysis to avoid biasing the low-mass slope.

We fit a power-law of the form $\text{SFR}=C\times M_*^{\beta}$ to the stacks at each redshift.
For the MOSDEF stacks, we exclude the highest-mass bin from the fitting due to potential bias (Sec~\ref{sec:stacking}).
At $z\sim3.3$, we also exclude the lowest-mass bin that is clearly biased high in SFR.
We retain the lowest-mass $z\sim2.3$ bin since the apparent flattening is only significant at the 1$\sigma$ level, but note that our results do not significantly change if this bin is also excluded from fitting.
We find the following best-fit relations, displayed as solid lines in Figure~\ref{fig:sfrmstar}:
\begin{equation}\label{eq:z0sfrmstar}
z\sim0\text{: } \text{log}\left(\frac{\text{SFR}}{\msun~\text{yr}^{-1}}\right) = (0.72\pm0.01)\times m_{10} + (0.22\pm0.01)
\end{equation}
\begin{equation}\label{eq:z2sfrmstar}
z\sim2.3\text{: } \text{log}\left(\frac{\text{SFR}}{\msun~\text{yr}^{-1}}\right) = (0.75\pm0.12)\times m_{10} + (1.44\pm0.04)
\end{equation}
\begin{equation}\label{eq:z3sfrmstar}
z\sim3.3\text{: } \text{log}\left(\frac{\text{SFR}}{\msun~\text{yr}^{-1}}\right) = (0.74\pm0.20)\times m_{10} + (1.62\pm0.06)
\end{equation}
where $m_{10}=\log(M_*/10^{10}\ \msun)$.
The slopes of the $z>2$ SFR-\mstar\ relations are consistent with the results of \citet{shi15} based on early MOSDEF data at $z=1.4-2.6$.

The parameterized SFR($M_*,z$) of \citet{spe14} at $z=2.3$ and $z=3.3$ is displayed in Figure~\ref{fig:sfrmstar}
 as dotted lines, color-coded by redshift.
Our best-fit SFR-\mstar\ relations at $z=2.3$ and $z=3.3$ closely match those of \citet{spe14}, indicating that the
 MOSDEF samples are representative of typical galaxies falling on the star-forming main sequence.
The lowest-mass $z\sim3.3$ stack is elevated 0.2 dex in SFR above our best-fit $z\sim3.3$ SFR-\mstar\ relation and that
 of \citet{spe14}.
At fixed \mstar, the SFR-dependence of O/H has been found to be $\Delta\log(\text{O/H})\sim-0.15\times\Delta\log(\text{SFR})$ for star-forming
 galaxies at $z\sim2.3$ \citep{san18}.
Accordingly, the SFR bias of the lowest-mass $z\sim3.3$ bin is expected to result in a bias of $\sim0.03$~dex in O/H.
The magnitude of this bias is less than the formal uncertainty in metallicity for this stack (0.04~dex).
We therefore retain the lowest-mass $z\sim3.3$ stack in our analysis of the mass-metallicity relation.
The highest-mass stacks are fully consistent with the best-fit relations and the \citet{spe14} parameterization.
We conclude that the MOSDEF sample does not display any large SFR biases over $10^{9.0-10.75}~\msun$ that would
 significantly affect the MZR derived from these data.

\section{Metallicity Derivations}\label{sec:metallicity}

The choice of metallicity calibrations is of critical importance to any metallicity scaling relation analysis.
As demonstrated by \citet{kew08}, the functional form and normalization of the local MZR varies significantly
 based on the metallicity indicator and calibration employed, such that comparing metallicities inferred from
 different indicators and calibrations can introduce severe biases.
In addition, the excitation properties of high-redshift star-forming galaxies suggest that metallicity calibrations
 evolve with redshift due to changes in the underlying physical properties of the ionized gas in \hii\ regions
 \citep[e.g.,][]{kew13,ste14,sha15,sha19,san16a,san20,san20b,kas17,kas19a,str17,str18}.
Accordingly, applying calibrations constructed for the local universe may yield biased metallicities at high redshifts
 and consequently bias the inferred metallicity evolution.

In this analysis, we address these issues by (1) using a uniform set of strong emission lines for samples at all redshifts,
 and (2) employing different metallicity calibrations at $z\sim0$ and $z>1$, all of which are empirically calibrated to
 direct-method metallicities, to reflect evolving ISM conditions.
At all redshifts, metallicities are estimated using line ratios of [O\ii], H$\beta$, [O\iii], and [Ne\iii].
This choice of emission lines is partly driven by observational limitations since these are the only strong optical
 emission lines that can be observed at $z>3$ with current facilities.
However, there is an advantage to using these particular lines.
This set of lines only contains $\alpha$-element metal species (O, Ne) that have the same production channel through
 core-collapse SNe and are thus more direct tracers of the gas-phase oxygen abundance than nitrogen-based metallicity
 indicators ([N\ii]/H$\alpha$, [O\iii]/[N\ii]) that are sensitive to N/O and the secondary production of N.

Our analysis at $z\sim0$ uses the composite spectra of AM13.
These composites have direct-method metallicities at log($M_*/\msun)<10.5$, but this direct-method subset does not fully sample
 the high-mass asymptotic metallicity region, requiring coverage up to log($M_*/\msun)\sim11.0$.
In order to cover the full AM13 mass range up to log($M_*/\msun)\sim11.5$, we 
fit relations between strong-line ratios and O/H using AM13 \mstar-binned stacks at
 $8.7<\log$($M_*/\msun)<10.5$ that have direct-method metallicities, spanning $8.4<12$+log(O/H$)<8.8$.
While this O/H range is sufficent for establishing the $z\sim0$ MZR over $8.7<\log$($M_*/\msun)<11.5$, we need
 to extend to lower metallicities to cover low-mass, high-SFR galaxies in the FMR.
For this purpose, we supplement the AM13 stacks with the \citet{ber12} ``combined select'' sample of dwarf
 galaxies from the \textit{Spitzer} Local Volume Legacy survey.
This sample comprises 38 galaxies falling on the star-forming main sequence,
 with $6.0\lesssim\log$($M_*/\msun)\lesssim9.0$ and
 direct-method metallicities extending down to 12+log(O/H$)=7.5$.
The dwarf galaxy sample is based on slit spectra of \hii\ regions and thus does not require DIG correction.
We calculate median line ratios and O/H of the \citet{ber12} dwarfs in 4 bins of O/H such that each bin contains an approximately
 equal number of galaxies and use these binned data for fitting.
Direct-method metallicities for both the AM13 and \citet{ber12} samples have been uniformly recalculated using
 PyNeb \citep{lur15} with the default set of atomic data.

\begin{figure*}
 \includegraphics[width=\textwidth]{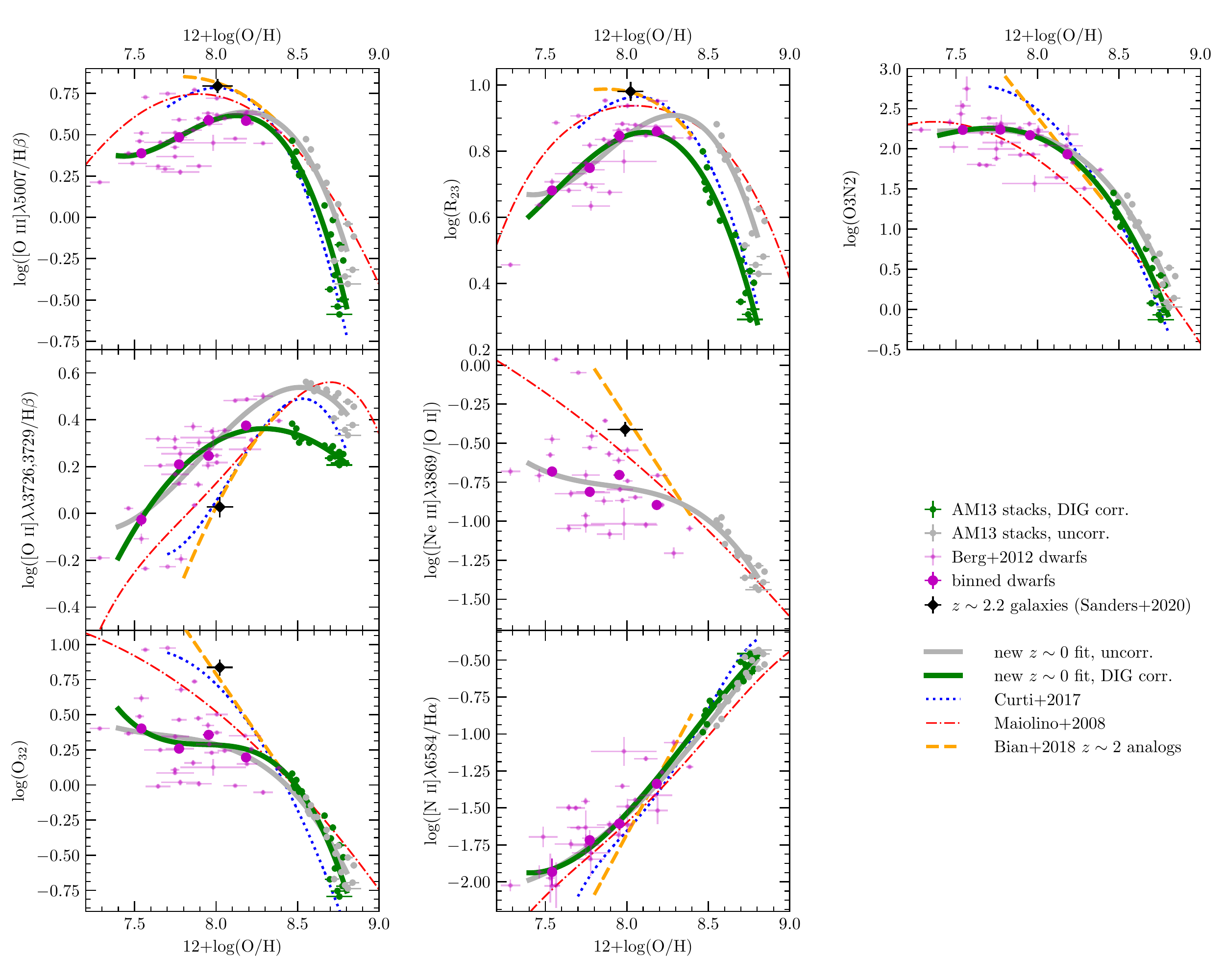}
 \centering
 \caption{
Calibrations between emission-line ratios and direct-method metallicity.
The calibrations used at $z\sim0$ are fit to the \citet[][AM13]{and13} \mstar-binned stacks of $z\sim0$ galaxies at high
 metallicities (green points, corrected for DIG) and representative dwarf galaxies from \citet{ber12} at low metallicities (pink points).
The solid green lines show the best-fit $z\sim0$ calibrations.
The gray points and solid gray lines show the form of the calibrations when the $z\sim0$ data are not corrected for DIG emission.
Best-fit coefficients are given in Table~\ref{tab:calibrations}.
Blue dotted and red dash-dotted lines display the $z\sim0$ calibrations of \citet{mai08} and \citet{cur17}, respectively. 
The orange dashed lines denote the high-redshift analog calibrations of \citet{bia18} that are used to derive metallicities for
 the high-redshift samples.
Each calibration is plotted over the range of metallicity of the empirical calibration sample.
The black point shows the median values of a sample of 18 galaxies at $z\sim2.2$ with direct-method metallicities from
 \citet{san20}, displaying close agreement with the high-redshift analog calibrations.
}\label{fig:calibrations}
\end{figure*}

We fit the line ratios [O\iii]$\lambda$5007/H$\beta$, [O\ii]$\lambda\lambda$3726,3729/H$\beta$,
 O$_{32}$=[O\iii]$\lambda$5007/[O\ii]$\lambda\lambda$3726,3729,
 R$_{23}$=([O\iii]$\lambda\lambda$4959,5007+[O\ii]$\lambda\lambda$3726,3729)/H$\beta$,
 [Ne\iii]$\lambda$3869/[O\ii]$\lambda\lambda$3726,3729,
 [N\ii]$\lambda$6584/H$\alpha$,
 and O3N2=([O\iii]$\lambda$5007/H$\beta$)/([N\ii]$\lambda$6584/H$\alpha$) as a function of O/H.
Figure~\ref{fig:calibrations} shows the results of fitting the AM13 and binned \citet{ber12} samples with
 cubic functions of the form
\begin{equation}\label{eq:calibrations}
\log(R) = c_0 + c_1 x + c_2 x^2 + c_3 x^3
\end{equation}
where $x=12$+log(O/H$)-8.69=\log(Z_{\text{neb}}/Z_{\odot})$.
Separate fits are carried out using the DIG-corrected AM13 stacks (green) and the uncorrected stacks (gray),
 where the proper set of calibrations is used for each case.
We do not fit [Ne\iii]/[O\ii] for the DIG-corrected stacks since the impact of DIG on [Ne\iii]
 was not characterized in \citet{san17}.
The best-fit coefficients are given in Table~\ref{tab:calibrations}, and these calibrations are used for the
 $z\sim0$ samples. 

\begin{table}
 \centering
 \caption{Best-fit calibrations between strong-line ratios and direct-method metallicities (Fig.~\ref{fig:calibrations}). Coefficients are given for the cubic function of equation~\ref{eq:calibrations}.
 }\label{tab:calibrations}
 \begin{tabular}{ l l l l l }
   \hline\hline
   line ratio ($R$) & $c_0$ & $c_1$ & $c_2$ & $c_3$ \\
   \hline
   \multicolumn{5}{c}{DIG-corrected $z\sim0$ data} \\
   \hline
   $[$O\iii$]$$\lambda$5007/H$\beta$ & $-0.143$ & $-3.16$ & $-4.06$ & $-1.49$ \\
   $[$O\ii$]$$\lambda$3727/H$\beta$ & $0.270$ & $-0.452$ & $-0.520$ & $0.0831$ \\
   O$_{32}$ & $-0.413$ & $-2.70$ & $-3.52$ & $-1.55$ \\
   R$_{23}$ & $0.469$ & $-1.51$ & $-1.75$ & $-0.508$ \\
   $[$N\ii$]$$\lambda$6584/H$\alpha$ & $-0.606$ & $1.28$ & $-0.435$ & $-0.485$ \\
   O3N2 & $0.461$ & $-4.40$ & $-3.37$ & $-0.761$ \\
   \hline\hline
   \multicolumn{5}{c}{Uncorrected $z\sim0$ data} \\
   \hline
   $[$O\iii$]$$\lambda$5007/H$\beta$ & $0.111$ & $-2.39$ & $-3.30$ & $-1.24$ \\
   $[$O\ii$]$$\lambda$3727/H$\beta$ & $0.498$ & $-0.479$ & $-1.55$ & $-0.654$ \\
   O$_{32}$ & $-0.388$ & $-1.91$ & $-1.74$ & $-0.570$ \\
   R$_{23}$ & $0.698$ & $-1.17$ & $-1.90$ & $-0.758$ \\
   $[$Ne\iii$]$$\lambda$3869/$[$O\ii$]$$\lambda$3727 & $-1.19$ & $-1.29$ & $-1.44$ & $-0.601$ \\
   $[$N\ii$]$$\lambda$6584/H$\alpha$ & $-0.663$ & $1.47$ & $0.215$ & $-0.102$ \\
   O3N2 & $0.772$ & $-3.86$ & $-3.33$ & $-0.939$ \\
   \hline\hline
   ratio & $\frac{[\text{O \textsc{iii}]}}{\text{H}\beta}$ & $\frac{[\text{O \textsc{ii}]}}{\text{H}\beta}$ & O$_{32}$ & R$_{23}$ \\
   $\sigma_{\text{cal}}$$^a$ & 0.10 & 0.13 & 0.19 & 0.08 \\
   \hline
   ratio & $\frac{[\text{Ne \textsc{iii}]}}{[\text{O \textsc{ii}]}}$ & $\frac{[\text{N \textsc{ii}]}}{\text{H}\alpha}$ & O3N2 & \\
   $\sigma_{\text{cal}}$$^a$ & 0.20 & 0.15 & 0.16 &  \\
   \hline\hline
 \end{tabular}
 $^{a}$ {Adopted logarithmic scatter in line ratio at fixed O/H.}
\end{table}

For the high-redshift samples, we employ the metallicity calibrations of \citet[][B18 hereafter]{bia18}\footnote{
Note that the calibration reported in \citet{bia18} for [O\iii]$\lambda$5007/H$\beta$ in fact uses the line ratio
 [O\iii]$\lambda\lambda$4959,5007/H$\beta$ (i.e., the sum of the two [O\iii] lines at 4959 and 5007~\AA.; F. Bian, private communication).
One should either use the correct [O\iii]$\lambda\lambda$4959,5007/H$\beta$ line ratio with the calibration coefficients given
 in \citet{bia18}, or else decrease the $y$-intercept of their equation~17 by log(3.98/2.98$)=0.126$~dex if using
 [O\iii]$\lambda$5007/H$\beta$ since [O\iii]$\lambda$5007/$\lambda4959=2.98$ \citep{sto00}.
We have adopted the latter correction in both \citet{san20} and this work.
}
 based on local analogs of $z\sim2$ galaxies.
B18 measured direct-method metallicities of stacked spectra for a sample of SDSS galaxies selected to lie on the $z\sim2$ star-forming galaxy
 sequence in the [N\ii] BPT diagram.
\citet{san20} found that a sample of 18 galaxies at $z\sim2.2$ with direct-method metallicities matches the B18 high-redshift analog calibrations
 on average for [O\iii]/H$\beta$, [O\ii]/H$\beta$, O$_{32}$, R$_{23}$, and [Ne\iii]/[O\ii] (displayed as the black diamond in
 Figure~\ref{fig:calibrations}).
Note that [N\ii] was not covered for the majority of the $z\sim2.2$ direct-method sample, thus \citet{san20} were unable to test
 [N\ii]/H$\alpha$ and O3N2 calibrations.
This comparison suggests that the B18 oxygen- and neon-based calibrations are appropriate to apply to $z>1$ galaxy samples.

The B18 $z\sim2$ analog calibrations typically have higher O/H at fixed line ratio relative to the DIG-corrected $z\sim0$ calibrations.
Until the number of high-redshift galaxies with direct-method measurements is large enough to independently produce calibrations,
 we must rely on local analogs for which sufficiently deep spectra are more easily obtained.
We note that the B18 calibration sample spans $7.8<12$+log(O/H$)<8.4$, such that we must extrapolate to cover the high-mass
 galaxies in the MOSDEF sample (the highest-mass $z\sim2.3$ stack has 12+log(O/H$)\approx8.6$).
Despite the uncertainty associated with extrapolating, we consider this approach to be more robust than applying $z\sim0$ calibrations
 to $z>1$ samples.
The identification of local analogs at higher metallicities should be pursued to extend these calibrations.

For comparison, we also show the calibrations of \citet[][M08, red]{mai08} and \citet[][C17, blue]{cur17} in Figure~\ref{fig:calibrations}.
These calibration sets are commonly employed in MZR and FMR studies at low and high redshifts
 \citep[e.g.,][]{man09,man10,yat12,tro14,ono16,suz17,cur20b,cur20a}.
While displaying general agreement with our new $z\sim0$ calibrations at high metallicity (12+log(O/H$)\gtrsim8.3$),
 the M08 and C17 calibrations diverge from our $z\sim0$ relations at lower O/H while more closely matching the B18 calibrations.
As discussed in \citet{san20}, both M08 and C17 calibration samples are composed entirely of individual SDSS galaxies with
 [O\iii]$\lambda$4363 detections at 12+log(O/H$)\lesssim8.3$.
Requiring detections of this weak auroral line selects a sample that is strongly biased towards high excitation, sSFR,
 and emission line equivalent width.
That M08 and C17 are similar to B18 at 12+log(O/H$)\sim8.0$ implies that these extreme $z\sim0$ galaxies have ISM properties similar
 to $z\sim2$ galaxies, and are thus not suitable to construct calibrations generally applicable in the local universe.

\citet{pat18} also tested strong-line calibrations for use at high redshift using a sample of $z>1$ galaxies with
 direct-method metallicities that significantly overlaps with that of \citet{san20}.
These authors found that, of the empirical strong-line calibrations they tested (M08; C17; B18; \citealt{jon15}),
 all performed reasonably well with none performing better than the others.
Even though all four are calibrated to $z\sim0$ samples, these calibration sets are based on highly biased local galaxies
 in the metallicity range probed by the $z>1$ sample (12+log(O/H$)\sim7.7-8.3$), as described above (\citealt{jon15} is
 also based on [O\iii]$\lambda$4363-detected galaxies in SDSS).
Thus, the results of \citet{pat18} agree with our conclusion that the M08, C17, and B18 calibrations appear to be more appropriate
 for applications at high redshifts than to typical $z\sim0$ galaxies in the low metallicity regime.
No strong-line calibrations have been tested at $z>1$ at higher metallicities (12+log(O/H$)>8.3$), thus it is unknown which
 calibrations perform best for metal-rich high-redshift galaxies.
Extending the dynamic range of the direct-method $z>1$ sample may be possible by detecting the low-ionization auroral line
 [O\ii]$\lambda\lambda$7320,7330 that is expected to be stronger than [O\iii]$\lambda$4363 at moderately high metallicities.

\begin{figure*}
 \includegraphics[width=\textwidth]{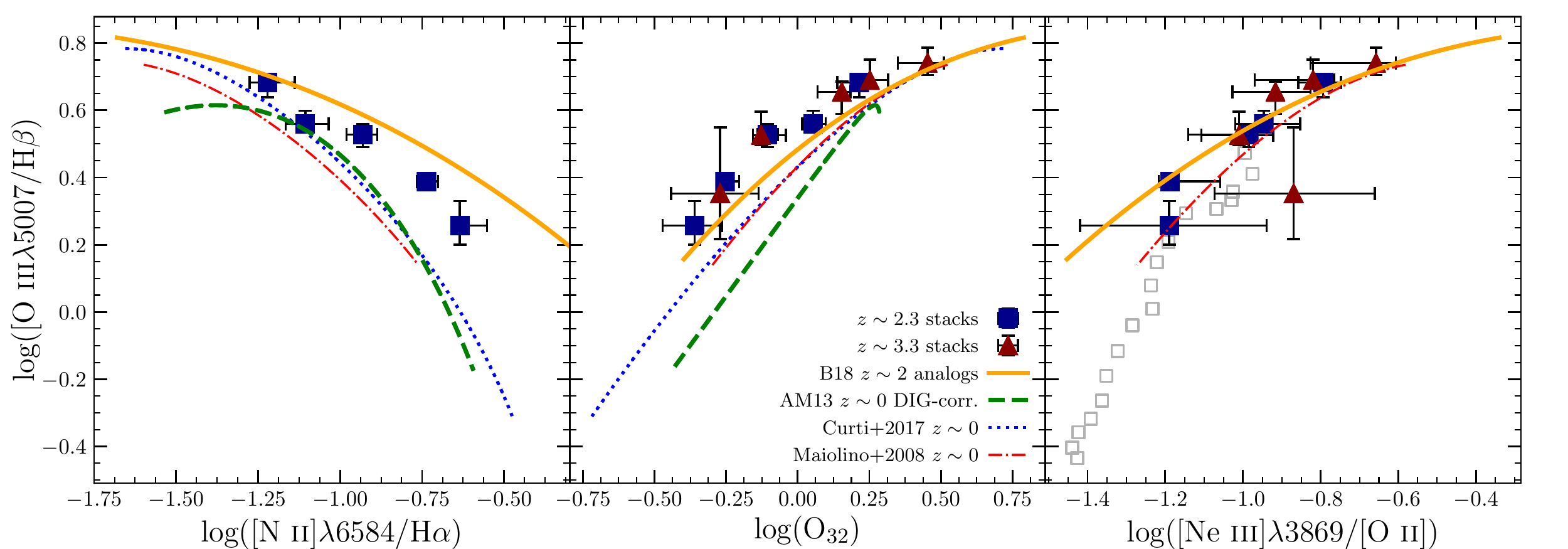}
 \centering
 \caption{
Excitation diagrams of [O\iii]$\lambda$5007/H$\beta$ vs.\ [N\ii]/H$\alpha$ ([N\ii] BPT, left), O$_{32}$ (middle),
 and [Ne\iii]$\lambda$3869/[O\ii]$\lambda\lambda$3726,3729 (right).
Stacked spectra of $z\sim2.3$ and $z\sim3.3$ galaxies (dark blue and red, respectively) are compared to the excitation
 sequences of strong-line calibrations evaluated over 12+log(O/H$)=8.0-8.7$.
The new DIG-corrected $z\sim0$ calibrations of this work are shown as green dashed lines.
The $z=0$ calibrations of \citet{mai08} and \citet{cur17} are presented in red dash-dotted and blue dotted lines, respectively.
The \citet{bia18} high-redshift analog calibrations are shown in solid orange, providing the closest match to the high-redshift stacks.
The new $z\sim0$ calibrations are not shown in the [Ne\iii]/[O\ii] diagram because DIG corrections for [Ne\iii] were not calibrated
 in \citet{san17}.
We instead display the uncorrected $z\sim0$ AM13 stacks as gray squares.
 DIG emission is expected to affect [O\ii] more strongly than [Ne\iii] such that the DIG-corrected [Ne\iii]/[O\ii] ratios should
 be larger at fixed O/H.
}\label{fig:calex}
\end{figure*}

Because emission-line ratios are closely tied to ISM physical conditions, a crucial requirement of any set of metallicity calibrations
 is that the excitation sequences in line ratio vs.\ line ratio
 diagrams of the calibrations match the sequences of the observed sample.
In Figure~\ref{fig:calex}, we show [O\iii]/H$\beta$ vs. [N\ii]/H$\alpha$ (left; [N\ii] BPT), O$_{32}$ (middle), and [Ne\iii]/[O\ii] (right).
In addition to the stacks of MOSDEF $z\sim2.3$ and $z\sim3.3$ galaxies, we show the line ratios calculated from the new $z\sim0$,
 B18 high-redshift analog, C17, and M08 calibrations over the metallicity range 12+log(O/H$)=8.0-8.7$.
In each diagram, the B18 high-redshift analog calibrations match the excitation sequences of the high-redshift samples more closely
 than any of the $z\sim0$ calibrations.
It is not a perfect match, however.
While the B18 calibrations show excellent agreement in [O\iii]/H$\beta$ vs.\ [Ne\iii]/[O\ii], they predict
 0.1~dex higher O$_{32}$ and 0.15~dex higher [N\ii]/H$\alpha$ at fixed [O\iii]/H$\beta$ than the high-redshift stacks on average.
The offset between the B18 calibrations and MOSDEF stacks in the [N\ii] BPT diagram is a result of the B18 selection, requiring
 galaxies to fall within 0.1~dex of the [N\ii] BPT sequence defined by the $z\sim2.3$ KBSS sample \citep{ste14} that is known to
 have a larger offset from the $z\sim0$ sequence than $z\sim2.3$ MOSDEF galaxies \citep{sha15}.
Until strong-line metallicity diagnostics directly calibrated to $z\sim2$ samples are available, it is worthwhile to refine methods
 of selecting local analogs that more closely match high-redshift galaxy properties.
In the meantime, we find that the B18 high-redshift calibrations provide a reasonable match to the $z>2$ data in all three panels.
We note that the B18 high-redshift calibrations provide a better match to the $z>2$ samples than all of the $z\sim0$ calibrations
 in the middle panel of Fig.~\ref{fig:calex} displaying [O\iii]/H$\beta$ vs. O$_{32}$.
Since the metallicities in this work are derived primarily from these two line ratios (see below), this agreement strongly suggests
 that the B18 calibrations are the most appropriate for our high-redshift samples.

Metallicities are calculated via a $\chi^2$ minimization over multiple line ratios simultaneously.
The best-fit metallicity is that which minimizes the expression
\begin{equation}
\chi^2 = \sum_i \frac{(R_{\text{obs},i} - R_{\text{cal},i}(x))^2}{(\sigma_{\text{obs},i}^2 + \sigma_{\text{cal},i}^2)}
\end{equation}
where the sum over $i$ denotes the set of line ratios used, $R_{\text{obs},i}$ is the logarithm of the $i$-th observed line ratio,
 $R_{\text{cal},i}(x)$ is the logarithmic $i$-th line ratio of the calibration at $x=12$+log(O/H),
 $\sigma_{\text{obs},i}$ is the uncertainty in the $i$-th observed line ratio,
 and $\sigma_{\text{cal},i}$ is the uncertainty in $i$-th line ratio at fixed O/H of the calibration.
Since our calibrations at both $z\sim0$ and $z>1$ are fit to stacked spectra, we cannot evaluate $\sigma_{\text{cal}}$ directly.
We instead take $\sigma_{\text{cal}}$ to be the average of the values reported for calibrations by M08, C17, and \citet{jon15},
 noting that these three works find similar scatter for each line ratio.
Our adopted values of $\sigma_{\text{cal}}$ are given in Table~\ref{tab:calibrations}.
When fitting stacks, $\sigma_{\text{cal}}$ is divided by $\sqrt{N}$, where $N$ is the number of galaxies in the stack.
Uncertainties on metallicity are estimated by perturbing the observed line ratios by their uncertainties and refitting 200 times,
 where the $1\sigma$ uncertainty is derived from the 68th percentile width of the resulting distribution.

As explained above, we only utilize line ratios of [O\ii], H$\beta$, [O\iii], and [Ne\iii] to derive metallicities at all redshifts.
This set of emission lines allows for three independent line ratios for fitting.
Here, we use O$_{32}$, [O\iii]/H$\beta$, and [Ne\iii]/[O\ii].
This set of line ratios is advantageous since it minimizes the number of line ratios that require dust correction.
We obtain similar results if we use different sets of independent ratios within the chosen set of emission lines
 (i.e., when using [O\ii]/H$\beta$ or R$_{23}$ instead of [O\iii]/H$\beta$ or O$_{32}$).
Since the calibrations between O$_{32}$ and O/H is monotonic, the minimum requirement to calculate metallicity is a detection
 of [O\iii] and [O\ii], while adding H$\beta$ and [Ne\iii] when available reduces uncertainties and improves the estimate.

Metallicities are determined for MOSDEF galaxies and stacked MOSDEF spectra using the B18 high-redshift analog calibrations.
We note that the stacked spectra have all four lines detected in every mass bin, thus three line ratios are used to
 infer metallicities for all high-redshift stacks.
For the $z\sim0$ samples, we use the new calibrations fit to the AM13 \mstar-binned stacks and given in Table~\ref{tab:calibrations}.
Since DIG corrections have not been calculated for [Ne\iii], we do not use [Ne\iii]/[O\ii] to derive metallicities for the
 AM13 stacks, basing the metallicities on O$_{32}$ and [O\iii]/H$\beta$ only.

\section{Results}\label{sec:results}

\subsection{Trends between line ratios and stellar mass}

\begin{figure}
 \includegraphics[width=\columnwidth]{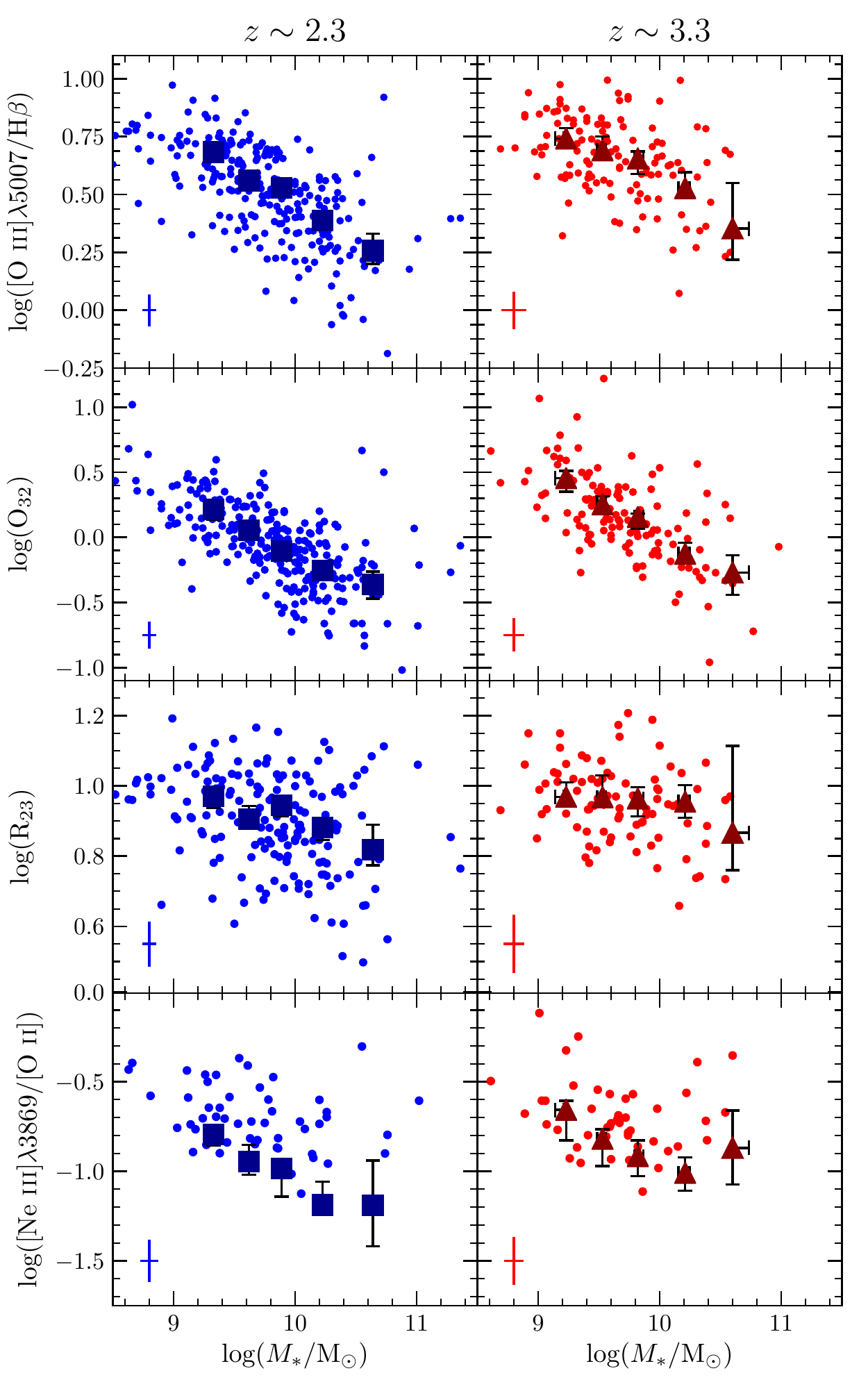}
 \centering
 \caption{
Reddening-corrected emission-line ratios vs.\ \mstar\ for $z\sim2.3$ (left) and $z\sim3.3$ (right) individual galaxies
 (small circles) and stacked spectra in bins of \mstar\ (large squares/triangles).
Individual galaxies with S/N$\ge$3 for each line in a particular ratio are shown.
The error bar in the lower left corner of each panel displays the median uncertainty of the individual galaxies.
}\label{fig:z2z3ratios}
\end{figure}

We first investigate empirical trends between optical emission-line ratios and \mstar.
Figure~\ref{fig:z2z3ratios} presents [O\iii]/H$\beta$, O$_{32}$, R$_{23}$, and [Ne\iii]/[O\ii] vs.\ \mstar\ for
 the MOSDEF $z\sim2.3$ (left) and $z\sim3.3$ (right) samples and composite spectra.
At both $z\sim2.3$ and $z\sim3.3$, we find that all four line ratios decrease with increasing \mstar, although
 R$_{23}$ is relatively flat over most of the mass range covered by these samples.
These trends are consistent with increasing metallicity as \mstar\ increases, with most galaxies lying on the
 higher-metallicity ``upper branch'' of [O\iii]/H$\beta$ and R$_{23}$ that are double-valued with O/H.
R$_{23}$ is known to saturate at metallicities of 7.8$\lesssim$12+log(O/H)$\lesssim$8.5 (Fig.~\ref{fig:calibrations};
 see also, e.g., \citealt{kew02,tre04,mai08}).
Thus, the flatness of R$_{23}$ (especially at $z\sim3.3$) suggests that much of our sample falls in this metallicity regime.
Trends are similar for individual galaxies and stacked spectra, although individual galaxies with [Ne\iii] detections fall almost
 entirely above the stacks in [Ne\iii]/[O\ii] because of the faintness of this line and the associated selection effects.

\begin{figure}
 \includegraphics[width=\columnwidth]{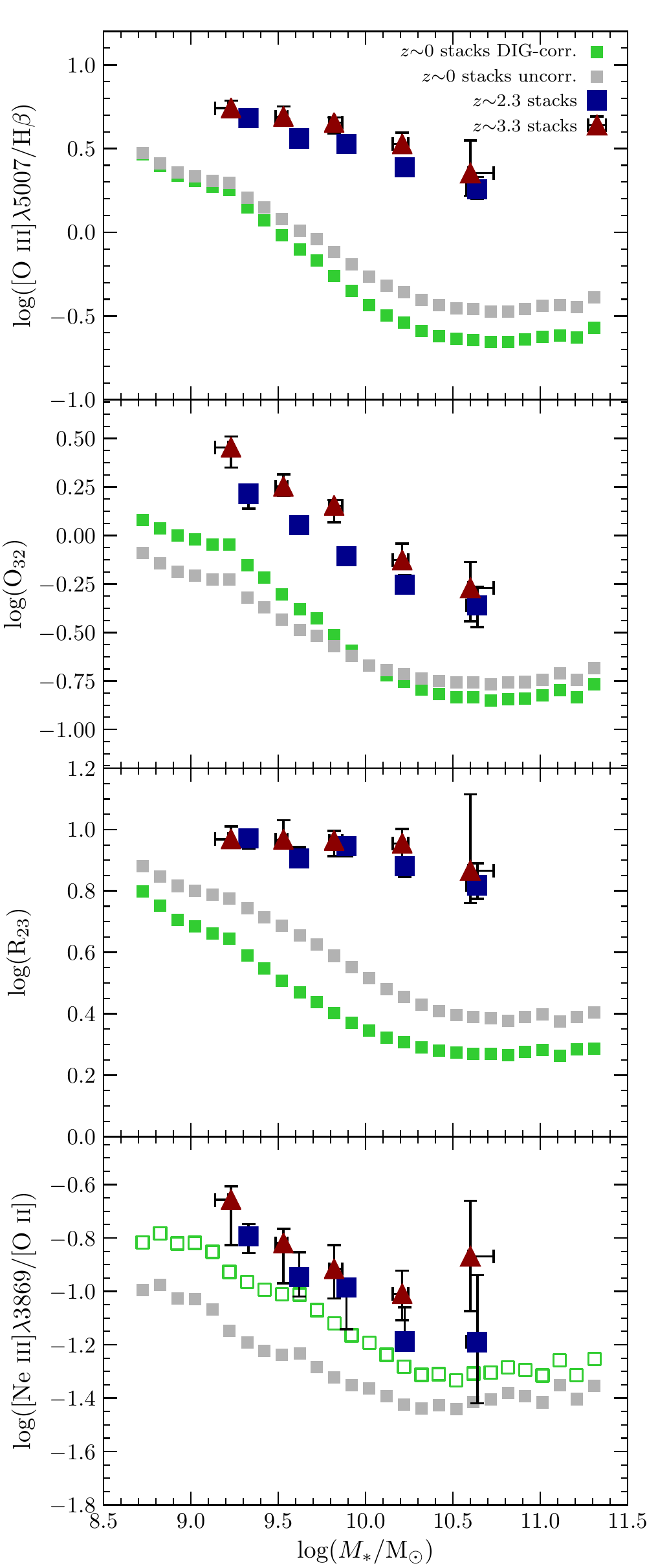}
 \centering
 \caption{
Reddening-corrected emission-line ratios vs.\ \mstar\ for stacked spectra at $z\sim0$, $z\sim2.3$ (blue), and $z\sim3.3$ (red).
Both DIG-corrected (green) and uncorrected (gray) stacks at $z\sim0$ are displayed.
The open green squares in the lower panel display the [Ne\iii]/[O\ii] ratios at $z\sim0$ after correcting only [O\ii] for DIG emission
 because a DIG correction for [Ne\iii] was not calibrated by \citet{san17}, thus representing an upper limit on the DIG-corrected
 [Ne\iii]/[O\ii] ratio.
}\label{fig:allratios}
\end{figure}

In Figure~\ref{fig:allratios}, we show the same four line ratios as a function of \mstar\ for samples at $z\sim0$, $z\sim2.3$, and $z\sim3.3$
 (only composites are displayed for the high-redshift samples).
The trends of decreasing [O\iii]/H$\beta$, O$_{32}$, R$_{23}$, and [Ne\iii]/[O\ii] with increasing \mstar\ are present in
 all samples, again suggesting that O/H increases with increasing \mstar\ at each redshift.
At fixed \mstar, all four line ratios are significantly higher at $z\sim2.3$ than at $z\sim0$.
In contrast, the line ratios only slightly increase at fixed \mstar\ from $z\sim2.3$ to $z\sim3.3$.
Collectively, these empirical trends represent a significant increase in excitation implying a
 large decrease in O/H at fixed \mstar\ between $z\sim0$ and $z\sim2.3$,
 but only a small change in O/H at fixed \mstar\ between $z\sim2.3$ and $z\sim3.3$.
At $z\sim0$, all line ratios flatten at high masses pointing towards a saturation in O/H at high \mstar.
No saturation at high \mstar\ is observed in the high-redshift samples, except for [Ne\iii]/[O\ii] of the highest-mass bins that
 may indicate low-level AGN activity is present in addition to star formation.
These empirical trends provide a qualitative picture of the MZR and its evolution,
 regardless of which strong-line metallicity calibrations are employed.

\subsection{The mass-metallicity relation at $z=0-3.3$}\label{sec:mzr}

\begin{figure*}
 \includegraphics[width=\textwidth]{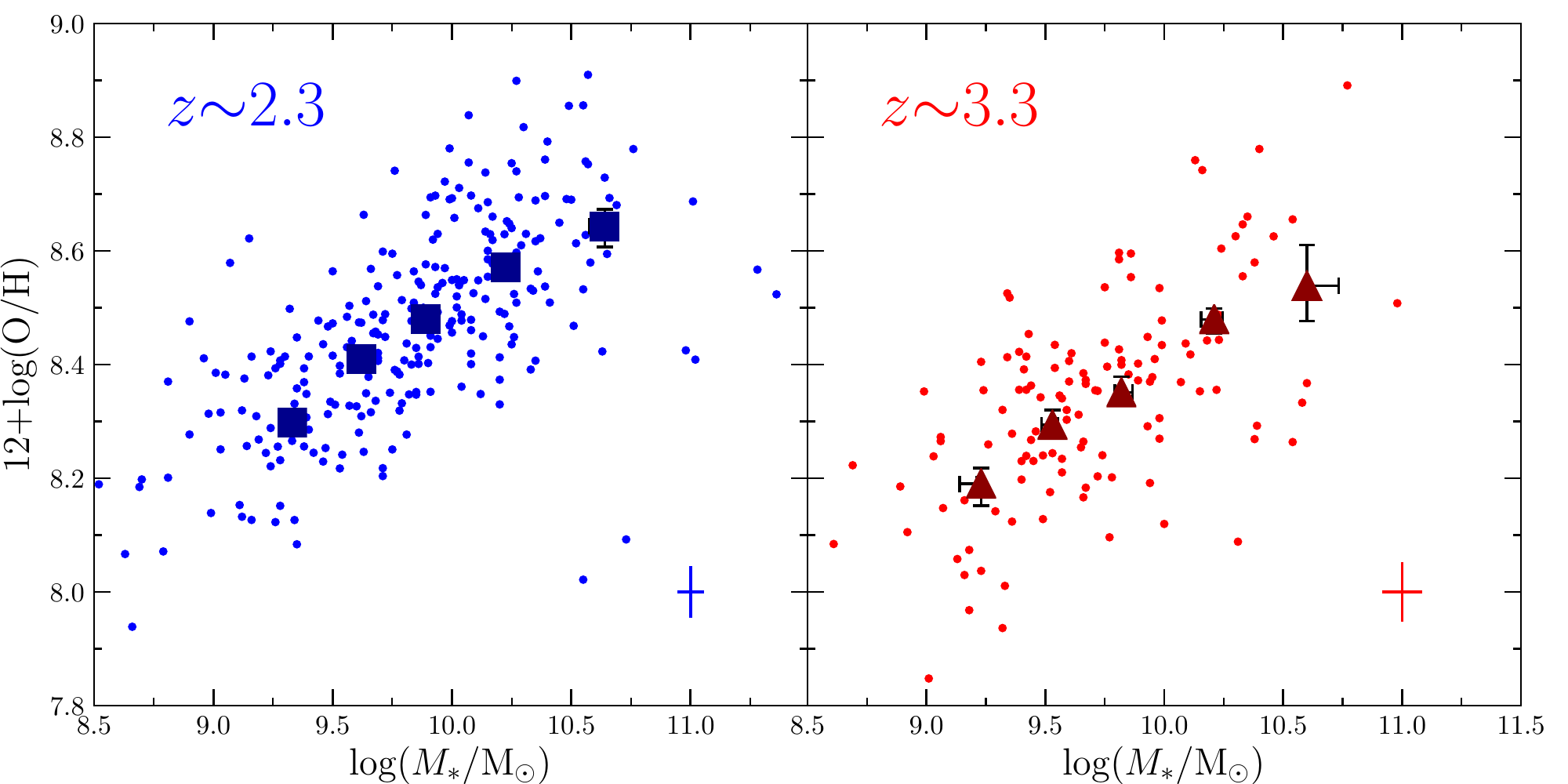}
 \centering
 \caption{
The MZR at $z\sim2.3$ (left) and $z\sim3.3$ (right) for individual galaxies (small circles)
 and stacked spectra in bins of \mstar\ (large squares/triangles).
The error bar in the lower right corner of each panel displays the median uncertainty of the individual galaxies.
}\label{fig:z2z3mzr}
\end{figure*}

We present the MZR at $z\sim2.3$ (left) and $z\sim3.3$ (right) in Figure~\ref{fig:z2z3mzr}, with O/H estimated
 as described in Section~\ref{sec:metallicity}.
We find a clear correlation between O/H and \mstar\ for both individual galaxies and composite spectra.
The $z\sim2.3$ and $z\sim3.3$ individual galaxy samples have Spearman correlation coefficients of 0.68
 and 0.56, respectively, with the p-value $\ll10^{-5}$ at both redshifts, indicating that the correlations
 between \mstar\ and metallicity are highly significant.
No obvious curvature in the relation is apparent at either redshift.
The $z\sim3.3$ sample displays slightly lower metallicity at fixed \mstar\ than the $z\sim2.3$ sample.

\begin{figure*}
 \includegraphics[width=0.8\textwidth]{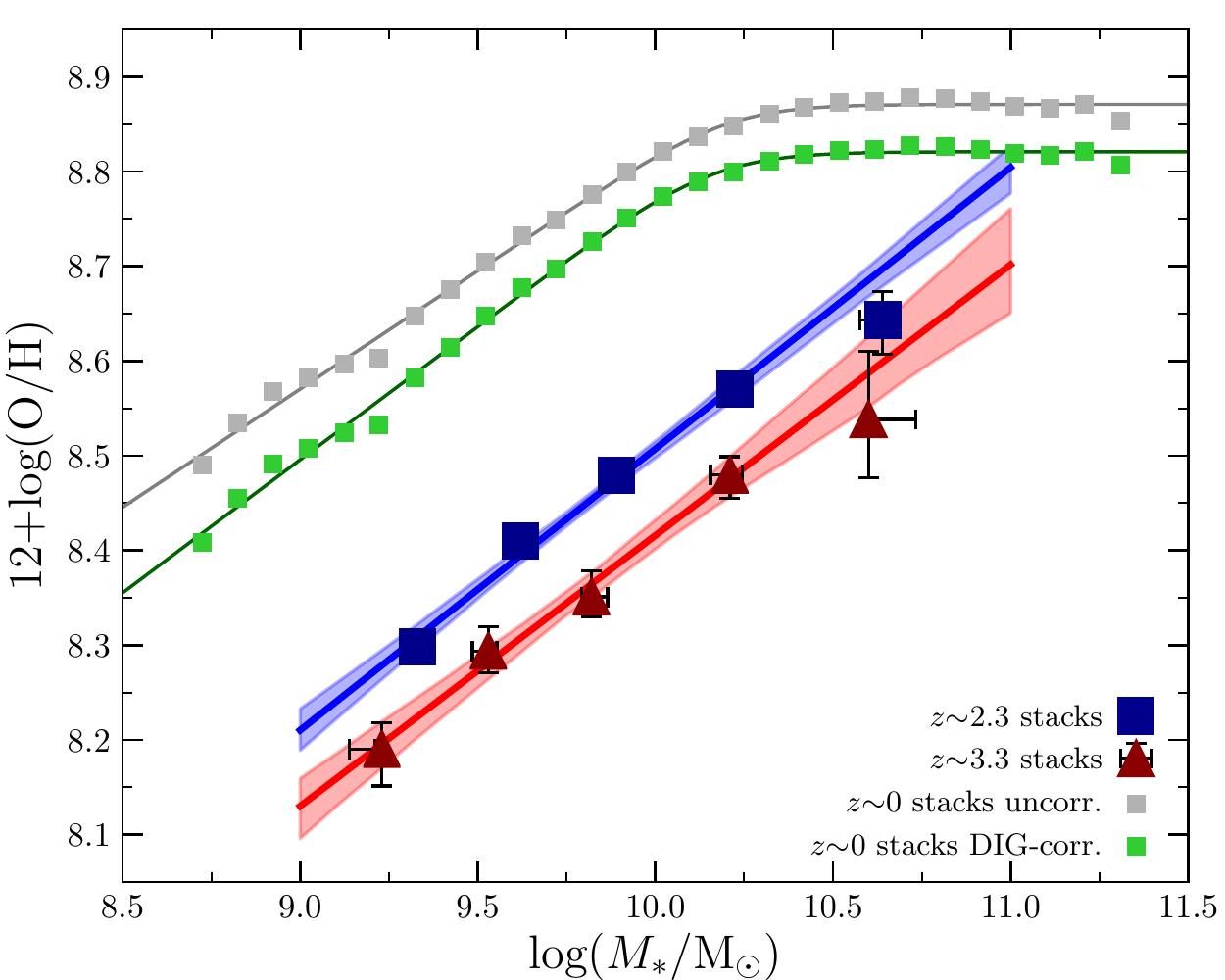}
 \centering
 \caption{
MZR for stacked spectra at $z\sim0$, $z\sim2.3$ (blue), and $z\sim3.3$ (red).
Both DIG-corrected (green) and uncorrected (gray) values are shown for the $z\sim0$ stacks.
The formal uncertainty in O/H for the $z\sim0$ stacks is typically smaller than the size of the points.
The solid lines represent the best-fit relations at each redshift, while the shaded regions display the $1\sigma$
 uncertainties of the fits at $z\sim2.3$ and $z\sim3.3$.
The $z\sim0$ data are fit using a smoothly broken power law (equation~\ref{eq:mzrbpl}).
At $z\sim2.3$ and $z\sim3.3$, the stacks (excluding the highest-mass bin) are fit with a power law (equation~\ref{eq:mzrlin}).
}\label{fig:allmzr}
\end{figure*}

The MZRs for stacked spectra at $z\sim0$, $z\sim2.3$, and $z\sim3.3$ are shown in
 Figure~\ref{fig:allmzr}.
Over the range of masses covered by our samples, we find a monotonic evolution towards lower metallicity
 with increasing redshift.
The evolution from $z\sim0$ to $z\sim2.3$ is markedly larger than the evolution from $z\sim2.3$ to $z\sim3.3$.
Because the high-redshift data do not obviously display a flattening at high mass,
 we fit the $z\sim2.3$ and $z\sim3.3$ stacks with a single power law of the form
\begin{equation}\label{eq:mzrlin}
12+\mbox{log(O/H)}=\gamma\times m_{10} + Z_{10}
\end{equation}
where $m_{10}=\log(M_*/10^{10}~\msun)$ and $Z_{10}$ is the metallicity at $10^{10}~\msun$.
We exclude the highest-mass bins at each redshift from this power law fit because of incompleteness in the MOSDEF sample at these masses
 (Sec.~\ref{sec:stacking}; \citealt{kri15}).
Accordingly, these fits are good over $9.0\le\log$($M_*$/$\msun)\le10.5$, though we extend the lines in Figure~\ref{fig:allmzr}
 to 10$^{11}~\msun$ for comparison to the highest-mass bins.
The best-fit relations and 1$\sigma$ uncertainties are shown in Figure~\ref{fig:allmzr},
 and the best-fit parameters are given in Table~\ref{tab:mzrbestfits}.

\begin{table}
 \centering
 \caption{Best-fit mass-metallicity relations to $z\sim0$, $z\sim2.3$, and $z\sim3.3$ composite spectra.
 }\label{tab:mzrbestfits}
 \begin{tabular}{ l l l l l }
   \hline\hline
   \multicolumn{5}{c}{$z\sim2.3$ and $z\sim3.3$ fits using equation~\ref{eq:mzrlin}} \\
   \hline
   sample & \multicolumn{2}{l}{$\gamma$} & \multicolumn{2}{l}{$Z_{10}$} \\
   \hline
   $z\sim2.3$ & \multicolumn{2}{l}{$0.30\pm0.02$} & \multicolumn{2}{l}{$8.51\pm0.02$} \\
   $z\sim3.3$ & \multicolumn{2}{l}{$0.29\pm0.02$} & \multicolumn{2}{l}{$8.41\pm0.03$} \\
   \hline
   \hline
   \multicolumn{5}{c}{$z\sim0$ fits using equation~\ref{eq:mzrbpl}} \\
   \hline
     & $Z_0$ & log$\left(\frac{M_{\text{TO}}}{\msun}\right)$ & $\gamma$ & $\Delta$ \\
   \hline
   \makecell{$z\sim0$ \\ DIG corr.} & 8.82$\pm$0.01 & 10.16$\pm$0.03 & 0.28$\pm$0.01 & 3.43$\pm$0.92 \\
   \makecell{$z\sim0$ \\ uncorr.} & 8.87$\pm$0.01 & 10.20$\pm$0.03 & 0.25$\pm$0.01 & 3.66$\pm$1.16 \\
   \hline
 \end{tabular}
\end{table}

We fit the $z\sim0$ MZR with the parameterization of \citet{cur20b}.
This function is a smoothly broken power law that approaches a constant slope $\gamma$ at masses below
 the turnover mass, \mto, and asymptotes to a constant metallicity $Z_0$  above \mto.
The functional form is
\begin{equation}\label{eq:mzrbpl}
12+\log(\mbox{O/H})=Z_0 - \gamma/\Delta\times\log\left[1+\left(\frac{M_*}{M_{\text{TO}}}\right)^{-\Delta}\right]
\end{equation}
where $\Delta$ is a smoothness parameter that dictates how sharp the transition between the two mass regimes is at \mto.
The transition region becomes smaller (i.e., higher curvature) as $\Delta$ increases.
Unlike earlier works that employed similar functional forms of the MZR that had a fixed curvature \citep[e.g.,][]{mou11,and13,zah14a},
 equation~\ref{eq:mzrbpl} allows for the curvature to be fit along with \mto, $Z_0$, and $\gamma$.
The best-fit parameters to both the DIG-corrected and uncorrected $z\sim0$ stacks are presented in Table~\ref{tab:mzrbestfits} and shown
 in Figure~\ref{fig:allmzr} by the green and gray lines, respectively.
The MZR shape is very similar between the two cases, but the normalization is $\approx0.05$~dex higher without correcting for DIG.
We consider the DIG-corrected data to be more accurate (Sec.~\ref{sec:z0sample}) and therefore adopt this as the fiducial case at $z\sim0$.
Our best-fit $z\sim0$ MZR is very similar to the best-fit relation of \citet{cur20b},
 fit to median values of individual SDSS galaxies
 binned by stellar mass instead of stacked spectra.
These authors find $\gamma=0.28\pm0.02$, $\log(M_{\text{TO}}/\msun)=10^{10.02\pm0.09}~\msun$, and $Z_0=8.793\pm0.005$, consistent with our values,
 although they infer a smaller curvature ($\Delta=1.2\pm0.2$).

\subsubsection{Low-mass slope, normalization, and scatter}\label{sec:mzr1}

At all redshifts, we find that the low-mass behavior of the MZR is consistent with a power law, with no evidence
 of the MZR slope either increasing or decreasing towards $10^9~\msun$.
The best-fit low-mass MZR slopes are remarkably consistent to high precision across all three redshifts, with
 $\gamma=0.28\pm0.01$ at $z\sim0$, $0.30\pm0.02$ at $z\sim2.3$, and $0.29\pm0.02$ at $z\sim3.3$.
This invariance of the MZR slope over 12~Gyr of cosmic time suggests that the same process sets the slope of the
 MZR over $z=0-3.3$.

At 10$^{10}~\msun$, the metallicities of the best-fit relations are 12+log(O/H$)=8.77\pm0.01$ at $z\sim0$,
 $8.51\pm0.02$ at $z\sim2.3$, and $8.41\pm0.03$ at $z\sim3.3$.
Thus, at log($M_*/\msun)=10.0$, we find an evolution of $-0.26\pm0.02$~dex in O/H from $z\sim0$ to $z\sim2.3$,
 and $-0.10\pm0.03$ between $z\sim2.3$ and $z\sim3.3$.
Because the low-mass slopes are almost identical, the offset in metallicity at fixed \mstar\ between $z\sim2.3$ and $z\sim3.3$
 is nearly constant below $10^{10.5}~\msun$.
Likewise, the O/H offset at fixed \mstar\ between the $z\sim0$ stacks and the high-redshift samples is constant below
 $\sim10^{10.2}~\msun$, decreasing at higher masses as the $z\sim0$ MZR flattens.
Given the median redshifts of our samples ($z_{\text{med}}=[0.08, 2.28, 3.24]$), the data are consistent with
 a uniform metallicity evolution of dlog(O/H)/d$z=-0.11\pm0.02$ below $10^{10.2}~\msun$ (the turnover mass at $z\sim0$).

We utilize the formal measurement uncertainties ($\sigma_{\text{meas}}$) on the metallicities and the scatter in the calibrations
 ($\sigma_{\text{cal}}$) to estimate the intrinsic scatter ($\sigma_{\text{int}}$) of the MZR at $z\sim2.3$ and $z\sim3.3$,
 assuming the observed scatter is $\sigma_{\text{obs}}^2=\sigma_{\text{int}}^2+\sigma_{\text{meas}}^2+\sigma_{\text{cal}}^2$.
Individual MOSDEF galaxies (Fig.~\ref{fig:z2z3mzr}) have an observed scatter of 0.14(0.17)~dex in O/H at fixed \mstar\ around
 the best-fit $z\sim2.3$(3.3) MZR.
The mean O/H measurement uncertainty is 0.04(0.05)~dex at $z\sim2.3$(3.3).
The B18 calibrations are based on stacked spectra and thus do not have measured calibration scatters.
We instead assume the same scatter in line ratio at fixed O/H as for the $z\sim0$ calibrations (Table~\ref{tab:calibrations}),
 convert to scatter in O/H at fixed line ratio using the slope of the calibrations at 12+log(O/H$)=8.4$ (the mean metallicity of
 the MOSDEF samples), and take the average of the calibration scatters among the set of line ratios used to derive the metallicities.
In this way, we estimate the calibration scatter to be 0.11~dex in O/H.
We infer the intrinsic $1\sigma$ scatter of the MZR to be 0.08~dex at $z\sim2.3$ and 0.11~dex at $z\sim3.3$, consistent
 with the intrinsic scatter of the $z=0$ MZR of $\approx0.1$~dex \citep{tre04,kew08,man10,yat12,cur20b}.

\subsubsection{Turnover mass and asymptotic metallicity}\label{sec:mzr2}

The $z\sim0$ MZR clearly flattens and approaches an asymptotic O/H at high masses.
Our best-fit $z\sim0$ MZR has a turnover mass of log(\mto/M$_{\odot})=10.16\pm0.03$ and a high-mass asymptotic metallicity of $Z_0=8.82\pm0.01$.
The high-mass flattening reflects the underlying physics that govern ISM metallicity, such that the differing behavior of the MZR
 at high-\mstar\ implies some fundamental change in metal production, dilution, and/or retention/removal \citep[e.g.,][]{tre04,zah14a,tor19}.
The turnover mass has been found to increase with increasing redshift out to $z\sim1.5$, while $Z_0$ displays little evolution
 over this range \citep{zah14a,zah14b}.
It is of interest to see if these trends continue at $z>2$.

The highest-mass bins at $z\sim2.3$ and $z\sim3.3$ fall below the best-fit MZR at each redshift,
 suggesting a possible flattening of the high-redshift MZR beginning at $\sim10^{10.5}~\msun$.
However, both highest-mass bins are $<2\sigma$ consistent with the single power-law fits.
These bins have the largest O/H uncertainties because they contain the lowest number of galaxies (see Table~\ref{tab:stacks}).
Furthermore, the highest-mass bins are potentially biased against red, dusty, metal-rich galaxies (see Sec.~\ref{sec:stacking}),
 which may explain why they fall below the power law that fits the lower-mass composites.
Due to these uncertainties and biases, we cannot place quantitative constraints on the value of \mto\ or $Z_0$ at $z\sim2.3-3.3$.
We can however say with confidence that the turnover mass at $z>2$ must be larger than \mto\ at $z\sim0$ ($10^{10.2}~\msun$)
 since we find no flattening in the four MOSDEF bins where the sample is complete that span up to log($M_*/\msun)=10.5$.
Our data suggest that the turnover mass at $z\sim2.3$ is larger than the value found at $z\sim1.4$ of log(\mto/M$_{\odot})\approx10.1$
 by \citet{zah14a}, and is thus consistent with their finding that \mto\ increases with increasing redshift.

Constraining the high-mass behavior of the MZR at $z>2$ and confirming whether the MZR flattens at all at these redshifts
 will require significantly larger and more complete samples of galaxies at log($M_*/\msun)>10.5$.
If the single power laws hold with no flattening, then the $z\sim2.3$(3.3) MZR would reach $Z_0(z=0)$ at log($M_*/\msun)\approx11.0$(11.5).
A robust investigation of the high-mass behavior should thus be well-sampled up to at least 10$^{11.0}~\msun$.
Given the rarity of such massive star-forming systems at high redshifts, assembling a sufficient sample will require a very wide
 area search exceeding that of existing deep legacy fields (e.g., CANDELS).
Constraining the high-mass MZR is crucial to understanding whether there are two regimes of metal processing in galaxies
 at high redshifts as at $z\sim0$.

\subsection{The fundamental metallicity relation at $z=0-3.3$}\label{sec:fmr}

\begin{figure*}
 \includegraphics[width=\textwidth]{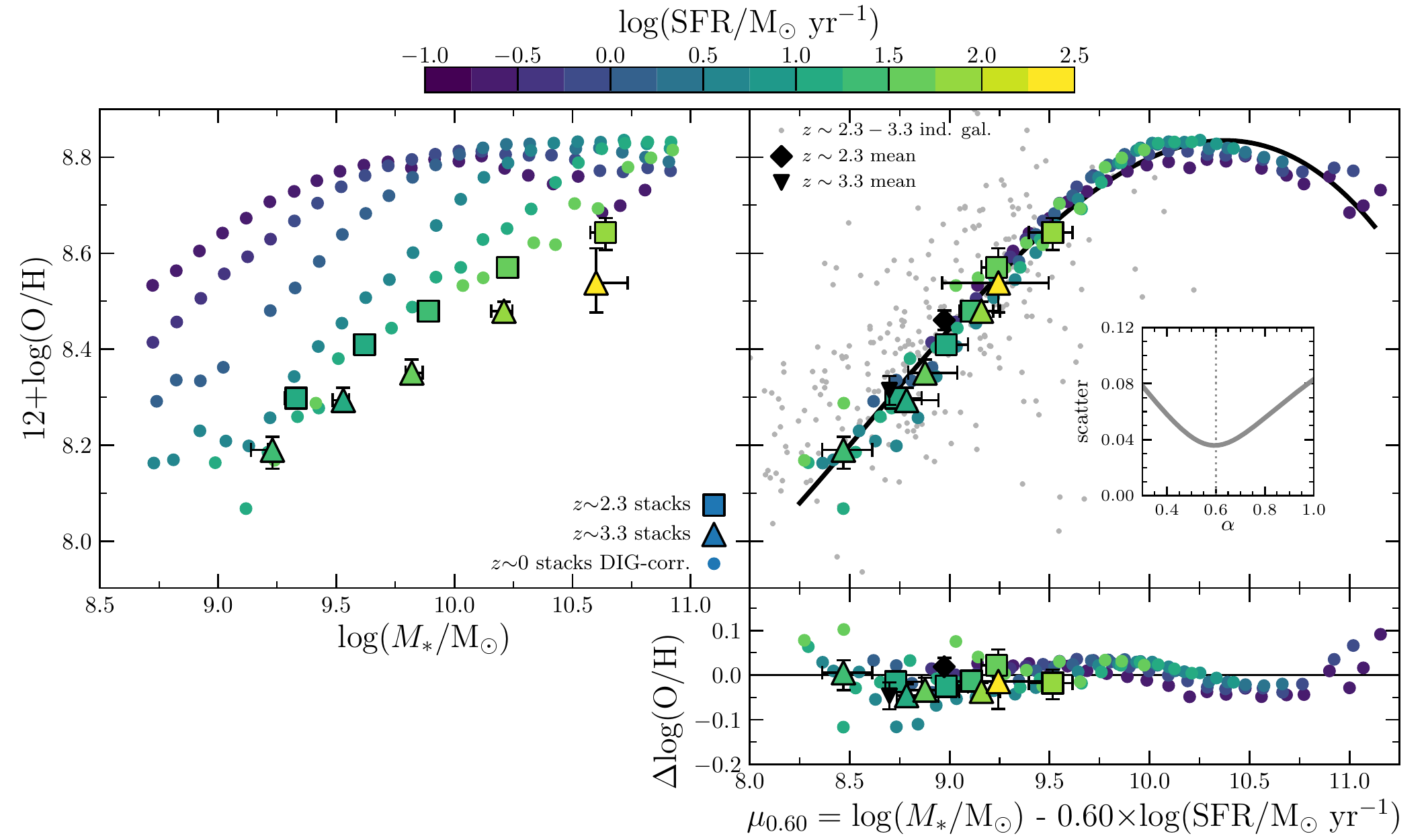}
 \centering
 \caption{
\textbf{Left:} O/H vs.\ \mstar\ for stacked spectra of galaxies at $z\sim0$ (circles), $z\sim2.3$ (squares),
 and $z\sim3.3$ (triangles).
Points are color-coded by SFR, where the SFR is determined using dust-corrected H$\alpha$ or H$\beta$ for all samples.
\textbf{Upper right:} Projection of the FMR as O/H
 vs.\ $\mu_{0.60}=\log(M_*/\text{M}_{\odot}) - 0.60\times\log(\text{SFR}/\text{M}_{\odot}\ \text{yr}^{-1})$,
 where the coefficient of the SFR, $\alpha=0.60$, is that which minimizes the scatter of the $z\sim0$ \mstar+SFR binned stacks.
The inset panel presents the $z\sim0$ scatter in O/H at fixed $\mu_{\alpha}$ as a function of $\alpha$.
The black line displays the best-fit cubic function to the $z\sim0$ stacks, given in equation~\ref{eq:fmr}.
Gray circles show $z\sim2.3$ and $z\sim3.3$ individual galaxies with metallicity measurements, while the black diamond
 and triangle show the mean values of these galaxies at $z\sim2.3$ and $z\sim3.3$, respectively.
\textbf{Lower right:} Residuals in O/H around the best-fit $z\sim0$ FMR projection (black line, upper right).
The high-redshift galaxies show excellent agreement with the $z\sim0$ FMR.
}\label{fig:fmr}
\end{figure*}

We now investigate the three-dimensional relation among \mstar, O/H, and SFR (i.e., the FMR), and whether this relation evolves with redshift.
We show O/H vs.\ \mstar\ color-coded by SFR at $z\sim0$ (circles), $z\sim2.3$ (squares), and $z\sim3.3$ (triangles) in Figure~\ref{fig:fmr},
 where the $z\sim0$ stacks are now those of AM13 binned in both \mstar\ and SFR.
We limit the $z\sim0$ sample to those \mstar-SFR bins containing at least 5 galaxies to ensure the stacks still represent
 a sample average.
This cut primarily limits the stacks at $<10^{10.0}~\msun$ with log(SFR/$\msun$ yr$^{-1})>1.5$ because of the rarity of
 galaxies with such extreme sSFRs in the local universe.
The high-redshift samples appear to show good agreement in O/H with the $z\sim0$ stacks where low-redshift stacks matched
 in \mstar\ and SFR exist.
There are no $z\sim0$ stacks closely matched to the highest-mass $z\sim2.3$ bin or the three highest-mass $z\sim3.3$ bins.
We note that the high-redshift stacks remain in close agreement with matched $z\sim0$ stacks without DIG correction because
 the DIG corrections to these high-sSFR $z\sim0$ stacks are small due to their large H$\alpha$ surface brightnesses.
However, the agreement is closer when a DIG correction is performed.

We parameterize the $z\sim0$ FMR using the method of \citet{man10}, where the value of $\alpha$ is identified that minimizes
 the scatter in O/H at fixed $\mu_{\alpha}$, where
\begin{equation}\label{eq:mu}
\mu_{\alpha} \equiv \log(M_*/\mbox{M}_{\odot}) - \alpha\times\text{log}\left(\frac{\text{SFR}}{\msun~\text{yr}^{-1}}\right) .
\end{equation}
While this simplistic functional form of the FMR can fail to capture the detailed flattening and turnover behavior
 at very high masses and low sSFRs \citep{yat12,cur20b}, we find that it is sufficient to describe the behavior over the
 range of masses and SFRs spanned by the $z\sim0$ stacks.
For a range of $\alpha$, we fit 12+log(O/H) vs.\ $\mu_{\alpha}$ of the $z\sim0$ stacks with a cubic function and calculate
 the residuals about the best-fit function.
We find that the scatter of the $z\sim0$ stacks is minimized at a value of $\alpha=0.60$ (right panel inset in Fig.~\ref{fig:fmr}).
This best-fit $\alpha$ is in close agreement with the values inferred using direct-method metallicities alone
 \citep[$\alpha=0.55-0.70$;][]{and13,san17}, and is also close to best fit for individual SDSS galaxies of $\alpha=0.55$
 found by \citet{cur20b}.
The best-fit $z\sim0$ FMR is shown by the black line in the right panel of Figure~\ref{fig:fmr}, with a functional form of
\begin{equation}\label{eq:fmr}
12+\log(\mbox{O/H})= 8.80 + 0.188 y - 0.220 y^2 - 0.0531 y^3
\end{equation}
where $y=\mu_{0.60} - 10$.
In this parameter space, the $z\sim2.3$ and $z\sim3.3$ stacks fall directly on the best-fit $z\sim0$ FMR, despite the high-redshift
 stacks not being included in the fitting process.

The lower panel of Figure~\ref{fig:fmr} displays the metallicity residuals at fixed $\mu_{0.60}$ about the best-fit $z\sim0$ FMR.
Collectively, the weighted-mean offset of all high-redshift stacks is $\Delta\log(\mbox{O/H})=-0.01\pm0.02$~dex,
 where the uncertainty reported here is the error of the weighted mean.
The individual MOSDEF galaxies with both metallicity and SFR detections (gray points, Fig.~\ref{fig:fmr}) have a mean offset in O/H of
 $0.04\pm0.02$ at $z\sim2.3$ and $0.02\pm0.03$ at $z\sim3.3$ (black points, Fig.~\ref{fig:fmr}).
We thus find that a single relation among \mstar, SFR, and O/H can describe the mean properties of galaxy samples over $z=0-3.3$
 with high precision.
In other words, the FMR does not evolve out to $z\sim3.3$.

The observed scatter of the O/H residuals of the individual galaxies,
 taken to be the standard deviation,
 is 0.16~dex at $z\sim2.3$ and 0.22~dex at $z\sim3.3$.
We perform the same scatter analysis as for the MZR (Sec.~\ref{sec:mzr1}), except here
 measurement errors account for uncertainty in both O/H and SFR since $\mu_{0.60}$ depends on SFR and the SFRs carry
 significant errors (typically $\sim0.2-0.3$~dex).
After removing the measurement uncertainty in $\Delta$log(O/H) at fixed $\mu_{0.60}$
 ($\sigma_{\text{meas}}=0.10$ (0.18)~dex at $z\sim2.3$ (3.3)) and the metallicity calibration scatter of $\sigma_{\text{cal}}=0.11$~dex,
 we find an intrinsic scatter around the best-fit FMR of 0.06~dex at both $z\sim2.3$ and $z\sim3.3$.
This intrinsic FMR scatter is lower than the intrinsic MZR scatter at $z\sim2.3$ (3.3) of 0.08 (0.12)~dex (Sec.~\ref{sec:mzr1}),
 indicating a second parameter dependence on SFR is present in the high-redshift data.
At $z=0$, the intrinsic scatter of the FMR is $\approx0.05$~dex \citep[e.g.,][]{man10,cre19,cur20b}, where the inclusion
 of SFR as an additional parameter has decreased the scatter from the value of 0.1~dex found for the MZR.
We thus find that the addition of SFR as a secondary parameter to the MZR results in a similar decrease
 in the intrinsic scatter in O/H at $z\sim2.3$ and $z\sim3.3$, from $\approx0.10$~dex around the MZR to 0.06~dex around the FMR.

The FMR projection in the right panel of Figure~\ref{fig:fmr} displays a flattening above $\mu_{0.60}=10.0$ where O/H has no dependence on
 SFR at fixed \mstar.
This flattening behavior at high-mass and low-SFR is a feature on which there is a consensus in the
 literature \citep[e.g.,][]{man10,yat12,and13,tel16,cre19,cur20b}.
The highest-mass $z\sim2.3$ stack has $\mu_{0.60}=9.5$, below the regime where the $z\sim0$ stacks begin to flatten.
Even at log($M_*/\msun)=11.0$, $z\sim2.3$ galaxies would only have $\mu_{0.60}\approx9.7$ assuming our best-fit SFR-\mstar\ relation
 holds (equation~\ref{eq:z2sfrmstar}), making it impractical to probe $\mu_{0.60}>10.0$ with samples at $z>2$.
It is of interest to more extensively test the FMR using high-mass and low-SFR galaxies at intermediate redshifts ($z\sim0.5-1.5$)
 to confirm whether the flattening at high $\mu_{0.60}$ remains beyond the local universe.

At $\mu_{0.60}<9.5$, the best-fit FMR can be described as a power law of the form $\mbox{O/H}\propto\mu_{0.60}^{0.45}$.
Accordingly, $\mbox{O/H}\propto\mbox{SFR}^{-0.27}$ at fixed \mstar\ in the best-fit FMR,
 based on the definition of $\mu_{0.60}$ (equation~\ref{eq:mu}).
For $\mbox{O/H}\propto\mbox{SFR}^{\nu}$, \citet{san18} found $\nu=-0.11$ to $-0.27$ for MOSDEF star-forming galaxies at $z\sim2.3$,
 where the strength of the SFR dependence varied with the choice of metallicity indicator and calibration.
We rederived metallicities for the \citet{san18} stacks that were binned in \mstar\ and offset from the $z\sim2.3$ SFR-\mstar\ relation
 (the ``\mstar-$\Delta$sSFR'' stacks) using the reported O$_{32}$ and [O\iii]/H$\beta$ ratios ([Ne\iii] was not covered in these stacks)
 and the B18 high-redshift analog calibrations.
The $z\sim2.3$ sample in this work has $\sim$80\% overlap with that of \citet{san18}, thus the stacks of \citet{san18} should be a fair
 representation of our sample.

\begin{figure}
 \includegraphics[width=\columnwidth]{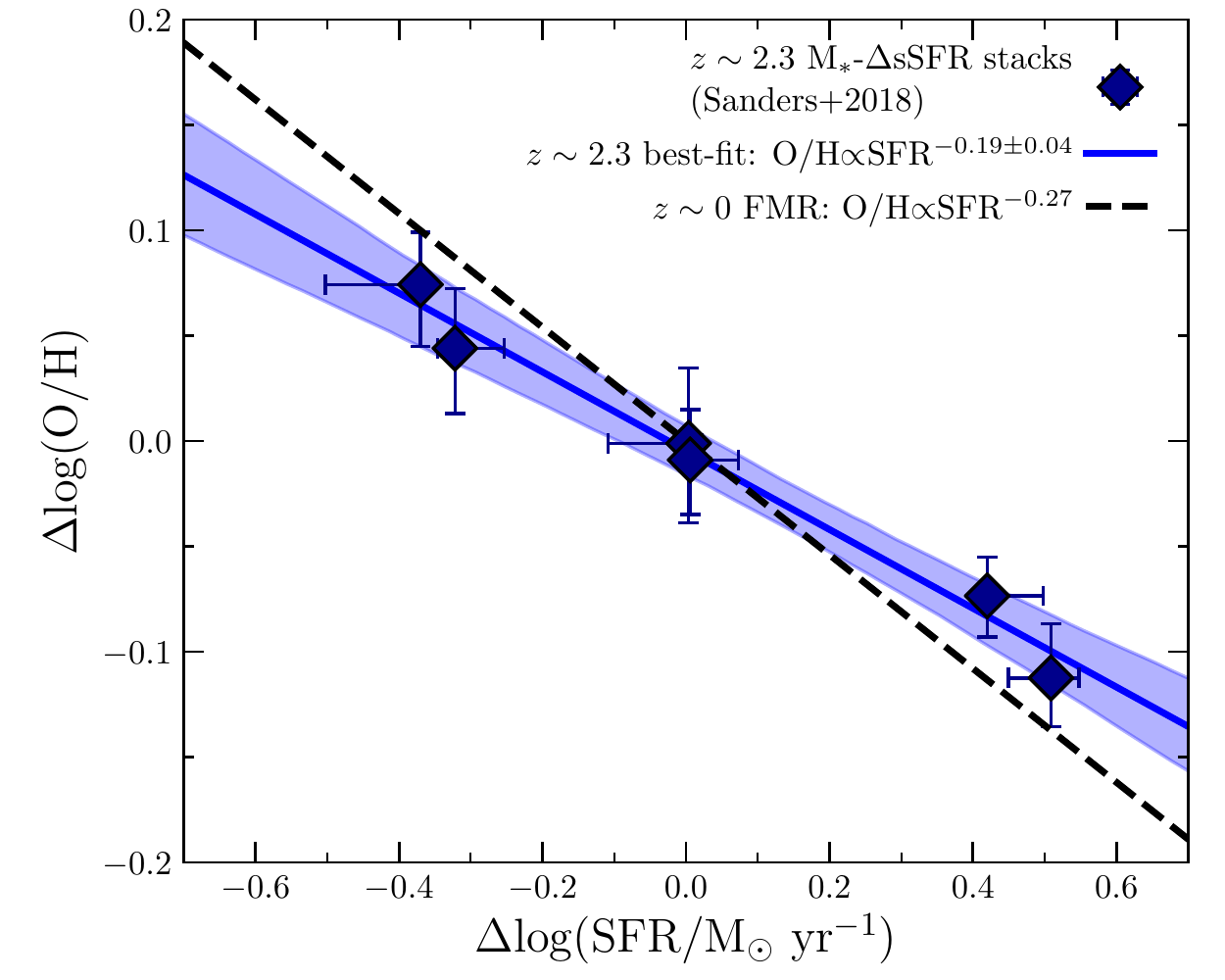}
 \centering
 \caption{
Residuals around the best-fit best-fit MZR ($\Delta$log(O/H)) vs.\ residuals around the best-fit SFR-\mstar\ relation
 ($\Delta$log(SFR)) for the $z\sim2.3$ \mstar-$\Delta$sSFR stacks of \citet[][blue diamonds]{san18}.
The metallicities of these stacks have been rederived using the methods in this work (Sec.~\ref{sec:metallicity}).
The blue line and shaded region shows the best-fit relation to the $z\sim2.3$ stacks, while the black dashed
 line displays the dependence of O/H on SFR at fixed \mstar\ predicted by the best-fit $z\sim0$ FMR.
}\label{fig:delohdelsfr}
\end{figure}

In Figure~\ref{fig:delohdelsfr}, we compare the residuals around the MZR at fixed \mstar\ ($\Delta$log(O/H)) to the residuals
 around the SFR-\mstar\ relation at fixed \mstar\ ($\Delta$log(SFR)) for the \mstar-$\Delta$sSFR stacks.
The best-fit power law to the $z\sim2.3$ stacks has $\nu=-0.19\pm0.04$ (blue line).
This relation is shallower than what is expected from the best-fit $z\sim0$ FMR ($\nu=-0.27$), but the offset is not
 statistically significant ($2\sigma$).
We thus find that the dependence of O/H on SFR at fixed \mstar\ internal to the $z\sim2.3$ sample is consistent with the expectation
 from the best-fit $z\sim0$ FMR.
Because of the smaller sample size and larger uncertainties on O/H and SFR, performing this exercise with stacks of the
 $z\sim3.3$ sample does not produce any useful constraints.
In summary, the secondary dependence of O/H on SFR is significantly detected at $z\sim2.3$ and is consistent at 2$\sigma$
 with the dependence measured at $z\sim0$.

Potential systematic biases remain in the comparison of the $z\sim0$ and $z>2$ FMR.
Based on the correlation between  high sSFR and extreme ISM conditions
 (see discussion in Sec.~\ref{sec:metallicity}), a different metallicity calibration
 may be needed for $z\sim0$ high-sSFR galaxies relative to $z\sim0$ main-sequence galaxies.
This approach will affect the comparison of low- and high-redshift galaxies in the FMR plane.
For example, if we apply the B18 calibrations to the $z\sim0$ \mstar-SFR stacks with log(sSFR/Gyr$^{-1})>-0.5$,
 it introduces a small offset of $-$0.05~dex in O/H between the $z>2$ stacks and the $z\sim0$ stacks most closely
 matched in \mstar\ and SFR.
More accurate derivations of the FMR and its evolution require a move beyond simple one-dimensional metallicity
 calibrations to multi-dimensional calibrations that take into account variations in ISM conditions across the galaxy population.
Such relations may be able to unify high-redshift and local calibrations into a single framework.
\citet{bro16} investigated the possibility of sSFR-dependent metallicity calibrations but did not include corrections
 for contributions from DIG emission that are highly sSFR dependent \citep[e.g.,][]{san17,val19}.
The combination of high-quality spectra and large sample sizes at $z>2$ has reduced the measurement uncertainties
 to a level where these finely detailed systematic effects need to be addressed in future FMR studies.

\subsection{Systematic effects on the high-redshift MZR}\label{sec:mzrsys}

\begin{figure*}
 \includegraphics[width=\textwidth]{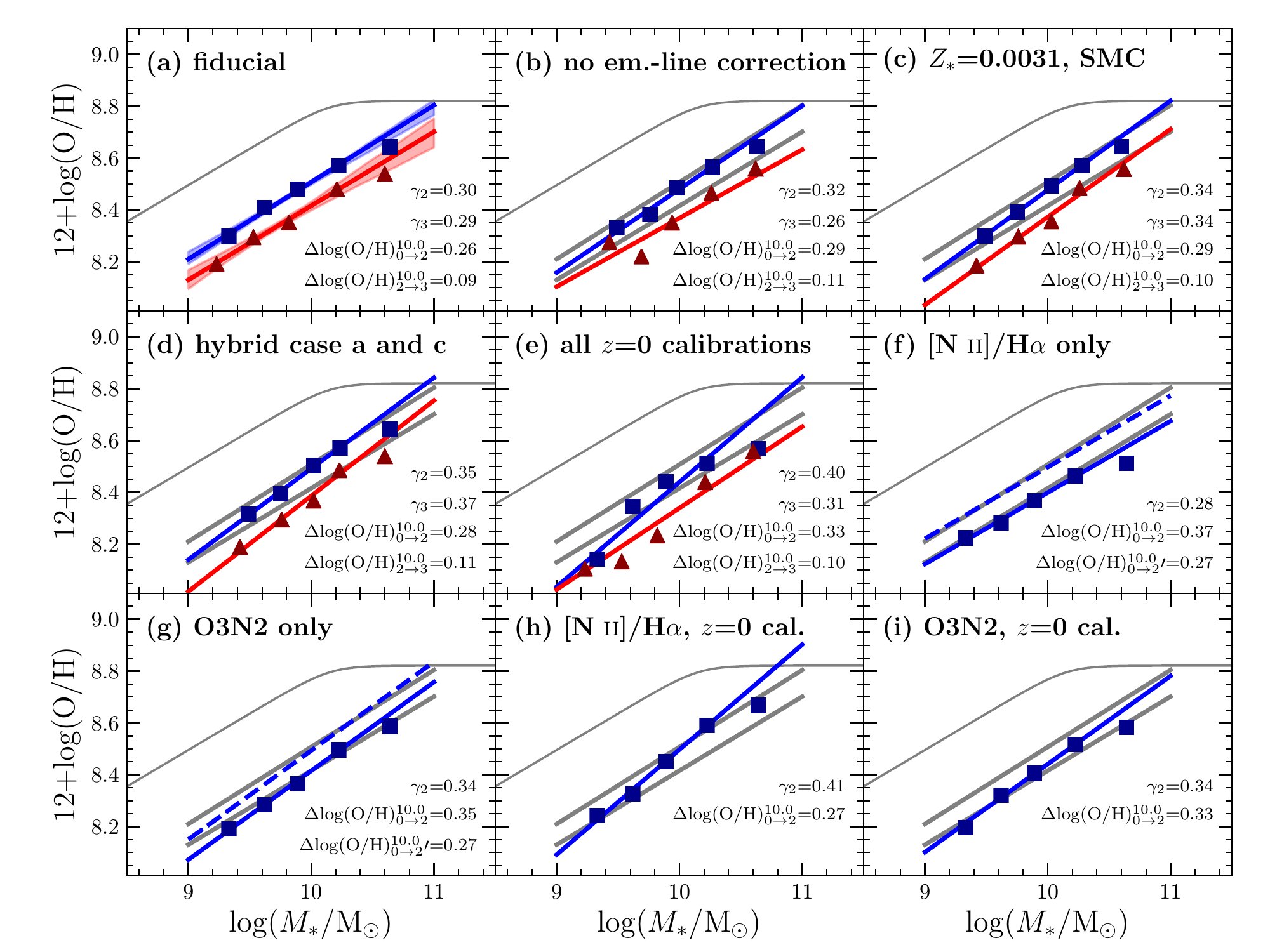}
 \centering
 \caption{
Each panel shows stacks and fits to the $z\sim2.3$ (blue) and $z\sim3.3$ (red)  MZR, where assumptions regarding SED fitting for stellar masses or
 calibrations for metallicity derivation are varied in each panel as described by the text at the top of each panel.
Panel (a) presents the $z\sim2.3$ and $z\sim3.3$ MZRs under our fiducial set of assumptions.
In each panel, the fiducial best-fit MZRs at $z\sim0$, $z\sim2.3$, and $z\sim3.3$ are shown as gray lines for comparison.
The text in the lower-right corner gives the best-fit MZR slopes at $z\sim2.3$ ($\gamma_2$) and $z\sim3.3$ ($\gamma_3$),
 as well as the offset in O/H at $10^{10}~\msun$ from $z\sim0$ to $z\sim2.3$ ($\Delta$log(O/H)$^{10.0}_{0\rightarrow2}$)
 and from $z\sim2.3$ to $z\sim3.3$ ($\Delta$log(O/H)$^{10.0}_{2\rightarrow3}$).
In panels (f) and (g), the dashed blue line displays the MZR inferred when shifting the B18 high-redshift analog calibrations
 0.15~dex lower in [N\ii]/H$\alpha$ and higher in O3N2 such that the B18 excitation sequences match that of the MOSDEF $z\sim2.3$
 sample in the [N\ii] BPT diagram.
}\label{fig:mzrsys}
\end{figure*}

We now investigate how assumptions for determining stellar masses and deriving metallicities from strong-line ratios
 systematically affect the shape and normalization of the MZR of the $z\sim2.3$ and $z\sim3.3$ samples.
Figure~\ref{fig:mzrsys}a displays the case for our fiducial assumptions for SED fitting to estimate stellar masses
 (Sec.~\ref{sec:sedfitting}) and the B18 high-redshift analog calibrations to infer metallicities from [O\ii], [Ne\iii],
 [O\iii], and H$\beta$ (Sec.~\ref{sec:metallicity}).
In this and the following panels of Figure~\ref{fig:mzrsys}, we display the high-redshift stacks, power-law fits to the stacks
 (excluding the highest-mass bin), and print the best-fit slope of the MZR ($\gamma_2$ ($\gamma_3$) at $z\sim2.3$ (3.3))
 as well as the O/H offset at log(\mstar/$\msun)=10.0$ from $z\sim0$ to $z\sim2.3$ ($\Delta$log(O/H)$^{10.0}_{0\rightarrow2}$)
 and from $z\sim2.3$ to $z\sim3.3$ ($\Delta$log(O/H)$^{10.0}_{2\rightarrow3}$).
The best-fit MZRs at $z\sim0$, $z\sim2.3$, and $z\sim3.3$ under our fiducial set of assumptions are shown by the gray lines in all panels.
We only show MZR variations for the high-redshift samples as the uncertainties pertaining to metallicity derivations and SED fitting
 are considerably larger at high redshift than at $z\sim0$.

We first vary our SED-fitting assumptions for determining stellar masses in panels (b)-(d) of Figure~\ref{fig:mzrsys}.
For each SED-fitting case, we remake the stacks according to the methods in Sec.~\ref{sec:stacking} using the
 new stellar masses to populate the mass bins.
In Figure~\ref{fig:mzrsys}b, we show the effect when the photometry is not corrected for the contribution from emission lines,
 resulting in a slightly steeper $z\sim2.3$ MZR with slightly lower normalization.
This effect arises because low-mass galaxies have higher emission-line equivalent widths \citep{red18} that contribute more
 strongly to the rest-optical photometry leading to an overestimate of \mstar, while at high-mass the equivalent widths are
 lower and do not significantly change the masses.
At $z\sim3.3$, galaxies have higher emission-line equivalent widths at fixed \mstar\ than at $z\sim2.3$ such that even the
 high-\mstar\ $z\sim3.3$ galaxy masses are biased by emission-line contaminated photometry, leading to a similar slope but
 systematically lower normalization compared to our fiducial case.

We assume a SMC extinction law \citep{gor03} and subsolar metallicity ($Z_*=0.0031$) in Figure~\ref{fig:mzrsys}c, as suggested to be
 appropriate for high-redshift galaxies by some works \citep{cap15,red15,red18b,shi20}, finding slightly steeper slopes and lower normalizations
 that only differ from our fiducial case by $\sim2\sigma$.
Recent studies at $z\sim2$ have suggested that the stellar attenuation curve steepens with decreasing \mstar\ and metallicity \citep{red18b,shi20}.
Motivated by these results, we show a hybrid of our fiducial case (a) and the SMC/sub-solar metallicity case (c) in Figure~\ref{fig:mzrsys}d,
 where we assume the
 fiducial assumptions (\citet{cal00} curve, solar metallicity) at log(\mstar/$\msun)>10.0$ and the SMC curve and sub-solar metallicity
 at log(\mstar/$\msun)<10.0$.
The result is that the highest two mass bins have not changed compared to the fiducial case, while the lower mass bins have slightly higher
 \mstar, again resulting in only slightly steeper slopes and marginally lower normalizations.

In panels (e)-(i) of Figure~\ref{fig:mzrsys}, we vary the metallicity calibration used to convert strong-line ratios to O/H.
Figure~\ref{fig:mzrsys}e shows the results when we use the same set of emission lines but apply the normal $z=0$ calibrations
 from this work (Fig.~\ref{fig:calibrations}, Table~\ref{tab:calibrations}) to the $z\sim2.3$ and $z\sim3.3$ samples.
We find overall lower normalization by $0.05$~dex at both redshifts and some change to the slope at $z\sim2.3$, but the relative offset between
 $z\sim2.3$ and $z\sim3.3$ remains unchanged.
Though not shown here, we found that the $z>2$ MZR normalization was slightly lower when using the $z\sim0$ M08 and C17 calibrations as well.
The relative offset between $z\sim2.3$ and $z\sim3.3$ was unchanged when employing C17, but was larger than our fiducial case when using M08.
Further discussion of discrepancies between our results and past studies that used M08 can be found in Sec.~\ref{sec:paststudies}.

To maintain a uniform set of emission lines over all redshifts, we have used only ratios of [O\ii], [Ne\iii], H$\beta$, and [O\iii]
 to estimate metallicities.
In panels (f)-(i), we investigate the use of calibrations based on ratios involving [N\ii]
 ([N\ii]/H$\alpha$ and O3N2=([O\iii]/H$\beta$)/([N\ii]/H$\alpha$)) and only show the $z\sim2.3$ sample since [N\ii] and H$\alpha$
 are not covered at $z>3$.
Panels (f) and (g) show the $z\sim2.3$ MZR using the [N\ii]/H$\alpha$ and O3N2 calibrations of the B18 high-redshift analogs.
We find almost the same slope as for the fiducial case, but offset 0.1~dex lower in normalization.
We chose to use the B18 calibrations because direct-method metallicities at $z\sim2$ match these calibrations on average
 \citep{san18}, but the $z\sim2$ direct-method sample did not have sufficient coverage to test [N\ii]-based indicators.
B18 selected high-redshift analogs to lie along the [N\ii] BPT sequence defined by the KBSS $z\sim2.3$ sample \citep{ste14},
 which displays a larger offset from the $z\sim0$ sequence in the [N\ii] BPT diagram \citep[Fig.~\ref{fig:calex};][]{sha15}.
The dashed blue line in panels (f) and (g) shows the resulting MZR if we shift the B18 calibration 0.15~dex lower (higher)
 in [N\ii]/H$\alpha$ (O3N2)
 at fixed O/H to match the [N\ii] BPT sequence of the MOSDEF $z\sim2.3$ stacks.\footnote{We could shift the B18 calibrations 0.1~dex
 lower in [O\iii]/H$\beta$ to bring them into agreement with the MOSDEF [N\ii] BPT sequence, but such a shift would result in worse
 agreement in the [O\iii]/H$\beta$ vs.\ O$_{32}$ and [Ne\iii]/[O\ii] diagrams (Fig.~\ref{fig:calex}) and with the
 $z\sim2.2$ direct-method metallicities (Fig.~\ref{fig:calibrations}).
For these reasons, we favor shifting [N\ii] alone, which could reflect differences in N/O between the two samples.}
After shifting the B18 calibrations to match the MOSDEF [N\ii] BPT sequence, we find a good match between our fiducial MZRs and
 those based on [N\ii] indicators.

The final two panels, (h) and (i), show the $z\sim2.3$ MZR derived using [N\ii]-based indicators and normal $z=0$ calibrations.
Similarly to case (e), we find slightly steeper slopes and a normalization that is $0.05-0.1$~dex lower than the fiducial case.
Panels (e), (h), and (i) collectively suggest that the primary effect of applying local calibrations to high-redshift samples is
 to underestimate O/H by $0.05-0.1$~dex relative to calibrations that are appropriate for the ISM conditions at $z\sim2$.
Applying typical local calibrations to high-redshift samples thus leads to larger inferred evolution of O/H at fixed \mstar,
 and would also lead us to infer an offset of $\sim0.1$ between $z>2$ galaxies and the $z\sim0$ FMR, as was reported using
 such methods in earlier works \citep{san15,san18}.

In summary, we find that assumptions regarding how stellar masses are derived and how metallicities are inferred from strong-line
 can affect the inferred slope and normalization of the high-redshift MZR.
However, these systematic effects are not severe, with the slope varying between $\gamma=0.28-0.41$ ($\gamma=0.30$ in the fiducial
 case) and the normalization varying no more than 0.05~dex in most cases, though offsets of up to 0.1~dex are possible when applying
 $z=0$ calibrations to high-redshift samples.
The latter effect carries important implications for the invariance of the FMR with redshift.
Of particular note is the fact that the relative offset in O/H at fixed \mstar\ between $z\sim2.3$ and $z\sim3.3$ is immune to
 the assumptions tested here, varying over only $0.10-0.12$~dex.
The evolution of the MZR slope and the relative offset in O/H between $z\sim0$ and high redshift are
 somewhat affected by these systematics, but typically at $\lesssim2\sigma$ relative to our fiducial case.
As high-redshift measurements improve, a careful treatment of SED fitting and metallicity calibration choices will become
 increasingly important to produce robust evolutionary studies of metallicity scaling relations.

\section{Analytic chemical evolution modeling}\label{sec:models}

We now turn to analytic galaxy evolution models to understand what physical processes set the slope and govern the evolution
 of the MZR over $z=0-3.3$.
We model our measured metallicities using the formalism of \citet[][hereafter PS11]{pee11}, which is more flexible than other models because it includes both
 mass and metal loading of accretion and outflows (i.e., accreting material need not be pristine, and outflowing material
 may have a metallicity different from that of the ISM).
In contrast, the gas-regulator model of \citet{lil13} assumes that the outflowing material has the same metallicity as the ISM,
 while the equilibrium model of \citet{dav12} also assumes $Z_{\text{out}}=Z_{\text{ISM}}$ and that the rate of change of the gas reservoir mass
 is zero such that galaxy metallicities have no explicit dependence on gas fraction or SFR
 (i.e., the FMR is not explicit in this formalism, as noted by \citealt{tor19}).

In the PS11 model, the metallicity of the ISM is expressed as
\begin{equation}\label{eq:ps11mzr}
\begin{multlined}
Z_{\text{ISM}}=\frac{y}{\zeta_{\text{out}} - \zeta_{\text{in}} + \alpha\mu_{\text{gas}} + 1}
\end{multlined}
\end{equation}
where $y$ is the nucleosynthetic stellar yield, and $\mu_{\text{gas}}\equiv M_{\text{gas}}/M_*$ is the gas fraction.
The coefficient to the gas fraction is
\begin{equation}
\begin{multlined}
\alpha\equiv(1-R)\left(\frac{d\log{M_{\text{gas}}}}{d\log{M_*}} + \frac{d\log{Z_{\text{ISM}}}}{d\log{M_*}}\right)
\end{multlined}
\end{equation}
where $R$ is the fraction of newly formed stellar mass that is returned to the ISM over time through stellar evolution
 processes, and $\alpha$ depends on the slope of \mugas($M_*$) and the MZR.
The other terms in the denominator of equation~\ref{eq:ps11mzr} are the metal loading factors of the outflowing
 galactic winds and inflowing gas accretion:
\begin{equation}
\begin{multlined}
\zeta_{\text{out}}\equiv \frac{Z_{\text{out}}}{Z_{\text{ISM}}}\times\frac{\dot{M}_{\text{out}}}{SFR}
\end{multlined}
\end{equation}
\begin{equation}
\begin{multlined}
\zeta_{\text{in}}\equiv \frac{Z_{\text{in}}}{Z_{\text{ISM}}}\times\frac{\dot{M}_{\text{in}}}{SFR}
\end{multlined}
\end{equation}
where \zout\ and \zin\ are the metallicities of the outflows and inflows, and \mdotout\ and \mdotin\ are the mass
 rates of the outflows and inflows.
The mass rates of gas flows are often parameterized as a ratio of the SFR in the mass-loading factors:
 $\eta_{\text{out}}=\dot{M}_{\text{out}}/SFR$ and $\eta_{\text{in}}=\dot{M}_{\text{in}}/SFR$.

In the PS11 framework, if the gas fraction and ISM metallicity are known (i.e., if
 \mugas($M_*$) and the MZR have been measured) and a return fraction and stellar yield are assumed, then the metal loading
 factors of the outflows and inflows can be solved for.
As is common, we make the simplifying assumption that \zetain\ is negligible so that we can uniquely solve for \zetaout.
This assumption does not require the inflows to be pristine, but simply that \zin$\ll$\zout.\footnote{
Note that, because launching sites of star-formation driven outflows are also production sites of elements and Type II SNe
 ejecta are highly enriched \citep[$Z_{\text{ej}}\sim7Z_{\odot}$;][]{woo95,nom06,nom13,rom10}, \zout$\ge$\zism.  The case of
 \zout=\zism\ is only reached if the outflow mass is dominated by entrained material over pure SNe ejecta.}
If this criterion is not true of real galaxies, then our determinations of \zetaout\ represent lower limits.
If $y/Z_{\text{ISM}} < 1 + \alpha \mu_{\text{gas}}$ (ignoring the \zetain\ term), then \zetaout\ is unphysically negative.
Thus, models with a low stellar yield cannot accomodate very high gas fractions.

There are only two terms in equation~\ref{eq:ps11mzr} that serve to increase metallicity:
 $y$, representing nucleosynthetic production through star formation; and \zetain, pertaining to accreted metals.
Ignoring \zetain, the stellar yield effectively sets a maximum ISM metallicity that is only reached if a system has no
 outflows and very little gas mass.
The actual ISM metallicity is set by the other two terms (\zetaout\ and \mugas) that serve to reduce metallicity through
 two distinct physical mechanisms.
The \mugas\ term represents the dilution mechanism whereby metals already present in the ISM and new metals from SNe
 are mixed into a larger hydrogen gas reservoir.
The \zetaout\ term encapsulates the metal removal mechanism in which metals are removed from the galaxy ISM by outflows.
Assuming $y$ does not strongly depend on \mstar\ and redshift, the slope and evolution of the MZR are determined by
 the dependence of both \zetaout\ and \mugas\ on \mstar\ and redshift.

In the following subsections, we apply the PS11 model to interpret our measurements of the MZR over $z=0-3.3$.
We first solve for \zetaout\ and constrain its scaling with \mstar\ by assuming a stellar yield,
 empirically-motivated gas fractions, and \zetain=0.
We then investigate the relative importance of dilution and metal removal in setting the slope of the MZR at each redshift,
 and governing the evolution of the MZR with redshift.

\subsection{Modeling the MZR at $z=0-3.3$}\label{sec:mzrmodeling}

We model the MZR at $z=0-3.3$ under the following fiducial set of assumptions.
We assume a stellar oxygen yield of $y_O=0.015$ as a mass fraction (12+log(O/H$)_y=9.2$ as a number fraction)
 and a return fraction of $R=0.30$\footnote{$R=0.25-0.45$ for standard IMFs \citep{vin16}. In practice, the derived \zetaout\ is
 not sensitive to $R$ over this range.}, values appropriate for a \citet{sal55} IMF
 with an upper mass cutoff of 100~$\msun$ \citep{vin16}.
Both $y_O$ and $R$ are assumed to be constant with \mstar\ and redshift.
At $z\sim0$, gas fractions as a function of \mstar\ are derived from the empirical \mugas($M_*, SFR$) relation of \citet{sai16}
 evaluated with our best-fit $z\sim0$ SFR-\mstar\ relation (equation~\ref{eq:z0sfrmstar}).
At $z>1$, we adopt the \mugas($M_*, z$) calibration of \citet{tac18} evaluated on the star-forming main sequence ($\delta\text{MS}=0$)
 at $z=2.3$ and $z=3.3$.
The \citet{sai16} \mugas\ relation includes the sum of the atomic and molecular gas masses, while the \citet{tac18} relation
 includes only molecular gas (see Sec.~\ref{sec:mugas} for further discussion).
With these assumptions, we invert equation~\ref{eq:ps11mzr} and solve for \zetaout\ for the stacked spectra,
 where $\alpha$ is calculated according to slopes of the best-fit MZR and assumed \mugas($M_*$) at each redshift.
This calculation allows us to examine implications of the observed MZR for metal loss via outflows.

\begin{figure}
 \includegraphics[width=\columnwidth]{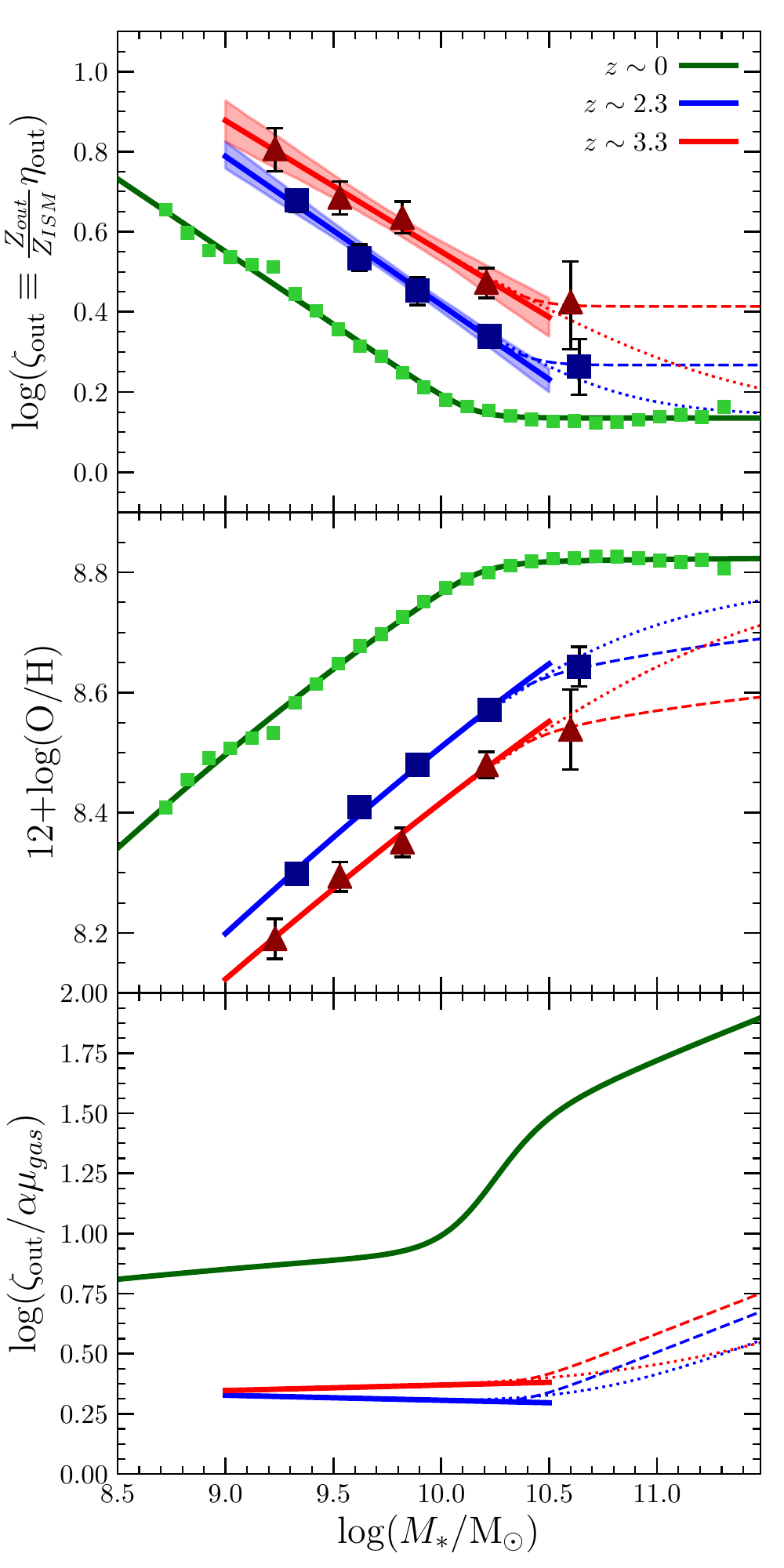}
 \centering
 \caption{
\textbf{Top:} Outflow metal loading factor, \zetaout, vs.\ \mstar\ for stacks of star-forming galaxies at $z\sim0$ (green squares),
 $z\sim2.3$ (blue squares), and $z\sim3.3$ (red triangles).
Best-fit relations are shown and given in equations~\ref{eq:z2zetaout} and~\ref{eq:z3zetaout} for $z\sim2.3$ and $z\sim3.3$, respectively.
The dashed line shows an extrapolation to high masses assuming \zetaout\ at $z=2.3-3.3$ turns over as quickly as the $z\sim0$ relation,
 while the dotted line displays an extrapolation assuming \zetaout\ asymptotes to the same $\zeta_{\text{out,0}}$ as the $z\sim0$ sample.
\textbf{Middle:} MZR at $z=0-3.3$, where the lines show the MZR inferred by applying the best-fit \zetaout($M_*$) and
 assumed \mugas($M_*$) relations in equation~\ref{eq:ps11mzr}, which match the observations by design.
\textbf{Bottom:} Ratio of \zetaout\ to $\alpha$\mugas at $z\sim0$, $z\sim2.3$, and $z\sim3.3$.
When log(\zetaout/$\alpha$\mugas$)>0$, the scaling of outflow efficiency with \mstar\ is more important for shaping the MZR
 than the scaling of gas fraction with \mstar.
}\label{fig:zetaout}
\end{figure}

The derived values of \zetaout\ vs.\ \mstar\ are shown in the top panel of Figure~\ref{fig:zetaout}.
The displayed errors on \zetaout\ take into account the measurement uncertainty in O/H, but do not include systematic
 uncertainty associated with the \mugas\ scaling relations.
The uncertainties in \mugas\ for the $z\sim0$ bins of \citet{sai16} result in an
 uncertainty in our derived $z\sim0$ \zetaout\ values of $<$0.01~dex, and are thus negligible.
The \mugas\ relation of \citet{tac18} has an error budget that is dominated by
 the uncertainty in the normalization of 0.15~dex, corresponding to a 1$\sigma$ systematic uncertainty of 0.06~dex in
 \zetaout\ at $z\sim2.3$ and $z\sim3.3$.
We have not included this uncertainty in the errors shown in Fig.~\ref{fig:zetaout} because 
 it would shift the $z\sim2.3$ and $z\sim3.3$ \zetaout\ values relative to $z\sim0$ while keeping the offset
 between the $z\sim2.3$ and $z\sim3.3$ points the same.
The uncertainty associated with the redshift and \mstar\ dependence of the \citet{tac18} relation has a much smaller effect
 on our inferred \zetaout\ values.

At all redshifts, we find that \zetaout\ decreases with increasing \mstar, with a significant flattening
 at high \mstar\ present at $z\sim0$.
At fixed \mstar, \zetaout\ increases with increasing redshift.
We fit the $z\sim2.3$ and $z\sim3.3$ stacks (excluding the highest-mass bin) with power laws, obtaining
\begin{equation}\label{eq:z2zetaout}
z\sim2.3\text{: } \log\left(\zeta_{\text{out}}\right) = (-0.37\pm0.04)\times m_{10} + (0.42\pm0.02)
\end{equation}
\begin{equation}\label{eq:z3zetaout}
z\sim3.3\text{: } \log\left(\zeta_{\text{out}}\right) = (-0.33\pm0.06)\times m_{10} + (0.55\pm0.02)
\end{equation}
where $m_{10}=\log(M_*/10^{10}\ \msun)$.

We fit \zetaout\ at $z\sim0$ with a smoothly broken power law of the form
\begin{equation}\label{eq:zetaoutbpl}
\log(\zeta_{\text{out}})=\log(\zeta_{\text{out,0}}) - \gamma/\Delta\times\log\left[1+\left(\frac{M_*}{M_{\text{TO}}}\right)^{-\Delta}\right] .
\end{equation}
The best-fit parameters to the $z\sim0$ data are
 [$\log(\zeta_{\text{out,0}})$, $\gamma$, log($M_{\text{TO}}/\msun$), $\Delta$]=[$0.14\pm0.01$, $-0.36\pm0.01$, $10.14\pm0.03$, $5.17\pm2.82$].
Similar to the best-fit MZRs, we find that the best-fit \zetaout($M_*$) displays a consistent slope across
 all three redshifts, with $\gamma=-0.36\pm0.01$, $-0.37\pm0.04$, and $-0.33\pm0.06$ at $z\sim0$, 2.3, and 3.3, respectively.
Taking the average of these slopes yields a universal scaling of \zetaout$\propto$$M_*^{-0.35\pm0.02}$.

Outflow mass and metal loading factors are often expressed as a function of the circular velocity, \vcirc, which is
 more closely related to the gravitational potential than \mstar.
Using the technique described in \citet{pee11} and the stellar mass-halo mass relation of \citet{mos13} yields
\zetaout$\propto$\vcirc$^{-1.96\pm0.01}$ at $z\sim0$, \zetaout$\propto$\vcirc$^{-1.82\pm0.17}$ at $z\sim2.3$,
 and \zetaout$\propto$\vcirc$^{-1.59\pm0.15}$ at $z\sim3.3$.
If we instead resample our average result of \zetaout$\propto$$M_*^{-0.35\pm0.02}$ into \vcirc\ using a stellar mass Tully-Fisher relation (sTFR)
 of $M_*\propto v_{\text{circ}}^{3.75}$ \citep{lel16}, then we find \zetaout$\propto$$v_{\text{circ}}^{-1.31\pm0.08}$, assuming
 the sTFR slope does not evolve.
The evolution of the sTFR zero-point leads to a larger evolution in \zetaout\ at fixed \vcirc\ than at fixed \mstar.
As demonstrated here, the inferred dependence of \zetaout\ on \vcirc\ depends on the method used to translate \mstar\ into \vcirc.

The middle panel of Figure~\ref{fig:zetaout} displays the model MZRs (solid lines) resulting from applying the
 assumed \mugas\ and best-fit \zetaout\ relations in equation~\ref{eq:ps11mzr}, which match the observations by design
 since we have derived \zetaout\ from the observed metallicities.
While we cannot constrain whether \zetaout\ flattens at high-masses at $z\sim2.3$ and $z\sim3.3$, we show for
 demonstration purposes two examples where \zetaout\ turns over quickly such that $\zeta_{\text{out,0}}$ evolves
 with redshift (dashed line) or \zetaout\ asymptotes to the same value of $\zeta_{\text{out,0}}$ as at $z\sim0$ (dotted line).

\subsection{What sets the slope of the MZR at $z\sim0$, 2.3, and 3.3?}\label{sec:mzrslope}

We first address the question of what process governs the slope of the MZR at each redshift.
According to equation~\ref{eq:ps11mzr}, if \zetain\ is negligible, the functional form of the MZR is primarily set by the scaling
 of either \zetaout\ or \mugas\ with \mstar, representing cases where lower-mass galaxies have lower metallicities
 because metals are more efficiently removed from low-\mstar\ galaxies by outflows or the high gas fractions of
 low-\mstar\ galaxies lead to stronger dilution of metals in the ISM.
If \zetaout$\gg$$\alpha$\mugas, then metal-enriched outflows determine the slope of the MZR and the low-mass behavior of the MZR
 is approximately $\text{O/H}\propto\zeta_{\text{out}}^{-1}$.
If $\alpha$\mugas$\gg$\zetaout, then the MZR is shaped by changing gas fractions and the low-mass slope
 is instead $\text{O/H}\propto\mu_{\text{gas}}^{-1}$.
If $\alpha$\mugas\ and \zetaout\ are approximately equal, then both outflows and gas fractions contribute significantly
 to the functional dependence of \zism\ on \mstar.

The lower panel of Figure~\ref{fig:zetaout} presents log(\zetaout/$\alpha$\mugas) vs.\ \mstar.
We find that \zetaout$\gg$$\alpha$\mugas\ at $z\sim0$ across log($M_*/\msun)\sim8.5-11.5$, indicating that the slope of
 the local MZR is set by the action of metal-enriched outflows whereby metals are more efficiently removed from low-mass galaxies.
The dominance of outflows at $z\sim0$ arises because the gas fractions are not large enough to drive the metallicities
 sufficiently low with our assumed oxygen yield (or indeed for any $y_O$ in the range produced by standard IMFs, $y_O=0.008-0.045$).
\citet{pee11} reached the same conclusion for the $z\sim0$ MZR, and many previous analyses have also concluded that
 outflows primarily shape the local MZR while gas fractions have negligible effect \citep[e.g.,][]{tre04,dal07,lil13,chi18}.
Because gas fractions are especially low in $z\sim0$ high-mass galaxies, the flattening of the local MZR at high masses ($M_*\gtrsim10^{10.2}~\msun$)
 requires \zetaout($M_*$) to flatten to an asymptotic value at high \mstar\ as seen in the top panel of Fig.~\ref{fig:zetaout},
 implying that there is a lower boundary to the
 global metal removal efficiency of star-formation driven outflows that is only reached at high mass at $z\sim0$.

At high redshifts, we find that \zetaout/$\alpha$\mugas\ is approximately constant over the range of \mstar\ probed by
 the samples, with \zetaout$\approx2-3\times$$\alpha$\mugas,
 with no significant difference between $z\sim2.3$ and $z\sim3.3$ when considering the uncertainties in \zetaout\ and \mugas.
We thus find that outflows remain the dominant mechanism that sets the MZR slope at $z\sim2.3$ and $z\sim3.3$.
While gas fraction carries more relative importance at high redshifts than at $z\sim0$ due to the increase in \mugas\ at fixed
 \mstar\ with increasing redshift, it still has only a minor effect on the MZR slope.
While keeping the normalization fixed at 10$^{10}~\msun$, changing the power-law slope of \mugas($M_*$) by $\pm$0.1 results
 in a change of the model MZR slope of only $\pm$0.02 at $z\sim2.3$ and $z\sim3.3$.
Thus, in this model, the reason that the slope of the MZR does not significantly change with redshift is because the slope of
 \zetaout($M_*$) does not significantly change with redshift.
The scaling of \zetaout\ with \mstar\ appears to be redshift invariant out to at least $z\sim3.3$,
 resulting in a constant low-mass MZR slope over the past 12~Gyr.

It is perhaps not surprising that the same mechanism sets the MZR slope in all three samples because the observed MZR slope
 is tightly constrained to be the same over $z=0-3.3$ with good precision.
If the parameter governing the MZR slope transitions from \zetaout\ to \mugas\ at some higher redshift, we expect to also observe
 a shift in the MZR slope unless \zetaout\ and \mugas\ have the same scaling with \mstar.

\subsection{What drives the evolution of the MZR normalization over $z=0-3.3$?}\label{sec:mzrnorm}

We now turn to the question of what process drives the evolution of the MZR normalization,
 leading to decreasing O/H at fixed \mstar\ with increasing redshift.
Gas fractions are observed to increase significantly at fixed \mstar\ with increasing redshift, with a
 scaling of \mugas$\sim$$(1+z)^{1.5}$ \citep[e.g.,][]{tac13,tac18,sco17,liu19}.
One possibility is thus that the evolution of the MZR normalization is caused by the underlying evolution in gas fractions,
 such that ISM metals are more heavily diluted at fixed \mstar\ with increasing redshift.
However, we have shown above that \zetaout\ also increases at fixed \mstar\ with increasing redshift (Fig.~\ref{fig:zetaout}),
 such that more efficient removal of metals by outflows may be the cause of lower metallicities at high redshifts.

\begin{figure}
 \includegraphics[width=\columnwidth]{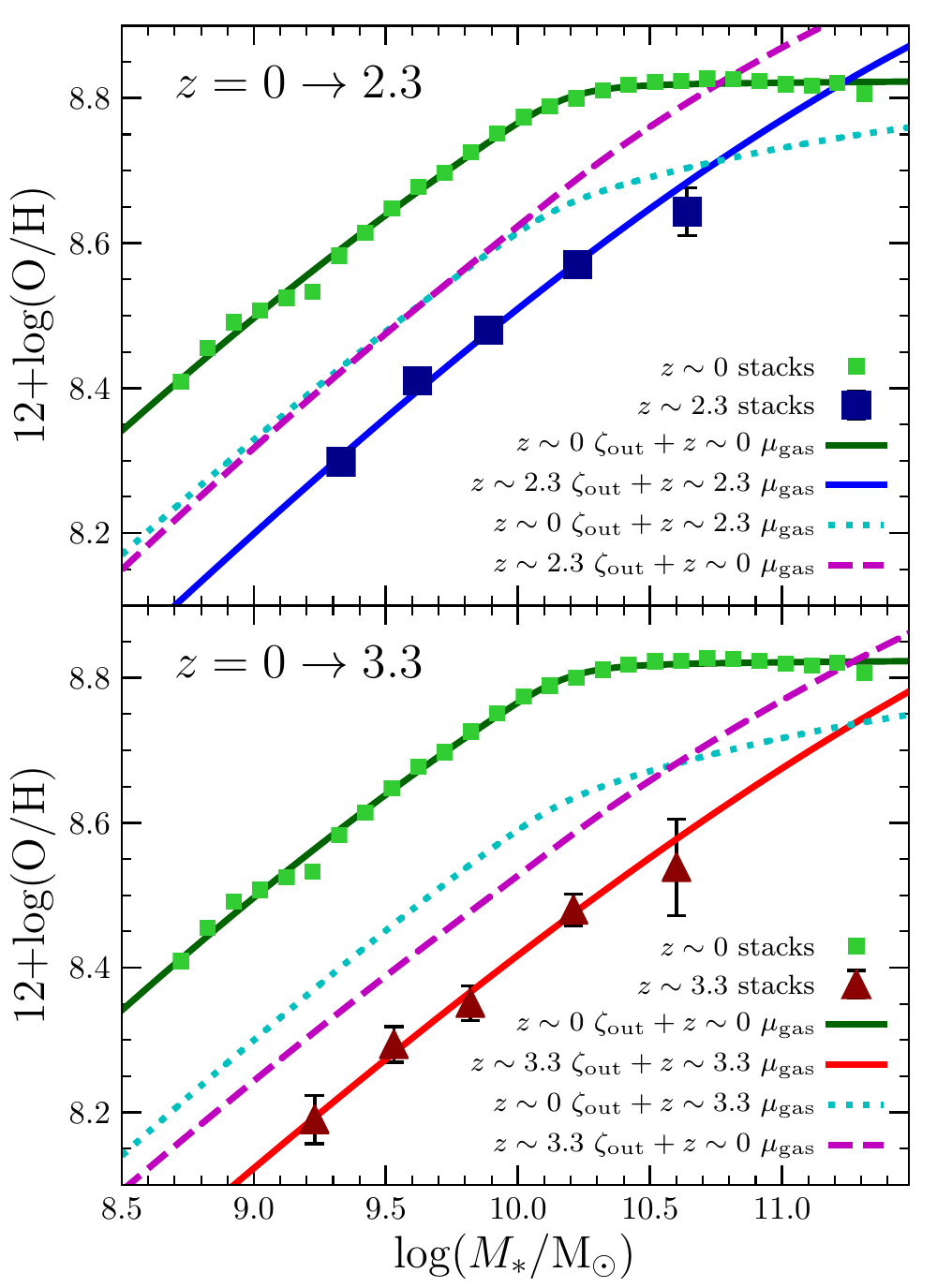}
 \centering
 \caption{
Models of MZR evolution from $z\sim0$ to $z\sim2.3$ (top) and $z\sim3.3$ (bottom),
 where the effects of evolving \mugas\ and \zetaout\ have been separated.
The solid lines represent our best-fit models at each redshift.
The cyan dotted line shows a model assuming the best-fit $z\sim0$ \zetaout\ and \mugas($M_*$) of $z\sim2.3$(3.3), isolating
 the contribution of evolving gas fractions to MZR evolution.
The magenta dashed line instead displays a model in which the high-redshift \zetaout($M_*$) is assumed while adopting
 $z\sim0$ gas fractions, isolating the impact of \zetaout\ evolution on MZR evolution.
Evolving \mugas\ and \zetaout\ each account for roughly half of the observed evolution in O/H at fixed \mstar\ over $z=0-3.3$.
}\label{fig:mzrevolution}
\end{figure}

In Figure~\ref{fig:mzrevolution}, we show the relative importance of evolving \mugas\ and \zetaout\ to the total
 MZR evolution between $z\sim0$ and $z\sim2.3$ (top panel) or $z\sim3.3$ (bottom panel).
In each panel, the dotted line shows the resulting MZR from a model with the best-fit $z\sim0$ \zetaout($M_*$) and high-redshift
 \mugas($M_*$) (i.e., only evolving the gas fraction with redshift),
 while the dashed line shows the MZR modeled with the best-fit high-redshift \zetaout($M_*$) and $z\sim0$ \mugas($M_*$)
 (i.e., only evolving the outflow metal loading factor with redshift).
We find that models evolving only \zetaout\ or only \mugas\ from $z\sim0$ yield metallicities that are $\sim0.1-0.15$~dex higher at
 fixed \mstar\ than the observed MZR at $z\sim2.3$ and 3.3, indicating that MZR evolution is not predominantly driven by evolution
 in either parameter alone.
Instead, evolution towards higher \mugas\ and higher \zetaout\ contribute roughly equally to the decreasing normalization
 of the MZR with increasing redshift.
This result indicates that high-redshift galaxies have lower metallicities than local galaxies at fixed \mstar\ because
 metals are more diluted due to their higher gas content \textit{and} metals are more efficiently removed from the ISM by outflows.
Both mechanisms remain important to MZR evolution out to $z\sim3.3$.

In Sec.~\ref{sec:mzrslope}, we showed that the slope of the MZR is primarily set by the functional form of \zetaout($M_*$)
 for samples at all three redshifts.
A natural question is how \mugas\ can be important to MZR normalization evolution when it has a subdominant effect on the
 slope at a given redshift.
At fixed log($M_*/\msun)=10.0$, the outflow metal-loading factor is log(\zetaout$)=0.20\pm0.01$, $0.43\pm0.05$, and $0.56\pm0.05$ at $z\sim0$,
 2.3, and 3.3, respectively.
Thus, we find that dlog(\zetaout)/d$z=0.10\pm0.03$ at fixed \mstar, or roughly \zetaout$\sim$$(1+z)^{0.5}$.
This evolution of \zetaout\ is weaker than that of \mugas$\sim$$(1+z)^{1.5}$ at fixed \mstar\ \citep{tac18}.
While \zetaout\ carries more weight at fixed redshift than \mugas, the relative importance of \mugas\ for setting \zism\ grows
 with increasing redshift, as reflected by the lower \zetaout/$\alpha$\mugas\ at $z\sim2.3-3.3$
 in the bottom panel of Fig.~\ref{fig:zetaout}.
If \mugas\ and \zetaout\ continue to evolve at these rates out to $z>4$, it implies that dilution of metals according to \mugas\ will
 become the dominant mechanism that controls MZR slope and normalization at $z\gtrsim5-6$.
Rest-optical spectroscopy of $z>4$ galaxies from $JWST$ will enable characterization of the MZR at these redshifts and
 test whether there is indeed a shift in the dominant physical mechanism.

\subsection{Can accreted metals be ignored?}

In the models above we have assumed that the metal loading of inflows, \zetain, is negligible compared to \zetaout.
This may not be true since some fraction of outflowing metals are believed to be re-accreted through galactic fountaining
 \citep{opp08,for14,ang17,mur17}
 and a significant amount of metals resides in the circumgalactic medium \citep{ste10,pee14,rud19}.
In the PS11 framework, the inflow and outflow mass loading factors are related according to the expression
\begin{equation}\label{eq:etain}
\eta_{\text{in}} - \eta_{\text{out}} = (1 - R)\mu_{\text{gas}}\left(1 - \frac{d\log{\mu_{\text{gas}}}}{d\log{M_*}}\right) - R + 1 .
\end{equation}
For reasonable ranges of return fraction ($R\sim0.25-0.45$) and the range of power law slopes of \mugas ($\sim0.3-0.5$),
 and over the entire mass range at $z\sim0$ and above $\sim$10$^{9.5}~\msun$ at $z\sim2-3$, all terms on the right hand side
 are of order unity.
In the scenario where outflow mass is predominantly entrained ISM such that \zout$\approx$\zism,
 \etaout\ is always greater than 1 based on the constraints on \zetaout\ presented in this work.
Accordingly, \etain$\approx$\etaout\ to first order and the relative importance of \zetain\ compared to that of
 \zetaout\ is simply determined by the ratio \zin/\zout.
Thus, if \zin$\ll$\zout, then \zetain\ can be safely ignored.

In general, including accreted metals increases the inferred \zetaout\ and would thus strengthen our conclusion that
 the slope of the MZR is set by metal-enriched outflows (not by gas fractions) and increase the relative importance of
 \zetaout\ to the evolving MZR normalization.
However, if \zetain\ increases significantly at fixed \mstar\ toward lower redshifts, then part of the MZR evolution could be explained by the
 changing importance of accreted metals.
Indeed, an increase in \zin/\zism\ with decreasing redshift was found to be the primary driver of MZR evolution in the models
 of \citet{dav11}; this effect can be understood as a result of gas recycling via galactic fountains.

With a large enough evolution of \zin/\zism, it is possible that the observed
 MZR evolution can be explained without any change in \zetaout\ with redshift.
This scenario requires a rough scaling of $(1-Z_{\text{in}}/Z_{\text{ISM}})\propto(1+z)^{0.5}$.
Assuming accreted gas is nearly pristine at $z\sim3.3$ (i.e., \zin/\zism$\approx$0), this scaling predicts \zin/\zism$\approx$0.5 at $z=0$, a
 lower limit since accretion at high-redshift may not be pristine due to vigorous outflow recycling \citep[e.g.,][]{ang17}.
Detections of Mg\ii\ absorption in gas thought to be inflowing onto $z\approx0.2$ galaxies suggests that low-redshift accreting gas
 is metal-enriched at some level \citep{ho17,mar19}.
High velocity H\one\ clouds around the Milky Way that are thought to be accreting onto the Galaxy have typical metallicities of
 $\sim$0.1\zism\ \citep{san08}.
If \zin/\zism$\approx$0.1 is typical at $z=0$, then accreted metals cannot play a major role in MZR evolution.

We conclude that the observed evolution of the MZR over $z=0-3.3$ is well explained by redshift evolution of both the outflow
 metal loading factor and gas fraction at fixed \mstar, but a significant effect from accreting metals cannot be ruled out.
There are currently insufficient observational constraints on the metallicity of accreting gas to stringently distinguish
 the influence that metal accretion has on metallicity scaling relations.
Constraints on the recycling timescales and importance of galactic fountains relative to IGM accretion are needed to evaluate the
 contribution of accreted metals to MZR evolution.

\subsection{Systematic effects of the assumed gas fractions}\label{sec:mugas}

The scaling of total gas mass ($M_{\text{gas}}=M_{\text{H~\textsc{i}}}+M_{\text{H}_2}$) of local star-forming galaxies
 with \mstar\ and SFR is well-characterized through extensive H\one\ 21~cm and CO emission surveys
 \citep[e.g.][]{cat18,sai17,bot14,cic17}.
The gas fractions at $z\sim0$ thus are not a source of significant systematic uncertainty in our models due to these robust
 empirical constraints.

The scaling of \mugas\ with \mstar\ is much more uncertain at high redshifts due to uncertainties in
 factors that convert CO or dust continuum measurements to molecular gas masses: $\alpha_{\text{CO}}$ and the dust-to-gas ratio, both of
 which are metallicity dependent and may evolve; the CO excitation ladder; and dust temperature.
With current facilities, samples are limited to massive galaxies (log$(M_*/\msun\gtrsim10.0$) at $z>1$ and thus do not span
 a wide dynamic range in mass.
Furthermore, the contribution by neutral hydrogen is unknown at high redshift, though there is observational evidence that H\one\ is not
 the dominant gas component at high redshift (see \citealt{tac18}, and references therein),
 unlike in the local universe where $M_{\text{H~\textsc{i}}}/M_{\text{H}_2}=3-10$ for $z\sim0$ star-forming galaxies \citep{sai16,sai17,cat18}.
Accordingly, the total gas mass is taken to be equivalent to M$_{\text{H}_2}$ at high redshift.
Below, we test high-redshift \mugas\ scaling relations from the literature against observations of our $z\sim2.3$ and $z\sim3.3$ samples
 where $M_{\text{gas}}$ has been estimated using the Kennicutt-Schmidt \citep[KS;][]{ken98b,sch59} relation.

\begin{figure*}
 \includegraphics[width=0.7\textwidth]{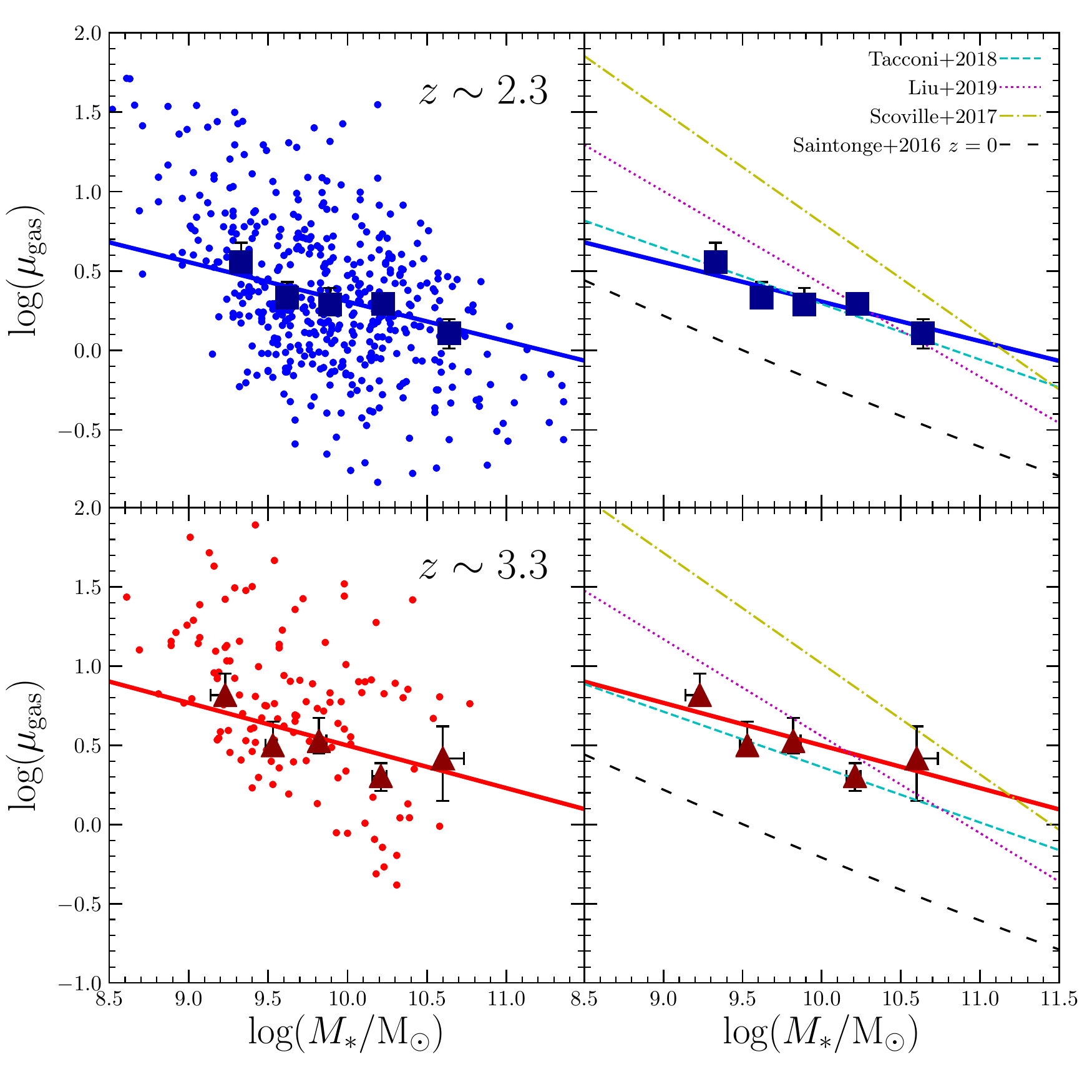}
 \centering
 \caption{
Gas fraction vs.\ \mstar\ for $z\sim2.3$ (top) and $z\sim3.3$ (bottom).
\textbf{Left}: Gas fraction for individual galaxies and stacked spectra (small circles and large squares/triangles, respectively) where the gas
 fraction has been estimated using SFR from dust-corrected Balmer lines, rest-optical effective radii \citep{van14}, and the $z\sim1.5$
 Kennicutt-Schmidt relation of \citet{tac13}.
The solid blue (red) line denotes \mugas($M_*$) found by combining our best-fit SFR-\mstar\ relation at $z\sim2.3$ (3.3) with
 the $R_{\text{eff}}(M_*, z)$ relation of \citet{van14} and the \citet{tac13} $z\sim1.5$ KS relation.
The resulting \mugas($M_*$) relations are given in equations~\ref{eq:z2ugas} and~\ref{eq:z3ugas}.
\textbf{Right:} Comparison of the stacks and \mugas($M_*$) relations from the left panels to \mugas\ scaling relations from the literature.
The relations of \citet[][cyan dashed line]{tac18}, \citet[][purple dotted line]{liu19}, and \citet[][yellow dash-dotted line]{sco17}
 evaluated for main sequence galaxies at $z=2.3$ and $z=3.3$ are shown.
The $z\sim0$ \mugas($M_*$) relation from \citet{sai16} for main-sequence galaxies is presented as the loosely dashed black line.
}\label{fig:z2z3gas}
\end{figure*}

We estimate $M_{\text{gas}}$ for the high-redshift samples using dust-corrected SFR(H$\alpha$), rest-optical effective radii ($R_{\text{eff}}$)
 from the catalog of \citet{van14}, and the $z\sim1.5$ molecular KS relation of \citet{tac13} derived from high-redshift CO measurements.
We assume $M_{\text{gas}}=M_{\text{H}_2}$ at $z>1$.
The median  $R_{\text{eff}}$ of galaxies in each \mstar\ bin was used in the calculation for the stacks.
The left panels of Figure~\ref{fig:z2z3gas} display \mugas\ for the $z\sim2.3$ and $z\sim3.3$ galaxies and stacks.
The blue and red solid lines show the relations resulting from combining our best-fit SFR-\mstar\ relations (eqs.~\ref{eq:z2sfrmstar}
 and~\ref{eq:z3sfrmstar}) with the $R_{\text{eff}}(M_*, z)$ relation of \citet{van14}, of the form
\begin{equation}\label{eq:z2ugas}
z\sim2.3\text{: } \text{log}\left(\mu_{\text{gas}}\right) = -0.25\ m_{10} + 0.31
\end{equation}
\begin{equation}\label{eq:z3ugas}
z\sim3.3\text{: } \text{log}\left(\mu_{\text{gas}}\right) = -0.27\ m_{10} + 0.50
\end{equation}
where $m_{10}=\log(M_*/10^{10}\ \msun)$.
If we were to use the $z=0$ molecular KS relation instead, we would infer \mugas\ values that are $\sim$0.3~dex higher at fixed \mstar.

In the right panels of Figure~\ref{fig:z2z3gas}, we compare to gas fraction scaling relations of \mugas($M_*, z$) for main-sequence galaxies
 from the literature \citep{tac18,sco17,liu19}.
The scaling relation of \citet{tac18}, employed in our fiducial model,
 is in closest agreement with our estimates at $z\sim2.3$ and $z\sim3.3$ using the KS relation across the full mass range of our
 samples.\footnote{
Note that the \citet{tac18} scaling relation is not independent of the data set used to calibrate the \citet{tac13} $z\sim1.5$ KS relation,
 since the PHIBSS1 CO measurements are included in \citet{tac18}.
However, the \citet{tac13} CO measurements make up $<25$\% of the $z>1$ sample used by \citet{tac18}, such that the agreement of
 the \citet{tac18} relation with our KS-based gas fractions is not entirely by construction.
}
Accordingly, there is no significant change in our results if we instead use the KS relation gas fractions for the $z\sim2.3$ and $z\sim3.3$ stacks.

Both the \citet{sco17} and \citet{liu19} relations are steep (power-law slope of $-0.7$) such that the gas fractions are extremely high at
 low masses, leading to a large disagreement with the KS \mugas\ values at log($M_*/\msun)\lesssim10.0$.
In contrast to our fiducial case, models using these \mugas($M_*$) relations yield a constant \zetaout\ with no \mstar\ dependence,
 and the MZR slope and normalization are
 predominantly set by the gas fractions due to the steep slope and high normalization of \mugas($M_*$).
However, such steep \mugas($M_*$) relations are not observed in numerical simulations across this mass and redshift range,
 which have power-law slopes of $\sim-0.2$ to $-0.5$ \citep{lag16,dav19,tor19}.
As mentioned earlier, CO and dust continuum samples at $z>1$ are entirely composed of galaxies at log($M_*/\msun)\gtrsim10.0$, with the
 majority at log($M_*/\msun)>10.5$.
We thus rely on extrapolations to compare to the low-mass half of our high-redshift samples.
The extreme gas fractions predicted at low masses by \citet{sco17} and \citet{liu19} suggest that extrapolations of these relations below
 $10^{10}~\msun$ are not reliable.

Since the high-redshift \mugas\ does not include neutral hydrogen, it is possible that the true gas fractions at $z\sim2.3-3.3$
 are larger than our estimates.
However, dynamical mass constraints do not allow for a large H\one\ contribution at $z\sim2$.
\citet{pri16} found that the median ratio of the dynamical and stellar masses for $z=1.4-2.6$ MOSDEF star-forming galaxies is
 0.36~dex, such that the typical \mugas\ can be at most $2-2.5$.
\citet{wuy16} found a similar dynamical-to-stellar mass ratio at $z\sim2$ using data from the KMOS$^{\text{3D}}$ survey.
At $z\sim2.3$ and $10^{10.0}~\msun$ (the median mass of our stacking sample), the typical molecular gas fraction using either the
 KS relation or the \citet{tac18} scaling relation is 2.0.
Accordingly, H\one\ can make up at most $\sim20$\% of the total gas mass on average, assuming dark matter is negligible within the baryonic disk.
Our results do not change significantly unless gas masses are $\gtrsim40$\% ($\gtrsim0.15$~dex) higher at fixed \mstar\ compared to our
 fiducial case.

\subsection{Systematic effects of the assumed stellar yield}\label{sec:yield}

We now address how assumptions about stellar yield affect our conclusions about \zetaout\ and the 
 mechanisms controlling the slope and evolution of the MZR.
We have assumed the oxygen yield is $y_O=0.015$ (12+log(O/H$)_y=9.2$), appropriate for a \citet{sal55} IMF with a 100~$\msun$
 high-mass cutoff.
However, the yield depends strongly on the IMF: the oxygen yields assuming the IMF of \citet{sal55}, \citet{kro93}, \citet{kro01}, or \citet{cha03}
 span $y_O=0.008-0.035$ (12+log(O/H$)_y=9.0-9.6$; \citealt{vin16}).
If the yield is higher than our assumed value (i.e., $y_O\approx0.030-0.035$ for a \citealt{cha03} or \citealt{kro01} IMF), then the inferred
 \zetaout($M_*$) would be larger at fixed \mstar\ than in our fiducial model but with a similar slope.
In this high-yield case, our conclusion that metal-enriched outflows set the slope of the MZR would be strengthened (\zetaout/$\alpha$\mugas\ would
 be larger at all redshifts), while evolution of \zetaout\ at fixed \mstar\ would be the dominant driver of MZR evolution, in contrast
 to our fiducial model in which \zetaout\ and \mugas\ contribute roughly equally to MZR evolution.
If we instead assume the lowest yield for a standard IMF ($y_O=0.008$ for the \citealt{kro93} IMF), \zetaout\ would be lower at each redshift
 such that gas fractions have an increased relative importance.
In the low-yield case, \zetaout\ still dominates the MZR slope at $z\sim0$, \zetaout\ and \mugas\ contribute equally to the MZR slope at $z\sim2.3-3.3$,
 and MZR evolution is primarily driven by evolving gas fractions at fixed \mstar.

We have assumed that the stellar yield is redshift invariant, but it is possible that the oxygen yield evolves.
In particular, the yield will evolve if the high-mass slope and/or upper mass cutoff of the IMF changes with redshift
 such that decreasing the slope or increasing the cutoff mass increases $y_O$.
There are theoretical expectations that the IMF may have both a higher upper mass cutoff and shallower high-mass slope
 in low-metallicity and high-SFR environments \citep{jer18,sch18,gut19}, suggesting that
 oxygen yield increases with redshift due to the evolution of the MZR and the SFR-\mstar\ relation.
If oxygen yields are higher at high redshifts than at $z\sim0$, then larger \zetaout\ is required at $z\sim2.3-3.3$ to drive metallicities
 down towards the observed values compared to our fiducial model.
Consequently, \zetaout\ becomes even more dominant over \mugas\ in setting the high-redshift MZR slope, and
 \zetaout\ would be the dominant driver of MZR evolution.
A scenario in which yields decrease with increasing redshift is not expected.

Observational constraints on the high-mass IMF at $z>1$ are needed to improve chemical evolution models of early galaxies.
The non-ionizing rest-UV contains features that are sensitive to the high-mass slope
 and upper mass cutoff of the IMF, including stellar wind features such as N~\textsc{v} and C~\textsc{iv} and nebular emission
 features such as He\ii\ that are sensitive to ionization from the most massive stars \citep{ste16,sen17}.
A systematic analysis of such IMF-sensitive rest-UV features for large samples at $z\sim2-3$ could determine whether the high-mass
 IMF changes appreciably at high redshift and thus reduce uncertainties in stellar yields at $z>1$.

\section{Discussion}\label{sec:discussion}

\subsection{Comparison to past studies}\label{sec:paststudies}

\subsubsection{Past studies of the MZR at $z\sim2.3$}

Our results do not differ drastically from past studies of the MZR at $z\sim2.3$, with a generally similar slope but
 smaller evolution in O/H at fixed \mstar\ relative to the local MZR than literature results
 \citep{erb06,wuy12,wuy16,bel13,kul13,hen13,cul14,mai14,ste14,kap15,kap16,san15,san18,san20,hun16a}.
The smaller evolution is due to our use of different calibrations at $z\sim0$ and $z>1$ to account for evolving ISM conditions.
If we apply local calibrations at high redshifts, as has been done by all previous studies, we find a $\sim0.1$~dex larger
 decrease in O/H at fixed \mstar\ ($\sim-0.33$~dex), in agreement with past work.

Using stacked spectra of 87 $z\sim2.3$ galaxies, \citet{erb06} found a sharp drop off in metallicity between
 $10^{10.0}~\msun$ and $10^{9.5}~\msun$, yielding a low-mass slope much steeper than our value.
The steepness of the \citet{erb06} MZR can be attributed to a sample bias where their low-mass bins are populated by galaxies
 falling up to $\sim1$~dex above the $z\sim2.3$ star-forming main sequence \citep{cul14}, resulting in below-average O/H at low \mstar.
Our more representative $z\sim2.3$ sample does not display such a steep low-mass slope.

\citet{ste14} investigated the MZR at $z=2.3$ using a sample of 242 star-forming galaxies from KBSS,
 inferring a much flatter MZR than ours, with a low-mass slope of $0.20\pm0.02$.
This shallow MZR could be a result of a bias in the KBSS sample against metal-rich, red, dusty galaxies at high masses.
The KBSS sample is primarily rest-UV selected, which could result in such a bias, although it is supplemented with galaxies
 meant to fill in the high-mass, red star-forming population \citep{ste14,str17}.
In contrast, MOSDEF is rest-optical selected \citep{kri15}.
A full comparison of the properties of the MOSDEF and KBSS samples would be useful in elucidating the origin of differing results
 in both the MZR and [N\ii] BPT diagram offset at $z\sim2.3$.

The $z\sim2.3$ MZR has been measured for MOSDEF data using a range of metallicity indicators and $z\sim0$ calibrations \citep{san15,san18}.
Using a simple power-law form, these early studies found the slope of the MZR at $z\sim2.3$ to be $0.26-0.34$ depending on the indicator used,
 generally consistent with the slope in this work ($0.30\pm0.02$).
Most of the indicators considered were nitrogen-based, but when inferring metallicities from O$_{32}$, a line ratio that overlaps
 with this work, \citet{san18} found an evolution of $-0.3$~dex in O/H at fixed \mstar\ over $z=0-2.3$ at 10$^{10}~\msun$, slightly
 larger than in the present work.
Though the sample in \citet{san18} has approximately 80\%\ overlap with the $z\sim2.3$ sample in this work,
 the results in this work are more robust since we account for evolving ISM conditions
 in our choice of metallicity calibrations.

\citet{san20} investigated the MZR using a sample of $\sim20$ galaxies at $z\sim2.2$ with direct-method
 metallicities, finding a low-mass slope of $-0.37\pm0.08$.
This value is consistent with our $z\sim2.3$ slope.
It is imperative to increase the number of $z>1$ galaxies with direct-method metallicities in order to
 improve constraints on metallicity calibrations appropriate for galaxies in the early universe.

\subsubsection{Past studies of the MZR at $z>3$}\label{sec:z3lit}

While our results show reasonable agreement with past studies of the $z\sim2.3$ MZR, we find that they differ
 significantly from earlier work at $z>3$.
\citet{mai08}, \citet{man09}, and \citet{tro14} analyze the AMAZE+LSD sample of 40 galaxies at $z\sim3.4$, deriving
 metallicities using the same set of emission lines as in this work ([O\ii], [Ne\iii], H$\beta$, and [O\iii]).
For both individual galaxies and composite spectra, all three works find a large evolution in O/H at $10^{10}~\msun$
 of $\sim-0.7$ to $-0.8$~dex from $z\sim0$ to $z\sim3.4$.
By comparing to the \citet{erb06} $z\sim2.3$ MZR, these authors find the metallicity evolution from $z\sim2.3$ to $z\sim3.4$
 ($0.3-0.4$ dex over only 1~Gyr) is as large as the evolution from $z\sim0$ to $z\sim2.3$ (0.3~dex over $\sim10$~Gyr).
These authors interpreted this sharp decrease in metallicity at fixed mass between $z\sim2.3$ and $z\sim3.4$
 as an indication that the mode of galaxy growth changed rapidly over this short time period.
They suggested that either galaxies at $z>3$ assemble from unevolved subcomponents and chemical enrichment mostly
 proceeds after merging into larger systems \citep{mai08}; have significantly larger gas inflow rates (from the IGM or gas-rich mergers)
 than galaxies at $z\sim2$, rapidly building up the gas reservoir while driving metallicities down \citep{man09} and potentially
 indicating a gas accumulation phase where gas flows are out of equilibrium \citep{dav12};
 or both gas inflow and outflow rates sharply rise from $z\sim2$ to $z>3$ while maintaining equilibrium \citep{tro14}.

\citet{ono16} investigated the MZR at $z\sim3.3$ using a sample of 41 galaxies at $z=3.0-3.7$ with MOSFIRE spectroscopy,
and \citet{suz17} expanded this sample with 10 [O\iii]-selected galaxies at the same redshifts.
Similarly to the AMAZE+LSD studies, these authors find that the MZR normalization decreases by $0.7$~dex between $z\sim0$ and $z\sim3.3$.
By applying analytic chemical evolution models \citep{lil13}, \citet{ono16} find the low metallicities at $z>3$ can be
 explained by a scenario in which the star-formation efficiency ($\epsilon\equiv\mbox{SFR}/\mbox{M}_{\text{gas}}$)
 evolves weakly with redshift at fixed \mstar, with a non-evolving $\epsilon$ providing a reasonable fit to their data.

These interpretations are in tension with recent observations and theoretical expectations.
Star-formation efficiency (the inverse of the depletion timescale) is observed to increase fairly strongly with increasing
 redshift as $\epsilon\propto(1+z)^{0.5-1.0}$ \citep[e.g.,][]{tac13,tac18,sco17,liu19}, though
 \citet{gen15} found a weaker $\epsilon\propto(1+z)^{0.34}$ that \citet{ono16} found plausible based on their data.
Furthermore, constant or weakly evolving $\epsilon$ models underpredict observed metallicities at $z\sim1.5-2.5$
 and predict stronger SFR dependence in the FMR than is observed at $z\sim0$ and $z\sim2.3$ \citep{man10,san18,cur20b}.
Modern numerical simulations of galaxy formation do not predict a break in the metallicity evolution above $z=3$, but
 instead find the MZR evolves smoothly up to very high redshifts ($z\gtrsim6$) at a rate of dlog(O/H)/d$z\approx-0.05$ to $-0.15$
 \citep{oka14,ma16,der17,dav17,dav19,tor19}, with the evolutionary rate actually slowing at $z\gtrsim3$ in some cases.
In this work, we find an approximately constant evolution at fixed \mstar\ of dlog(O/H)/d$z=-0.11$ out to $z\sim3.3$, in close agreement
 with simulations based on hierarchical galaxy formation and consistent with models where star formation efficiency increases with redshift.

It is of interest to understand why our results differ from earlier MZR studies at $z>3$.
We present a comparison of the sample properties of the MOSDEF, AMAZE+LSD, and \citet{ono16} $z>3$ samples in Appendix~\ref{app:z3lit}.
Briefly, we find that the AMAZE+LSD and \citet{ono16} samples are not biased in SFR compared to the MOSDEF $z\sim3.3$ sample, but
 the AMAZE+LSD sample displays a significant bias towards higher excitation (thus, lower O/H) manifested as higher [O\iii]/H$\beta$
 and O$_{32}$ ratios at fixed \mstar.
The \citet{ono16} sample is not obviously biased in excitation.
We have shown that the dust correction methods used in \citet{ono16} based on either \betauv\ or assuming \ebvgas=\ebvstars\ underestimates
 the true nebular reddening (see Appendix~\ref{app:dust}), biasing line ratios and leading to lower inferred metallicities.
In the stacking method of \citet{ono16}, they both normalize spectra by [O\iii]$\lambda$5007 and apply inverse variance weighting when
 combining spectra.
Including both of these steps gives high-SFR galaxies more weight in the stacked emission lines, which biases the metallicities of
 stacks low according to the FMR.
 
A major difference between this work and the analyses of \citet{mai08}, \citet{man09}, \citet{tro14}, and \citet{ono16} is the
 set of metallicity calibrations used to derive metallicities.
All four of these earlier studies use the calibrations of \citet{mai08} that are based on theoretical photoionization models at high metallicities
 (12+log(O/H$)\gtrsim$8.3) and empirical direct-method metallicities at low metallicities.
Theoretical calibrations are known to yield metallicities that are $\sim0.25$ dex higher than direct-method calibrations \citep[e.g.,][]{kew08}.
The $z\sim0$ samples lie almost entirely in the high-metallicity regime calibrated to photoionization models,
 while the $z>3$ samples lie in the direct-method calibration regime for the \citet{mai08} relations.
The mixing theoretical and empirical metallicities in the \citet{mai08} calibration sample thus artificially introduces $\sim0.25$~dex of
 additional MZR evolution at $z>3$.
In this work, we use calibrations based purely on the direct-method at all redshifts.

In Sec.~\ref{sec:mzrsys}, we found that $z=0$ calibrations underestimate the metallicities of high-redshift galaxies by $\sim0.1$~dex.
Thus, the use of $z=0$ calibrations at all redshifts in past studies also contributes to the larger observed evolution over $z=0-3.3$.
An additional systematic effect may arise from mixing different metallicity indicators at different redshifts, where in past studies
 metallicites at $z\lesssim2.5$ primarily depended on nitrogen-based indicators while metallicities at $z>3$ are based solely on oxygen-based indicators.
The redshift evolution of the star-forming sequence in the BPT diagram suggests that nitrogen- and oxygen-based $z=0$ calibrations will not produce
 consistent metallicities when applied at $z\gtrsim2$ \citep{ste14,san15,sha15}.

We thus find that earlier studies of the MZR at $z>3$ were impacted by a combination of effects that bias metallicity low,
 including biases in sample excitation properties, reddening correction, stacking techniques, and metallicity calibrations.
This analysis supersedes these earlier works with a sample that is representative of typical $z\sim3.3$ galaxies and a factor
 of four times larger, a dust correction method that is more robust,
 and metallicity derivations that use a uniform set of lines across all redshifts while properly accounting for evolving ISM conditions.

\subsubsection{Past studies of the FMR at $z>2$}

Early investigations of the FMR at high redshifts yielded conflicting results regarding whether O/H secondarily
 depends on SFR at $z>1$ and whether the FMR evolves
 \citep[e.g.,][]{chr12a,wuy12,wuy14,wuy16,bel13,hen13,sto13,cul14,mai14,ste14,zah14b,sal15,san15,yab15}.
More recent work based on larger samples and more uniform analyses of metallicity and SFR have demonstrated that
 O/H does carry a secondary dependence on SFR at fixed \mstar\ at $z\sim2$ \citep{san18} and ruled out strong
 evolution of the FMR, with O/H evolving $\le0.1$~dex at fixed \mstar\ and SFR out to $z\sim2.5$ \citep{san18,cre19,cur20b}.
The strength of the SFR dependence in high redshift samples remains poorly constrained (O/H$\propto$SFR$^{-0.1\text{ to }-0.3}$; \citealt{san18}),
 and larger samples spanning a wide dynamic range in sSFR are needed to improve this measurement.

In \citet{san15} and \citet{san18}, we found that $z\sim2.3$ galaxies have $\sim0.1$~dex lower O/H compared to $z\sim0$ galaxies
 matched in \mstar\ and SFR.
We used local metallicity calibrations for both the low- and high-redshift samples in those works.
In the present analysis, we now use an appropriate calibration at $z>2$ that yields $\sim0.05-0.1$~dex higher metallicities than local
 calibrations (Sec.~\ref{sec:mzrsys}), effectively eliminating the 0.1~dex offset from the $z=0$ FMR observed in earlier MOSDEF studies.
We now find excellent agreement with the $z\sim0$ FMR out to $z\sim3.3$, with $\Delta$log(O/H)$<$0.04~dex on average.
While earlier works found that $z>3$ galaxies fell $\sim0.3-0.6$~dex below the local FMR \citep{man10,tro14,ono16}, improvements in
 sample size, representativeness, and metallicity derivation techniques have seen this offset from the FMR at $z>3$ disappear
 (see Sec.~\ref{sec:z3lit}).

In Section~\ref{sec:fmr}, we fit a new parameterization of the $z\sim0$ FMR using our new direct-method $z\sim0$ calibrations
 (Fig.~\ref{fig:calibrations}).
\citet{cur20b} have recently fit the $z\sim0$ FMR using SDSS data and the $z\sim0$ direct-method calibrations of C17.
We find that our high-redshift samples do not display significant evolution relative to the $z\sim0$ total SFR FMR parameterization
 of \citet{cur20b}, with both the $z\sim2.3$ and $z\sim3.3$ stacks and means of the individual galaxies offset by $<$0.05~dex in O/H.
Note that this result is based on applying the high-redshift B18 metallicity calibrations to our $z>2$ samples.
If we instead use the $z\sim0$ C17 calibrations (the same set used by \citealt{cur20b}), we find that the $z>2$ galaxies are offset
 $\approx$0.10~dex lower in O/H than the \citet{cur20b} FMR.
This comparison again emphasizes the importance of accounting for evolving metallicity calibrations when studying MZR and FMR evolution
 over a wide redshift range.
Insofar as the relation among \mstar, SFR, and O/H reflects the interplay of gas flows and star formation, the non-evolution
 of the FMR suggests that galaxies remain near the equilibrium condition through the smooth baryonic growth process since $z\sim3.3$.

\subsection{The evolution of the outflow mass loading factor over $z=0-3.3$}

In Section~\ref{sec:mzrmodeling}, we constrained the mass-scaling of the outflow metal loading factor, \zetaout$\propto$$M_*^{-0.35\pm0.02}$, and
 found that this scaling holds over $z=0-3.3$ while the normalization increases with increasing redshift.
We now consider the implications for the scaling and normalization of the outflow mass loading factor, \etaout$\equiv$\mdotout/SFR.
\zetaout\ is the product of \etaout\ and the ratio of the outflow metallicity to that of the ISM (\zout/\zism).
If outflows are predominantly composed of entrained ISM material such that swept up ISM gas dominates the outflow mass over pure SNe ejecta,
 then \zout/\zism$\approx$1 and is constant with \mstar.
Consequently, \zetaout$\approx$\etaout.
Thus, based on the scaling we found between \zetaout\ and \mstar, \etaout$\propto$$M_*^{-0.35\pm0.02}$$\sim$\vcirc$^{-1.3~\text{to}~-1.8}$
 at all redshifts, \etaout$>$1 at all masses and redshifts, the MZR slope is set by \etaout($M_*$) at all redshifts,
 and MZR evolution is partially driven by an increase in \etaout\ at fixed \mstar\ with increasing redshift.

\subsubsection{Scaling of \etaout\ with \mstar}\label{sec:etaoutslope}

Our inferred \etaout$\propto$$M_*^{-0.35\pm0.02}$ scaling is in good agreement with observational constraints.
\citet{chi17} find \etaout$\propto$$M_*^{-0.43\pm0.07}$ and \etaout$\propto$\vcirc$^{-1.56\pm0.25}$ using measurements of rest-UV
 absorption lines for seven $z=0$ galaxies spanning log($M_*/\msun)\sim7-10.5$.
\citet{hec15} find \etaout$\propto$\vcirc$^{-0.98}$ ($\sim M_*^{-0.3}$) for the ``strong outflow'' subset of their $z\sim0$ sample using similar
 techniques, though their data are consistent with \etaout$\propto$\vcirc$^{-1\text{ to }-2}$ ($\sim M_*^{-0.3\text{ to }-0.6}$) when
 including the ``weak outflow'' objects as well.\footnote{Outflow properties inferred from observations of low-ionization rest-UV absorption lines or broad rest-optical emission
 lines only probe the warm ionized phase of outflows, while outflows are thought to additionally comprise neutral,
 molecular, and hot phases.
Comparing \etaout\ as inferred from chemical evolution models or numerical simulations to observational constraints on \etaout\ for the warm ionized phase
 implicitly assumes that \etaout\ scales similarly with \mstar\ across all outflow phases.}
\citet{cul19} measured the stellar mass-stellar metallicity relation (MZ$_*$R) for star-forming galaxies at $z\sim3.5$ (well matched in
 redshift to our sample at $z\sim3.3$) and found a MZ$_*$R slope that is similar to that of our $z\sim3.3$ gas-phase relation.
These authors found that models with \etaout$\propto$$M_*^{-0.4}$ provide a good fit to the $z\sim3.5$ MZ$_*$R, consistent with our findings.
\citet{lee19} model the MZ$_*$R for quiescent galaxies at $z\sim0.5$,
 tracing $\alpha$ elements via [Mg/H]$_{\text{stars}}$, and infer \etaout$\propto$$M_*^{-0.21\pm0.09}$.
\citet{for19} find \etaout$\propto$$M_*^{-0.1\pm0.2}$ from broadened emission lines in stacked spectra of 600 galaxies at $z=0.6-2.7$,
 consistent with our scaling within the large uncertainty.

We also find good agreement with \etaout\ scalings in cosmological hydrodynamic simulations.
In the the FIRE cosmological zoom-in simulations, \citet{mur15} find \etaout$\propto$$M_*^{-0.35\pm0.02}$ and
 \etaout$\propto$\vcirc$^{-1.0}$.
These authors do not find any strong redshift evolution in \etaout($M_*$), but likely would be unable to resolve the weak evolution
 of dlog(\etaout)/d$z\approx0.10$ implied by the constant \zout/\zism\ scenario because of the small number of low-redshift galaxies
 in FIRE.
Outflow metallicities probed in FIRE at 25\% of the virial radius are found to be a constant near-unity fraction of the ISM metallicity,
 \zout$\approx$1.2\zism\ with no \mstar\ dependence \citep{mur17}, in agreement with our
 assumption that outflows are almost entirely composed of entrained material.
In IllustrisTNG, \citet{nel19} find that \etaout$\approx$50 at 10$^{7.5}~\msun$ and 4 at 10$^{10.5}~\msun$ for outflowing material
 with $v_{\text{rad}}>0$~km~s$^{-1}$ at 10~kpc galactocentric radius, implying \etaout$\propto$$M_*^{-0.37}$.
These authors also find that \zout$\approx$\zism.
\citet{mit20} find \etaout$\propto$\vcirc$^{-1.5}$ for stellar feedback in the EAGLE simulations, and
 \etaout$\propto$$M_*^{-0.3\text{ to }-0.4}$ at $z\sim2-3$.

\subsubsection{Normalization of \etaout}\label{sec:etaoutnorm}

We showed in Figure~\ref{fig:zetaout} that the normalization of \zetaout($M_*$) increases with increasing redshift as
 dlog(\zetaout)/d$z=0.10\pm0.03$ at fixed \mstar, while its slope remains constant over $z=0-3.3$.
With \zout$\approx$\zism and \zetaout$\approx$\etaout, the outflow mass loading factor at $10^{10}~\msun$ (\etaoutten)
 is log(\etaoutten$)=0.20\pm0.01$ at $z\sim0$, $0.43\pm0.05$ at $z\sim2.3$, and $0.56\pm0.05$ at $z\sim3.3$ in our fiducial model.
As discussed in Section~\ref{sec:yield}, the normalization of \zetaout\ is sensitive to the assumed stellar yield, which varies
 by a factor of $\sim2$ lower or higher than our assumed value of $y_O=0.015$ over the range of standard IMFs.
Increasing (decreasing) the assumed yield by a factor of 2 results in an increase (decrease) of \zetaout\ normalization by a
 factor of $\sim3$, such that the systematic uncertainty in our inferred \etaout\ values associated with the stellar
 yield is $\pm0.5$~dex.
Note that changing the yield has no significant effect on the inferred slope of \zetaout($M_*$) or \etaout($M_*$).

The \etaout\ normalization above agrees well with results from recent cosmological numerical simulations.
\citet{nel19} find log(\etaoutten$)\approx0.7$ at $z\sim0$ in IllustrisTNG.
This value is a factor of 3 larger than our \etaoutten\ at $z\sim0$, but IllustrisTNG has a higher stellar yield
 ($y_Z=0.050$, corresponding to $y_O\approx0.30-0.35$; \citealt{tor19}) than we assume by a factor of $\sim2$ that
 accounts for the difference in \etaout\ normalization. 
In the EAGLE simulations, log(\etaoutten)$\approx$0.1 at $0.0<z<0.3$ and log(\etaoutten)$\approx$0.35 at $2.4<z<3.4$, in reasonable
 agreement with our results considering uncertainties in supernova yields \citep{mit20}.
Using cosmological zoom-in simulations, \citet{chr16} find log(\etaoutten)$\approx$0.2 with little change over $z=0-2$, in good agreement with
 our results at $z\sim0$ but slightly lower than we find at $z\sim2.3$.
\citet{mur15} find log(\etaoutten)=0.55 in the FIRE simulations for a sample of galaxies predominantly at $2.0<z<4.0$, in excellent agreement
 with our values at $z\sim2.3-3.3$.

Studies observationally constraining \etaout\ normalization generally find \etaoutten$\sim$$0.1-1$ at $z\sim1-2$ for the
 low-ionization or warm ionized outflow phases \citep{mar13,jon18,davies19,for19}.
These values represent lower limits because they are based on measurements of only one phase of the multiphase outflowing gas.
As such, these obervational constraints are consistent with both our inferred \etaout\ normalization and those found in simulations,
 and suggest that mass outflow rates are not orders of magnitude different from galaxy SFRs.

We thus find that our inferred \etaout($M_*$) agrees well with numerical simulations in both normalization and slope (Sec.~\ref{sec:etaoutslope}).
Because the physics of SNe energy injection occur on subgrid scales, the IllustrisTNG and EAGLE simulations input mass loading and
 velocity scalings at injection that depend on \vcirc\ or \mstar.
Both simulations are in reasonable agreement with our results, such that we cannot distinguish between their different feedback prescriptions.
Out of these simulations, only FIRE (with which we find the closest agreement) has sufficient resolution to implement ISM-scale stellar
 feedback models including stellar winds, radiation pressure, photoionization and photoelectric heating, and Type Ia and core-collapse SNe such that
 the loading of outflows is entirely emergent from these physical processes \citep{hop14}.
Our close agreement with FIRE suggests that their simulations capture the most important physical mechanisms for star-formation driven outflows.

\subsubsection{Redshift evolution of \etaout}\label{sec:etaoutevo}

A key feature of our fiducial model is that \zetaout, and consequently \etaout, increases at fixed \mstar\ with increasing redshift.
In EAGLE, \citet{mit20} find that \etaout\ increases at fixed \mstar\ with redshift by $\sim0.2-0.4$~dex between $z\sim0$ and $2.4<z<3.4$,
 in good agreement with our inferred evolution of dlog(\etaout)/d$z=0.10\pm0.03$.
In contrast, \etaout\ is constant or slightly declining with redshift at fixed \mstar\ in IllustrusTNG \citep{nel19}.
Understanding why EAGLE and IllustrisTNG display this differing behavior in the evolution of \etaout\ may yield insight
 into which subgrid feedback prescriptions are more realistic.
\citet{mur15} do not report any redshift dependence of \etaout($M_*$) for FIRE, but likely would not be able to resolve the
 slow \etaout\ evolution that we infer because their sample is primarily made up of simulated galaxies at $2.0<z<4.0$ with
 only a handful of galaxies at $z<0.5$ to set an evolutionary baseline.

There is empirical evidence that \etaout\ may be larger in typical high redshift galaxies than in $z\sim0$ galaxies at the same \mstar.
Observations suggest \etaout\ positively correlates with $\Sigma_{\text{SFR}}$ \citep[e.g.,][]{new12,arr14,dav19},
 such that $z\sim2-3$ galaxies may have larger \etaout\ than
 similar-mass $z\sim0$ galaxies because of their $\sim2$ orders of magnitude larger $\Sigma_{\text{SFR}}$ \citep{sha19}.
Seemingly in conflict with this trend is the finding that \etaout\ and SFR are anti-correlated in local samples \citep[e.g.,][]{hec15}.
However, at fixed redshift, the trend between \etaout\ and SFR carries an imprint of the anti-correlation between \etaout\ and
 \mstar\ or \vcirc\ due to the SFR-\mstar\ relation.
Our model can be explained if instead \etaout\ increases with SFR \textit{at fixed stellar mass}.

\begin{figure}
 \includegraphics[width=\columnwidth]{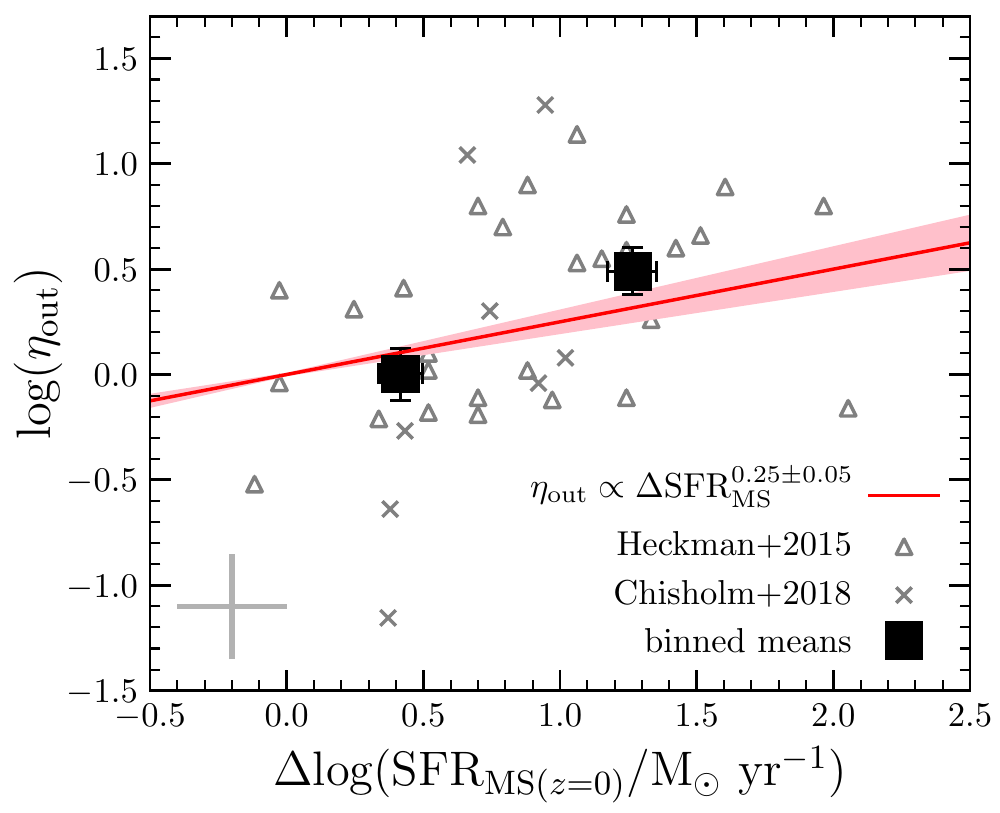}
 \centering
 \caption{
The outflow mass loading factor as a function of offset from the $z=0$ star-forming main sequence.
Gray points display the $z\sim0$ samples of \citet[][triangles]{hec15} and \citet[][x-shaped points]{chi17,chi18}, while the
 gray error bar shows the mean uncertainty.
Black squares denote median values in two bins of main sequence offset.
The red line and shading represents the relation required to match the evolution of \etaout\ with redshift in our models based
 on the offset of the $z\sim2.3$ and $z\sim3.3$ samples from the $z=0$ SFR-\mstar\ relation, arbitrarily normalized for
 comparison to the $z\sim0$ data.
}\label{fig:z0delsfms}
\end{figure}

To search for this trend, we use samples of $z\sim0$ galaxies with \etaout\ derivations based on rest-UV absorption profiles
 from \citet{hec15} and \citet{chi17,chi18}.
We take \etaout\ and SFR(UV) values as tabulated in these sources, and calculate the offset in SFR from the $z=0$ star-forming main
 sequence, $\Delta$SFR$_{\text{MS}(z=0)}$, using the $z=0$ SFR(UV)-\mstar\ relation of \citet{coo14}.
In Figure~\ref{fig:z0delsfms}, we show \etaout\ vs.\ offset from the star-forming main sequence for these galaxies.
We find a loose but 3$\sigma$ significant correlation between \etaout\ and $\Delta$SFR$_{\text{MS}(z=0)}$, with a Spearman correlation
 coefficient of 0.46 and a p-value of 0.003.
This trend implies that \etaout\ increases with increasing SFR at fixed \mstar.
Since SFR increases at fixed \mstar\ with increasing redshift, this trend is qualitatively consistent with the results of our modeling.

At fixed \mstar\ in our models, \etaout\ increases by $0.23\pm0.05$~dex between $z\sim0$ and $z\sim2.3$, and by $0.36\pm0.05$~dex over $z=0-3.3$.
Likewise, at fixed \mstar, SFR increases by 1.22~dex and 1.4~dex over these same redshift intervals, respectively (Fig.~\ref{fig:sfrmstar} and
 equations~\ref{eq:z0sfrmstar}-\ref{eq:z3sfrmstar}).
If we operate under the assumption that \etaout\ is connected to SFR independent of redshift,
 our models predict a rough scaling of \etaout$\propto$$\Delta$SFR$_{\text{MS}(z=0)}$$^{0.25\pm0.05}$, shown by the red line in
 Figure~\ref{fig:z0delsfms}.
This scaling is consistent with binned means of the individual galaxies (black squares).
If this relation between \etaout\ and SFR at fixed \mstar\ is redshift invariant, then it can explain the increasing normalization of
 \zetaout($M_*$) and \etaout($M_*$) with increasing redshift.
Thus, a redshift invariant relation between \etaout\ and SFR at fixed \mstar\ can explain the increasing normalization of
 \zetaout($M_*$) and \etaout($M_*$) with increasing redshift.

Theoretical work is required to understand why \etaout\ increases with SFR at fixed \mstar.
One clear difference in the ISM of low- and high-SFR galaxies of the same mass is the gas fraction:
 \mugas$\propto$SFR$^{0.5}$ at fixed \mstar\ \citep[e.g.,][]{sai16,gen15,tac18}.
A higher \etaout\ with increasing $\Delta$SFR$_{\text{MS}(z=0)}$ implies that a larger mass of ISM gas is swept into outflows
 per unit SFR (i.e., per SN) in high-SFR, gas-rich galaxies.
This scenario agrees with the analytic theory work of \citet{hay17}, in which larger gas fractions correspond to higher mass-loading
 factors in a turbulent ISM.
In their framework, turbulence creates gas patches with a range of surface densities, where low-density patches are more easily blown
 out of the galaxy by SNe feedback to form large-scale outflows.
The fraction of the ISM below the critical escape density increases with increasing \mugas, leading to more efficient winds and larger
 \etaout\ at higher gas fractions.
For simplicity, we have treated the actions of gas fraction (metal dilution) and outflows (metal removal) as two independent mechanisms,
 but this theoretical work demonstrates that \mugas\ and \etaout\ (and, in turn, \zetaout) are not decoupled but are instead physically linked.
This connection is natural since gas content and SFR are tightly linked (e.g., the KS relation; \citealt{ken98b}), and star formation is the source
 of energy injection to drive outflows.

\subsection{On the non-evolution of the FMR}

In Section~\ref{sec:fmr}, we found that the FMR shows no sign of evolving out to $z\sim3.3$ with good precision (Fig.~\ref{fig:fmr}).
That is, galaxies with the same \mstar\ and SFR have the same O/H on average at all redshifts over $z=0-3.3$.
In our model framework, a simple way to produce a non-evolving FMR is to have \mugas($M_*,SFR$) and \etaout($M_*,SFR$) be
 redshift invariant.
We can test this scenario by seeing whether $z\sim0$ galaxies matched in \mstar\ and SFR to $z\sim2$ galaxies also have gas fractions
 similar to those of the high-redshift sample. 
Since we have shown that such a matched set of galaies have similar O/H, the PS11 theoretical framework then implies
 that \zetaout\ and, by extension, \etaout\ are also the same.

Empirically, gas fraction scales approximately as \mugas$\propto$SFR$^{0.5}$ at fixed \mstar\ at $z\sim0$ \citep{gen15,sai16,tac18}.
At fixed \mstar, SFR increases by 1.22~dex between $z\sim0$ and $z\sim2.3$ (Fig.~\ref{fig:sfrmstar}).
Accordingly, local galaxies matched in \mstar\ and SFR to our $z\sim2.3$ sample have \mugas\ that is $\sim0.6$~dex higher
 than $z\sim0$ main-sequence galaxies of the same mass.
At log($M_*/\msun)=10.5$ (the mass of the $z>1$ \mugas\ calibration samples) and for galaxies on the main sequence,
 the difference in \mugas\ between $z\sim2.3$ and $z\sim0$ is 0.53~dex based on our assumed scaling relations (Fig.~\ref{fig:z2z3gas}).
Thus, galaxies matched in \mstar\ and SFR at both redshifts have similar gas fractions.
Since they have nearly the same metallicity, they consequently must have similar \etaout\ based on our models.

The theoretical picture of the FMR is usually explained by variations in gas fraction based on gas accretion rates:
 at fixed \mstar, high-SFR galaxies have low metallicities because recent accretion of metal-poor gas has increased the gas fraction and
 diluted the metals in the ISM while driving up the SFR \citep[e.g.,][]{man10,yat12,day13}.
Our results instead suggest that changes in \mugas\ alone cannot fully account for the observed O/H variation in the FMR.
At fixed \mstar, galaxies with higher SFR must also have larger \etaout\ to explain their lower O/H.
In Figure~\ref{fig:z0delsfms}, we showed that measurements of \etaout\ in local star-forming galaxies positively correlate
 with SFR at fixed \mstar, supporting this scenario.

We thus find that the FMR is not only driven by metal dilution due to variations in gas fractions that reflect accretion rates,
 but also requires variations in metal removal efficiency as a function of SFR.
To explain the observed $z\sim0$ FMR, galaxies above the local star-forming main sequence must have both higher gas
 fractions and more efficient metal removal through winds (i.e., higher \zetaout\ and \etaout) than main-sequence galaxies at the same \mstar.
At fixed \mstar, the lower O/H of $z\sim2.3$ and $z\sim3.3$ galaxies relative to the $z\sim0$ MZR reflects their larger \mugas\ and
 higher \zetaout, in accordance with their higher SFRs, than local main-sequence galaxies.
The reason that the FMR does not evolve is because galaxies that have the same \mstar\ and SFR have both similar \mugas\ and
 \etaout\ over $z=0-3.3$, consequently yielding similar ISM metallicity.
Stated another way, the FMR is redshift invariant out to $z\sim3.3$ because \mugas($M_*,SFR$) and \etaout($M_*,SFR$) do
 not significantly evolve over this redshift range.

\subsection{A constant \zout\ with \mstar\ is disfavored}

In the discussion above, we have assumed that outflows are dominated by entrained ISM material such that \zout$\approx$\zism.
In contrast, \citet{chi18} find that \zout\ is roughly solar metallicity (\zout=$1.0\pm0.6$Z$_{\odot}$) and independent of \mstar\
 over $10^7-10^{10.5}~\msun$
 in the same seven local galaxies analyzed in \citet{chi17}, such that \zout/\zism$\propto$\zism$^{-1}$$\propto$$M_*^{-0.4}$.
Combined with their scaling of \etaout$\propto$$M_*^{-0.4}$ \citep{chi17},
 this leads to \zetaout$\propto$$M_*^{-0.8}$ (\vcirc$^{-3.4}$),
 much steeper than our best-fit $z\sim0$ \zetaout($M_*$) and unable to fit the observed $z\sim0$ MZR with a low-mass slope of $-0.3$.
The steep \zetaout($M_*$) of \citet{chi18} may be due to the properties of the lowest-mass galaxies in their sample.
The mass dependence is anchored by two galaxies at 10$^7~\msun$ that both lie $\sim1$~dex
 above the $z=0$ star-forming main sequence \citep{coo14} and thus may have unrepresentative outflow properties
 that lead to a steep inferred \zetaout($M_*$).
\citet{pee11} inferred a similarly steep \zetaout$\propto$\vcirc$^{-3.5}$
 and argued for a relatively flat \zout-\mstar\ relation in order to explain observed MZRs with steep low-mass slopes
 \citep[e.g.,][]{tre04,kew08}.
However, modern determinations of the $z\sim0$ MZR yield shallower low-mass slopes that are inconsistent with this scenario
 (e.g., this work; \citealt{bla19}; \citealt{cur20b}).

If we assume \zout\ is a constant 1.0~$Z_{\odot}$ with no dependence on \mstar\ or redshift,
 then \zout/\zism$\propto$\zism$^{-1}$$\propto$$M_*^{-0.30}$ for our best-fit MZR.
Since \zetaout$\propto$$M_*^{-0.35\pm0.02}$ (Fig.~\ref{fig:zetaout}), this then implies that \etaout\ is nearly constant with very
 little \mstar\ dependence (\etaout$\propto$$M_*^{-0.05}$).
Neither observations nor theory support a scenario in which \etaout\ is nearly constant with \mstar.
Theoretically, momentum- and energy-driven winds are expected to have \etaout\ scale as \vcirc$^{-1}\sim M_*^{-1/3}$ and
 \vcirc$^{-2}\sim M_*^{-2/3}$, respectively \citep{dek86,mur05}.
A scenario where \zout\ is roughly independent of \mstar\ is therefore disfavored because it requires \etaout\ to be nearly
 independent of \mstar\ or \vcirc, in conflict with observations and theoretical models of star-formation driven winds.

\section{Summary \& Conclusions}\label{sec:summary}

We have investigated the evolution of the mass-metallicity relation (MZR) and the fundamental metallicity relation (FMR)
 using representative samples of $\sim300$ galaxies at $z\sim2.3$ and $\sim150$ galaxies at $z\sim3.3$ from the MOSDEF survey.
Our analysis improves upon past studies by utilizing a larger and more representative high-redshift sample,
 employing a dust corrrection method calibrated to H$\alpha$/H$\beta$ measurements at $z\sim2$ for improved SFRs and dereddened line ratios
 for galaxies lacking either H$\alpha$ or H$\beta$ detections,
 and applying a uniform metallicity derivation that accounts for evolving ISM conditions.
Specifically, we use the same set of emission lines ([O\ii], [Ne\iii], H$\beta$, and [O\iii]) to estimate metallicities
 at all redshifts, and apply different calibrations at $z\sim0$ and $z>1$, where the $z\sim0$ relations are calibrated
 to typical $z\sim0$ star-forming galaxies while the calibrations used at high redshifts are calibrated to local
 analogs of high-redshift galaxies that have similar ionization conditions to those in galaxies observed at $z\sim2$.
Despite this careful approach, there is still a non-negligible systematic uncertainty associated with the choice of
 metallicity calibration which can be addressed by future work.
Larger samples of galaxies at $z>1$ with direct-method metallicities and multi-dimensional approaches at $z\sim0$ capable of
 taking into account the wide dynamic range of galaxy properties (e.g., sSFR) are needed to make further progress.
Our main conclusions are summarized as follows.

\begin{enumerate}
\item The line ratios [O\iii]/H$\beta$, O$_{32}$, R$_{23}$, and [Ne\iii]/[O\ii] decrease with increasing \mstar (Fig.~\ref{fig:z2z3ratios}),
 as expected if O/H is positively correlated with \mstar.
These four line ratios evolve significantly from $z\sim0$ to $z\sim2.3$ at fixed \mstar, but only show a small change from $z\sim2.3$
 to $z\sim3.3$ (Fig.~\ref{fig:allratios}), suggesting that the evolution in O/H at fixed \mstar\ is smaller over $z=2.3-3.3$ than over $z=0-2.3$.
\item Stellar mass and O/H are significantly correlated at $z\sim2.3$ and $z\sim3.3$ (Fig.~\ref{fig:z2z3mzr}).
Individual galaxies follow a tight sequence around the mean MZRs defined by stacked spectra with an intrinsic scatter of $\approx$0.1~dex in O/H,
 similar to the scatter of the $z\sim0$ MZR.
\item The low-mass power law slope of the MZR does not evolve out to $z\sim3.3$, with a value of $\gamma=0.28\pm0.01$ at $z\sim0$,
 $0.30\pm0.02$ at $z\sim2.3$, and $0.29\pm0.02$ at $z\sim3.3$ (Fig.~\ref{fig:allmzr}).  This remarkable invariance of the MZR slope
 suggests that the same physical process (i.e., the scaling of \zetaout\ with \mstar, see point 7 below)
 controls the slope of the MZR at low and high redshifts.
\item At fixed \mstar, O/H smoothly decreases with increasing redshift as dlog(O/H)/d$z=-0.11\pm0.02$ out to $z\sim3.3$.
This evolution rate is consistent over log($M_*/\msun)=9.0-10.5$.
The offsets in O/H at $10^{10}~\msun$ are $-0.26\pm0.02$~dex from $z\sim0$ to $z\sim2.3$,
 and $-0.10\pm0.03$~dex between $z\sim2.3$ and $z\sim3.3$.
This gradual metallicity evolution that is uniform across \mstar\ is consistent with MZR evolution in modern cosmological numerical
 simulations.
\item The FMR does not evolve out to $z\sim3.3$ (Fig.~\ref{fig:fmr}).
Galaxies at $z\sim2.3$ and $z\sim3.3$ fall on the FMR defined by $z\sim0$ galaxies, with an average offset of $<$0.04~dex in O/H
 for stacked spectra and individual galaxies.
The intrinsic scatter of individual $z>2$ galaxies around the FMR is 0.06~dex in O/H, smaller than the MZR scatter and comparable to
 the FMR scatter at $z\sim0$.
\item Using analytic chemical evolution models \citep{pee11}, we infer the outflow metal loading factor,
 $\zeta_{\text{out}}\equiv \frac{Z_{\text{out}}}{Z_{\text{ISM}}}\times\frac{\dot{M}_{\text{out}}}{SFR}$,
 which parameterizes
 the efficiency with which winds remove metals from the ISM (Fig.~\ref{fig:zetaout}).
At all redshifts, \zetaout\ decreases with increasing \mstar.
The scaling of \zetaout\ with \mstar\ is consistent across $z=0-3.3$, with \zetaout$\propto$$M_*^{-0.35\pm0.02}$.
At fixed \mstar, \zetaout\ increases with increasing redshift as dlog(\zetaout)/d$z=0.10\pm0.03$.
\item The slope of the MZR is primarily set by the scaling of \zetaout\ with \mstar\ at all redshifts over $z=0-3.3$.
Increasing gas fractions with decreasing \mstar\ do not play a major role in setting the low-mass MZR slope.
Our models suggest that the low-mass MZR slope is invariant out to $z\sim3.3$ because the metal removal efficiency of
 winds scales similarly with \mstar\ over this entire redshift range.
\item The evolution of the normalization of the MZR towards lower O/H at fixed \mstar\ with increasing redshift
 is driven by both an increase in gas fraction and an increase in \zetaout\ at fixed \mstar\ toward high redshift.
Evolution of \mugas\ and \zetaout\ each account for roughly half of the observed metallicity evolution (Fig.~\ref{fig:mzrevolution}).
Thus, compared to low-redshift galaxies of the same mass, high redshift galaxies have lower metallicity because metals are more
 heavily diluted in the gas-rich ISM \textit{and} metals are more efficiently removed from the ISM through outflows.
\item If the dominant mass component of outflows is entrained ISM gas, then \zout/\zism$\approx$1 and the
 outflow mass-loading factor, \etaout$\equiv$$\dot{M}_{\text{out}}$/SFR, scales as \etaout$\propto$$M_*^{-0.35\pm0.02}$.
This scaling is in agreement with observations of ionized outflows and recent numerical simulations.
At fixed \mstar, \etaout\ increases with increasing redshift.
Observational constraints on \etaout\ from rest-UV absorption lines suggest that \etaout\ increases with increasing SFR
 at fixed \mstar\ (Fig.~\ref{fig:z0delsfms}),
 consistent with our model when the evolution of the SFR-\mstar\ relation is taken into account.
This model implies that, at fixed \mstar, both \mdotout\ and \mdotin\ increase relative to SFR with increasing redshift.
\item The FMR does not evolve out to $z\sim3.3$ because \mugas($M_*,SFR$) and \etaout($M_*,SFR$) do not evolve with redshift.
Over the range $z=0-3.3$, galaxies at fixed \mstar\ and SFR have similar O/H and gas fractions, leading us to infer that they must
 also have similar outflow mass and metal loading factors.
The dependence of O/H on SFR at fixed \mstar\ not only reflects dilution of ISM metals from recent accretion, but is also driven
 by variations in the metal removal efficiency of outflows.
Variations in \etaout\ and \mugas\ correlate, such that galaxies with higher gas fractions have higher outflow mass loading.
This picture is in agreement with theoretical work in which such a link between \etaout\ and \mugas\ arises naturally
 from density variations in a turbulent ISM.
\end{enumerate}

The redshift invariant MZR slope and FMR, and the gradual evolution of the MZR over $z=0-3.3$ point to
 a picture in which galaxies remain close to equilibrium between inflows, outflows, and internal gas
 processing (star formation and gas reservoir growth) without rapid changes in the mode of galaxy assembly
 over the past 12~Gyr of cosmic history.
While it is not feasible to probe gas-phase metallicity evolution at $z>4$ with current facilities,
 \textit{JWST} will provide rest-optical spectra of galaxies at $z=4-10$ in
 the near-future, opening the door to chemical evolution studies in the earliest epoch of galaxy formation.
If gas fractions continue rising rapidly beyond $z=4$, we expect that the shaping of the MZR
 transitions from outflow-dominated at $z\lesssim3$ to gas-dominated at very high redshifts, which may manifest in
 a change in slope or evolution rate.
Early galaxies may also be out of equilibrium as the gas reservoir is rapidly built up for the first time \citep{dav12}.
Extending gas-phase abundance studies beyond $z=4$ is crucial to understanding the formation of the first
 generation of galaxies.

\acknowledgements Support for this work was provided by NASA through the NASA Hubble Fellowship grant \#HST-HF2-51469.001-A awarded
 by the Space Telescope Science Institute, which is operated by the Association of Universities for Research
 in Astronomy, Incorporated, under NASA contract NAS5-26555.
We acknowledge support from NSF AAG grants AST-1312780, 1312547, 1312764, and 1313171,
 archival grant AR-13907 provided by NASA through the Space Telescope Science Institute,
 and grant NNX16AF54G from the NASA ADAP program.
We also acknowledge a NASA
 contract supporting the ``WFIRST Extragalactic Potential
 Observations (EXPO) Science Investigation Team'' (15-WFIRST15-0004), administered by GSFC.
We additionally acknowledge the 3D-HST collaboration
 for providing spectroscopic and photometric catalogs used in the MOSDEF survey.
We wish to extend special thanks to those of Hawaiian ancestry on
 whose sacred mountain we are privileged to be guests. Without their generous hospitality,
 the work presented herein would not have been possible.



\appendix

\section{A. New nebular reddening correction calibration}\label{app:dust}

Deriving reddening from the observed flux ratio of hydrogen recombination lines widely separated in wavelength
 (e.g., H$\alpha$/H$\beta$) is considered the gold standard for the dust correction of nebular emission spectra.
This analysis includes a sample at $z=2.9-3.8$ for which H$\alpha$ is not covered in the spectral bandpass and
 H$\beta$ is typically the only H recombination line detected.
Furthermore, H$\beta$ is typically one of the weaker lines and is not detected for all $z\sim2.3$ sources that have H$\alpha$
 detections.
To maximize our sample sizes and avoid biasing the samples by requiring detections of weak lines, we require a dust correction
 technique that does not rely on detections of certain sets of emission lines.
All galaxies in our sample have extensive photometry from which stellar properties (e.g, \mstar, SFR(SED), and \ebvstars) were derived
 through SED fitting (see Sec.~\ref{sec:sedfitting}).
We calibrated a relation between best-fit properties from SED fitting and \ebvgas\ derived from the Balmer decrement using
 a sample of 326 star-forming galaxies at $2.04\le z\le2.65$ from the MOSDEF survey that have detections of both H$\alpha$
 and H$\beta$ at S/N$\ge$3. 

\begin{figure*}
 \includegraphics[width=\textwidth]{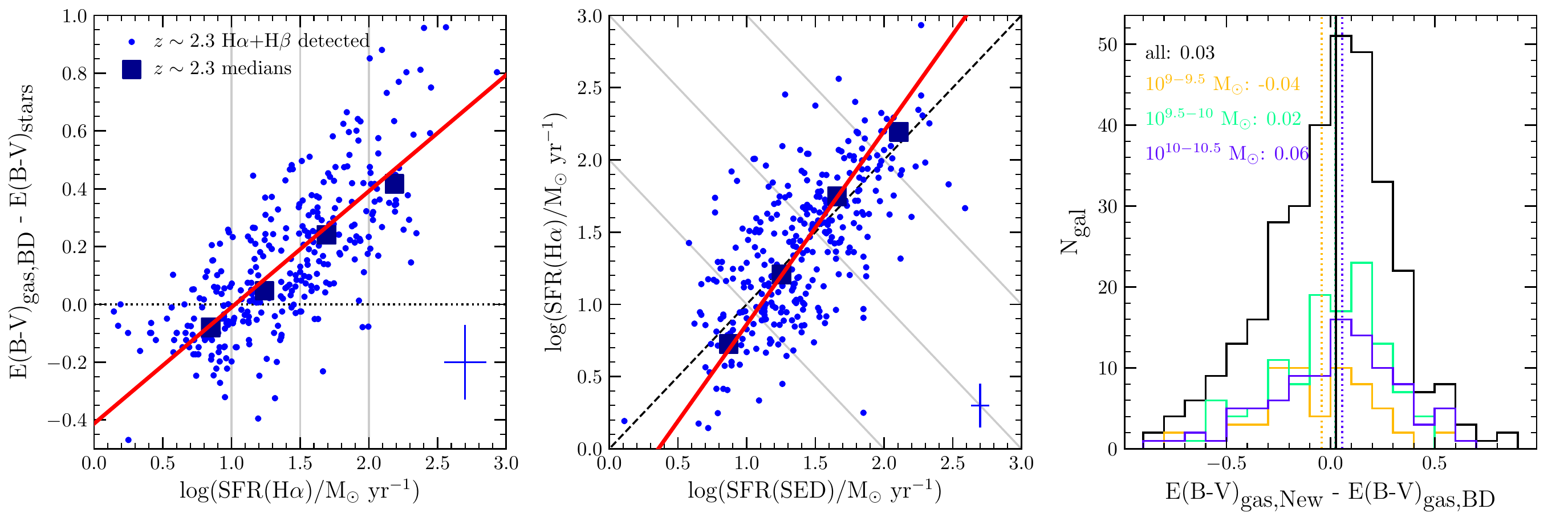}
 \centering
 \caption{
\textbf{Left:} The difference between nebular reddening (\ebvgas) and stellar reddening (\ebvstars) as a function of SFR(H$\alpha$)
 for star-forming galaxies at $z\sim2.3$ with detections of both H$\beta$ and H$\alpha$, where \ebvgas\ is derived using the Balmer decrement.
Individual galaxies are shown as blue circles, while medians in bins defined by the gray lines are presented as dark blue squares.
The error bar in the lower right corner denotes the median uncertainty of the individual galaxies.
The red line shows the best-fit relation.
\textbf{Middle:} SFR(H$\alpha$) vs.\ SFR(SED), with points and lines as in the right panel.
\textbf{Right:} Histogram of the difference between \ebvgas\ derived via our new calibration (equation~\ref{eq:newebvgas1})
 and \ebvgas\ inferred from the Balmer decrement.
The black line shows the distribution for the entire $z\sim2.3$ sample, while the colored histograms display the distributions for
 subsets in stellar mass.
Vertical lines show the median offset of each histogram and are given in the text in the top-left corner.
}\label{fig:newdust}
\end{figure*}

\citet{red15} noted that the difference between \ebvgas\ and \ebvstars\ is a function of SFR(H$\alpha$), where the difference
 between the reddening of the two components increases with increasing SFR.
This relation is displayed in the left panel of Figure~\ref{fig:newdust} for our $z\sim2.3$ Balmer decrement sample.
The best-fit linear relation to the individual galaxies is
\begin{equation}\label{eq:deltaebv}
\begin{multlined}
\text{E(B-V)}_{\text{gas}} - \text{E(B-V)}_{\text{stars}} = \\ 0.402\pm0.019\times\log\left(\frac{\text{SFR(H}\alpha)}{\text{M}_{\odot} \text{yr}^{-1}}\right) - 0.413\pm0.028
\end{multlined}
\end{equation}
where slope and intercept have a covariance of $\rho=-0.94$.
Calculating SFR(H$\alpha$) requires a dust correction, therefore we fit the relation between SFR(H$\alpha$) and SFR(SED),
 displayed in the middle panel of Figure~\ref{fig:newdust}, obtaining
\begin{equation}\label{eq:sfrhasfrsed}
\begin{multlined}
\log\left(\frac{\text{SFR(H}\alpha)}{\text{M}_{\odot} \text{yr}^{-1}}\right) = \\ 1.338\pm0.069 \times \log\left(\frac{\text{SFR(SED)}}{\text{M}_{\odot} \text{yr}^{-1}}\right) - 0.477\pm0.098
\end{multlined}
\end{equation}
with a covariance of $\rho=-0.97$ between slope and intercept.

Combining equations~\ref{eq:deltaebv} and~\ref{eq:sfrhasfrsed} yields
\begin{equation}\label{eq:newebvgas1}
\begin{multlined}
\text{E(B-V)}_{\text{gas}} = \text{E(B-V)}_{\text{stars}} - 0.604 \\ + 0.538\times\log(\text{SFR(SED)}) .
\end{multlined}
\end{equation}
Because SFR increases with increasing redshift at fixed \mstar while \ebvstars\ is approximately constant with redshift at fixed \mstar\ 
 \citep{whi17,mcl18,cul18}, equation~\ref{eq:newebvgas1} implies that \ebvgas\ increases with redshift at fixed \mstar.
Instead, observations based on the Balmer decrement suggest that \ebvgas\ at fixed \mstar\ does not significantly evolve out to $z\sim2.3$
 \citep[][Shapley et al., in prep.]{the19}.
At log($M_*/\msun)=9.9$ (the median mass of our metallicity samples), we find that the mean SFR(SED) for MOSDEF star-forming galaxies
 at $z_{\text{med}}$=[1.53, 2.29, 3.27] is $\langle$log(SFR(SED)/$\msun$ yr$^{-1})\rangle$=[1.26, 1.41, 1.62], implying
 dlog(SFR(SED))/dlog($z$)$\approx$0.20.
We add a redshift term, normalized to $z=2.3$, to equation~\ref{eq:newebvgas1} to account for the evolution of the SFR(SED)-\mstar\ relation
 such that \ebvgas\ will not evolve at fixed \mstar.
In this way, we obtain the final expression for dust correction in the absence of a Balmer decrement measurement:
\begin{equation}\label{eq:newebvgas2}
\begin{multlined}
\text{E(B-V)}_{\text{gas}} = \text{E(B-V)}_{\text{stars}} - 0.604 \\ + 0.538\times[\log(\text{SFR(SED)}) - 0.20\times(z - 2.3)] .
\end{multlined}
\end{equation}
This calibration is valid over $z\sim1-4$ and log($M_*/\msun)\sim9.0-11.0$,
 and only for stellar properties derived from SED fitting with a similar set of assumptions to ours (Sec.~\ref{sec:sedfitting}).
Outside of this redshift range, the redshift term may require modification to properly trace the evolution of the
 star-forming main sequence.

Recent work has suggested that a steep attenuation law similar to the SMC curve is more appropriate for high-redshift
 galaxies, especially at low \mstar\ and low metallicity \citep[e.g.,][]{cap15,red18b,shi20}.
If we instead assume subsolar stellar metallicity ($Z_*=0.0031$) and the SMC extinction curve of \citet{gor03}, the calibration
 differs:
\begin{equation}\label{eq:newebvgassmc}
\begin{multlined}
\text{E(B-V)}_{\text{gas}} = \text{E(B-V)}^{\text{SMC}}_{\text{stars}} - 0.645 \\ + 0.933\times[\log(\text{SFR(SED)}^{\text{SMC}}) - 0.20\times(z - 2.3)].
\end{multlined}
\end{equation}
Equation~\ref{eq:newebvgas2} should be applied if the \citet{cal00} curve is assumed for SED fitting, while
 equation~\ref{eq:newebvgassmc} should be used in the case of the steeper SMC curve.
Note that we obtain consistent nebular reddening estimates for our samples in either case
because the SED-derived properties are calibrated to the same Balmer decrement measurements regardless of SED fitting assumptions.

\begin{figure*}
 \includegraphics[width=\textwidth]{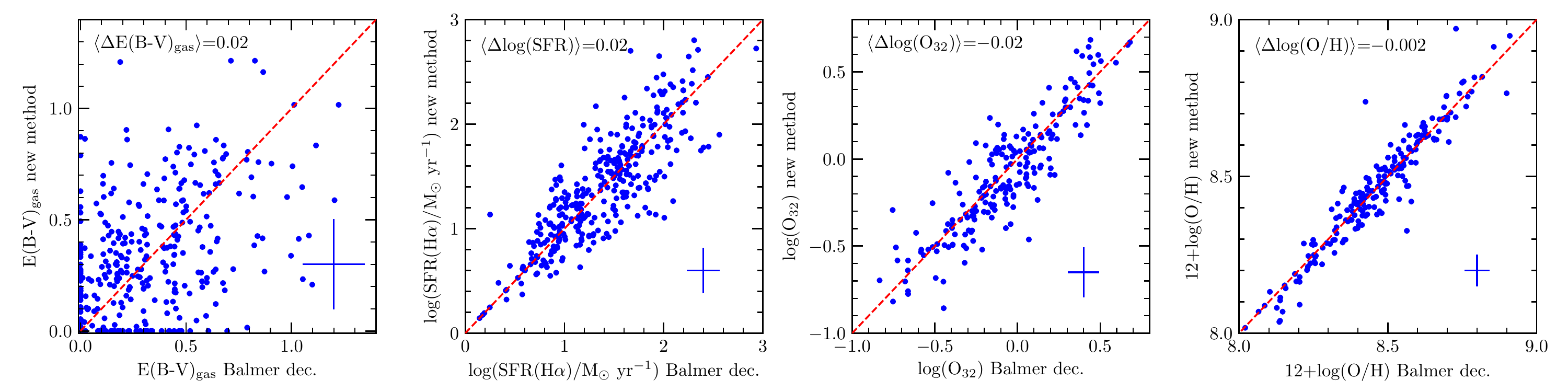}
 \centering
 \caption{
Comparison of \ebvgas\ (left), SFR(H$\alpha$) (left-middle), dust-corrected O$_{32}$ (right-middle), and O/H (right) derived using
 the new \ebvgas\ calibration (equation~\ref{eq:newebvgas1}; vertical axis) and \ebvgas\ based on the Balmer decrement (horizontal axis).
Red dashed lines show a one-to-one relation, and the mean vertical offset is given in the upper-left corner of each panel.
}\label{fig:newdustcomparison}
\end{figure*}

The right panel of Figure~\ref{fig:newdust} displays a histogram of the difference between the new method and Balmer decrement derived
 \ebvgas.
While there is no significant offset on average between \ebvgas\ derived using either method, there is significant scatter between the two.
The intrinsic scatter between the new and Balmer decrement \ebvgas\ is 0.23 magnitudes after accounting for measurement uncertainties.
This calibration scatter dominates over the formal uncertainties of the best-fit coefficients and is taken into account when
 estimating SFR, line ratio, and O/H uncertainties (Sec.~\ref{sec:reddening}).
The offset remains small with similar scatter in different bins of \mstar\ (colored histograms in Figure~\ref{fig:newdust}), demonstrating
 that the new reddening calibration will not bias the inferred SFR-\mstar\ relation and MZR.
In Figure~\ref{fig:newdustcomparison}, we compare \ebvgas, SFR(H$\alpha$), O$_{32}$, and O/H derived using this new reddening method
 to those values obtained using the Balmer decrement.
The average offset is small across all four properties, and no bias is present across the entire dynamic range in each panel.

\begin{figure}
 \centering
 \includegraphics[width=0.6\columnwidth]{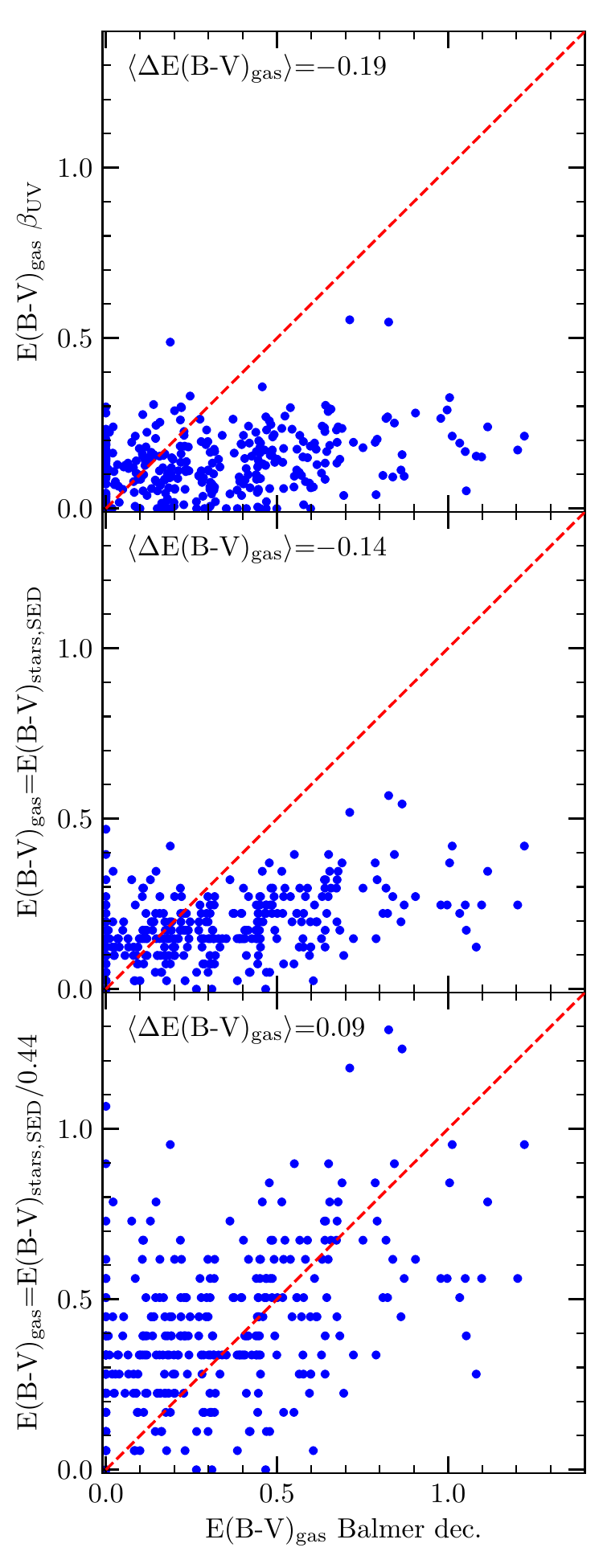}
 \centering
 \caption{
\ebvgas\ derived from \betauv (top), assuming \ebvgas=\ebvstars\ (middle), and assuming \ebvgas=\ebvstars/0.44 (bottom)
 vs.\ \ebvgas\ inferred from the Balmer decrement.
Red dashed lines show a one-to-one relation, and the mean vertical offset is given in the upper-left corner of each panel.
}\label{fig:otherdustcomparison}
\end{figure}

In Figure~\ref{fig:otherdustcomparison}, we compare \ebvgas\ derived from the Balmer decrement to three commonly adopted methods
 in high-redshift galaxy studies: (1) deriving \ebvgas\ from the rest-frame 1600~\AA\ slope, \betauv; (2) \ebvgas=\ebvstars; and
 (3) \ebvgas=\ebvstars/0.44.
In the first case, we estimated \betauv\ from the photometry \citep{red15,red18}, converted \betauv\ to A$_{\text{UV}}$
 using the relation of \citet{cal00}, translated A$_{\text{UV}}$ to \ebvstars\ assuming a \citet{cal00} attenuation curve,
 and then assumed \ebvgas=\ebvstars\ derived in this way.
In the second and third cases, we use \ebvstars\ from the best-fit SED model (Sec~\ref{sec:sedfitting}).
All three methods show a significant average offset, underestimating \ebvgas\ in cases 1 and 2 and overestimating it in case 3.
The first two cases perform well at low reddening (\ebvgas$<$0.25), but drastically underestimate \ebvgas\ with increasing
 severity as \ebvgas increases, leading to an underestimate of both SFR and metallicity (when based on [O\ii], [O\iii], and H$\beta$)
 for high-mass, dusty objects.
Using these two methods, both the SFR-\mstar\ relation and MZR are artificially flattened and have a lower normalization.
Assuming \ebvgas=\ebvstars/0.44 performs the best out of these methods, but still overestimates reddening for low-mass, dust-poor objects.
This tension alleviated if a steeper SMC-like curve is instead assumed for low-mass and low-metallicity galaxies \citep{shi20}.
In summary, our new \ebvgas\ calibration (equations~\ref{eq:newebvgas2} and~\ref{eq:newebvgas}) performs significantly better than all three
 of these commonly utilized methods, yielding unbiased SFR and O/H on average.

\section{B. Comparison to literature samples at $z>3$}\label{app:z3lit}

\begin{figure}
 \centering
 \includegraphics[width=\columnwidth]{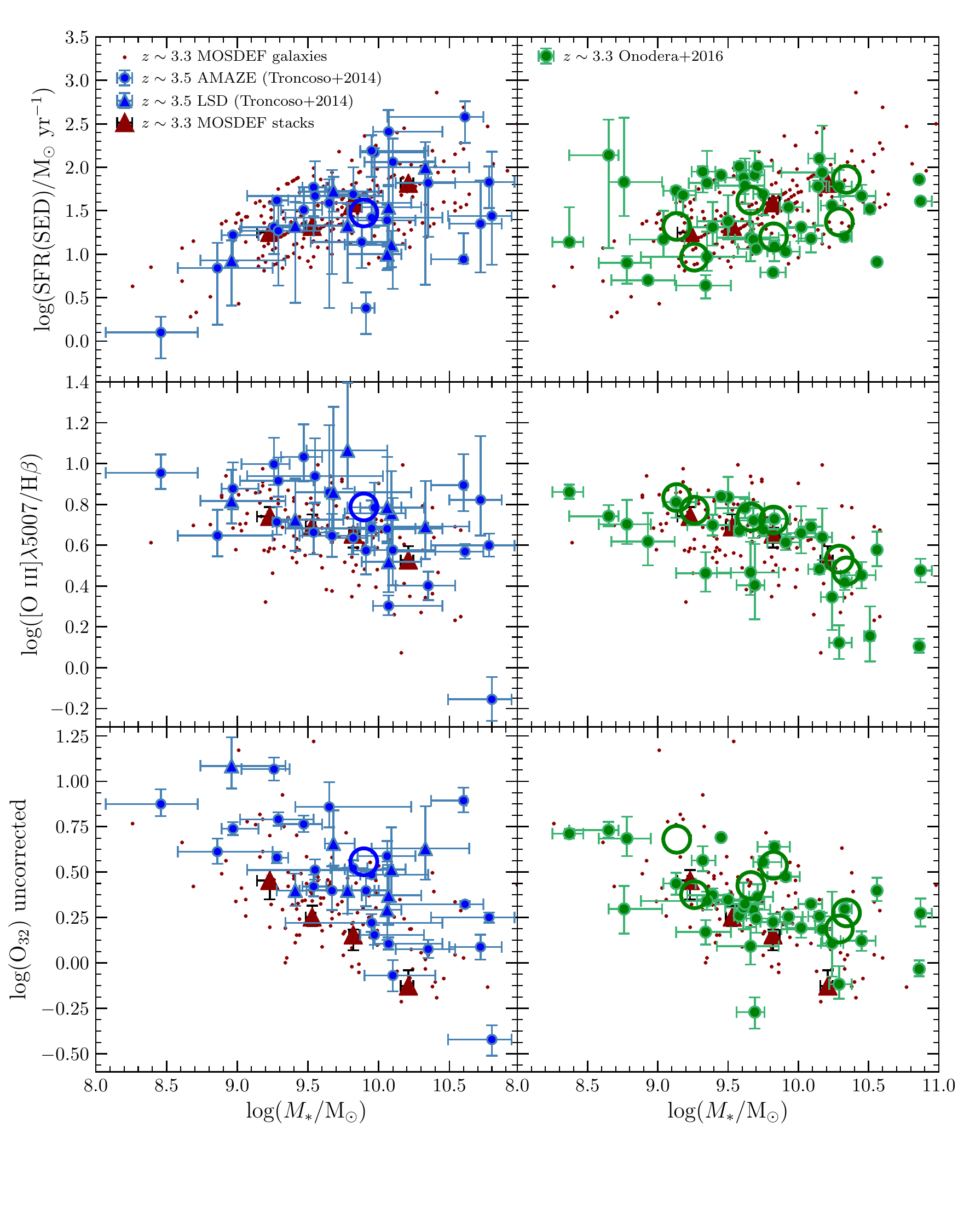}
 \centering
 \caption{
Comparison of the AMAZE+LSD $z\sim3.4$ sample \citep[left column, blue;][]{mai08,man09,tro14} and \citet{ono16} $z\sim3.3$ sample
 (right column, green) to the MOSDEF $z\sim3.3$ star-forming galaxies (red).
The panels display SFR(SED) (top), [O\iii]$\lambda$5007/H$\beta$ (middle), and O$_{32}$ uncorrected for reddening (bottom)
 vs.\ \mstar.
The open circle in the left column shows the composite spectrum of all AMAZE+LSD galaxies from \citet{tro14}.
The open circles in the right column displays the compoiste spectra in bins of \mstar\ and SFR from \citet{ono16}.
}\label{fig:z3lit}
\end{figure}

In Figure~\ref{fig:z3lit}, we compare the properties of the MOSDEF $z\sim3.3$ star-forming galaxy sample (red, both columns)
 with those of the AMAZE+LSD $z\sim3.4$ sample \citep[blue, left column;][]{mai08,man09,tro14} and the
 $z\sim3.3$ sample of \citet[][green, right column]{ono16}.
The top panel displays SFR(SED) vs.\ \mstar.
We choose to compare SFR(SED) instead of SFR derived from nebular lines because the latter depends on the method used to infer \ebvgas.
Both literature samples are well-matched to the SFR(SED)-\mstar\ relation defined by the MOSDEF galaxies, scattering around
 the MOSDEF stacks and displaying no obvious biases in SFR.
The middle and bottom panels show [O\iii]/H$\beta$ and O$_{32}$, respectively, as a function of \mstar.
These line ratios are sensitive to excitation and metallicity such that higher [O\iii]/H$\beta$ and O$_{32}$ corresponds
 to higher excitation and lower metallicity.
Here, O$_{32}$ is uncorrected for reddening in order to avoid differences in the inference of \ebvgas\ between the samples.
The \citet{ono16} sample generally follows the sequences described by the MOSDEF $z\sim3.3$ sample.
In contrast, the AMAZE+LSD sample has much higher [O\iii]/H$\beta$ and O$_{32}$ at fixed \mstar\ than the MOSDEF sample.
This offset cannot be driven by dust since [O\iii]/H$\beta$ is insensitive to reddening.
A consequence is that the AMAZE+LSD sample will have lower O/H at fixed \mstar\ than the other two samples when applying
 the same calibrations to all.

The origin of the high-excitation nature of the AMAZE+LSD sample is unclear, but is likely related to sample selection.
The AMAZE+LSD sample is rest-UV selected using a standard Lyman Break technique \citep{ste03}, and should be representative of the
 Lyman Break Galaxy population at $z\sim3$.
The \citet{ono16} sample, on the other hand, is primarily selected based on photometric redshifts using rest-UV to rest-NIR photometry,
 with an additional requirement that the expected H$\beta$ flux is $>5\times10^{-18}$~erg~s$^{-1}$~cm$^{-2}$.
Interestingly, the $z\sim2.3$ KBSS sample \citep[primarily rest-UV selected;][]{ste14} displays a larger offset than the
 $z\sim2.3$ MOSDEF sample (rest-optical selected) in the [N\ii] BPT diagram despite having similar SFR-\mstar\ distributions,
 suggesting that the mean excitation properties of LBGs may be different than rest-optical selected galaxies.
These differences highlight the impact that selection effects can have on determinations of the MZR at high redshifts.

\bibliography{ms}


\end{document}